\documentstyle[epsfig,12pt]{article}

\begin{document}

\onecolumn

\begin{titlepage}
\begin{center}
{\LARGE \bf Exact Solutions to the Motion of Two Charged Particles in Lineal 
Gravity}
\\ \vspace{2cm}
R.B. Mann\footnotemark\footnotetext{email: 
mann@avatar.uwaterloo.ca},
D. Robbins\footnotemark\footnotetext{email: 
dgr@gpu.srv.ualberta.ca} 
\\
\vspace{0.5cm} 
Dept. of Physics,
University of Waterloo
Waterloo, ONT N2L 3G1, Canada\\
\vspace{1cm}
T. Ohta \footnotemark\footnotetext{email:
t-oo1@ipc.miyakyo-u.ac.jp}\\
\vspace{0.5cm} 
Department of Physics, Miyagi University of Education,
Aoba-Aramaki, Sendai 980, Japan\\
\vspace{1cm}
and \\
M. Trott\footnotemark\footnotetext{email:mrtrott@physics.utoronto.ca}\\
\vspace{0.5cm} 
Dept. of Physics,
University of Toronto
Toronto, ONT, Canada\\
\vspace{1cm}
\vspace{2cm}
PACS numbers: 
13.15.-f, 14.60.Gh, 04.80.+z\\
\vspace{2cm}
\today\\
\end{center}

\begin{abstract}
We extend the canonical formalism for the motion of $N$-particles in  lineal 
gravity to include charges. Under suitable coordinate conditions and 
boundary conditions the Hamiltonian is defined as the spatial integral of 
the second derivative of the dilaton field which is given as a solution to the 
constraint equations. For a system of two particles the determining equation 
of the Hamiltonian (a kind of transcendental equation) is derived from the 
matching conditions for the dilaton field at the particles' position. The 
canonical equations of motion are derived from this determining equation.

For the equal mass case the canonical equations in terms of the proper time 
can be exactly solved in terms of hyperbolic and/or trigonometric functions. 
In electrodynamics with zero cosmological constant the trajectories for 
repulsive charges exhibit not only bounded motion but also a countably 
infinite series of unbounded motions for a fixed value of the total energy 
$H_{0}$, while for attractive charges the trajectories are simply 
periodic. When the cosmological constant $\Lambda$ is introduced, the motion 
for a given $\Lambda$ and $H_{0}$ is classified in terms of the 
charge-momentum diagram from which we can predict what type of the motion is 
realized for a given charge. 
In general the cosmological constant acts on the particles as a 
repulsive ($\Lambda>0$) or an attractive ($\Lambda<0$) potential.
But for a certain range of $\Lambda<0$, small $q$ and small mass the 
trajectory shows a double peak structure due to an interplay among the induced 
momentum dependent $\Lambda$ potential, the gravitational attraction and the 
relativistic effect. 
For $\Lambda>0$, depending on the value of charge, only bounded motion or 
the infinite series of unbounded motion or both are realized. 

Since in this theory the charge of each particle appears in the form
$e_{1}e_{2}$ in the determining equation, the static balance condition in 
1+1 dimensions turns out to be identical with the condition in Newtonian 
theory. We generalize this condition to non-zero momenta, obtaining the first
exact solution to the static balance problem that does not obey the 
Majumdar-Papapetrou condition.

\end{abstract}
\end{titlepage}
\onecolumn

\section{INTRODUCTION}

The problem of determining the motion of a system of $N$ particles mutually
interacting through specified forces is one that has attracted attention
since the dawn of physics. We continue in this paper our previous
explorations of this problem in two spacetime dimensions when the specified
interactions are both gravitational and electromagnetic.

Our work here represents the first exact solution to include both
interactions in a relativistic framework. Although an exact solution is
known in three spatial dimensions for pure Newtonian gravity in the $N=2$
case, dissipation of energy in the form of gravitational radiation has
obstructed progress toward obtaining exact solutions for the motion of $N$
bodies in the general theory of relativity, even when $N=2$. \ However by
reducing the number of spatial dimensions this obstruction disappears, at
least in the vacuum. Apart from the absence of gravitational radiation, most
(if not all) of the remaining conceptual features of relativistic gravity
are retained, and so lower dimensional theories of gravity offer the hope of
garnering insight into the nature of classical and quantum gravitation in a
wide variety of physical situations.

For these reasons we have been investigating the $N$-body problem in two
dimensional gravity for the past 3 years. We have chosen to work with a 2D
theory that models 4D general relativity in that it sets the Ricci scalar
equal to the trace of the stress-energy of prescribed matter fields and
sources. Hence matter governs the evolution of spacetime curvature which
reciprocally governs the evolution of matter \cite{r3}. This theory
(sometimes referred to as $R=T$ theory) has a consistent Newtonian limit 
\cite{r3}, a problematic limit in a generic $(1+1)$-dimensional theory of
gravity theory \cite{jchan}. When the stress-energy is that of a
cosmological constant, the theory reduces to Jackiw-Teitelboim (JT) theory 
\cite{JT}.

Working in the canonical formalism \cite{OR}, we have so far been able to
obtain exact solutions to the two body problem in the absence \cite{2bd} and
presence \cite{2bdcossh} of a cosmological constant. In this paper we extend
these considerations to include charged bodies. \ Specifically, we formulate
the charged $N$-body problem in relativistic gravity by taking the matter
action to be that of $N$ charged point-particles minimally coupled to
gravity and electromagnetism. We extend the previous canonical formalism we
developed for this action in $R=T$ lineal gravity \cite{OR} to include this
case. When $N=2$ we obtain exact solutions for the motion of two bodies of
unequal (and equal) charge and mass. In the slow motion, weak field limit
the Hamiltonian we obtain coincides with that of Newtonian gravity with
electromagnetism in $(1+1)$ dimensions. \ We are also able to extend our
solutions to include a cosmological constant $\Lambda $, so that in the
limit where all bodies are massless and neutral, spacetime has constant
curvature (ie the JT theory is obtained). Our solution is the first
non-static exact solution to the charged 2-body problem in any relativistic
theory of gravity. 

Our exact solution to the $N=2$ case can be formulated in several ways. We
derive an exact solution for the Hamiltonian as a function of the proper
separation and the centre-of-inertia momentum of the charged bodies. We are
also able to construct a solution in which the proper separation of the two
charged point masses is given as a function of their mutual proper time in
the equal mass case. If the masses are not equal our exact solution is given
in terms of a time-coordinate that is not the proper time.

A scalar (dilaton) field must be included in the action \cite{BanksMann}
since the Einstein action is a topological invariant in (1+1) dimensions.
Canonically reducing the action, we find that the Hamiltonian is given in
terms of a spatial integral of the second derivative of the dilaton field,
which is a function of the canonical variables of the particles (coordinates
and momenta) and is determined from the constraint equations. The matching
conditions of the solution to the constraint equations yield an equation
which determines the Hamiltonian in terms of the remaining degrees of
freedom of the system when $N=2$ . We refer to this transcendental equation
as the determining equation, since it allows one to determine the
Hamiltonian in terms of the centre of inertia momentum and proper separation
of the bodies. From this Hamiltonian we can derive the canonical equations
of motion. In the equal mass case we find that the separation and momentum
are given by differential equations in terms of the proper time, and can be
exactly solved in terms of hyperbolic and/or trigonometric functions.

Several different types of motion are expected in the 2 body case, depending
upon the signs and magnitudes of the masses, charges, energy and other
parameters (e.g. gravitational coupling constant, cosmological constant). If
the charges are of opposite sign the particles will remain bounded, whereas
if they are of the same sign either bounded or unbounded motion can occur
for the same value of the total energy. For a given set of parameters there
is in this case a countably infinite series of unbounded motions labelled by
an integer $n$. A balance condition exists between the bounded and the
unbounded cases, and reduces to the expected (Newtonian-like) static balance
condition in the absence of particle momenta. We shall analyze these various
states of motion, and discuss the transitions which occur between them. A
cosmological constant can qualitatively change the motion, rendering bound
states unbound and vice-versa. We find several surprising situations,
including the diverging separation of the two bodies at finite proper time
in the repulsive case, even for $\Lambda =0$. In the $\Lambda <0$ case the
motion shows a double maximum behavior for a certain range of the
parameters, which is a characteristic effect of the induced momentum
dependent $\Lambda $ potential. For classification of the motion we utilize
a charge-momentum diagram from which we can easily predict what type of the
motion is realized for a given charge.

In the unequal mass case the proper time is no longer the same for the two
particles, and a more careful analysis is necessary in order to describe the
motion. We obtain phase space trajectories from the determining equation and
explicit solutions for the proper separation in terms of a transformed time
coordinate which reduces to the mutual proper time in the case of equal mass.

In Sec.II we describe the outline of the canonical reduction of the theory
when the charges are included and define the Hamiltonian for the $N$- body
system. In Sec.III we solve the constraint equations for the two-body case
and get the determining equation of the Hamiltonian, from which the
canonical equations of motion are explicitly derived. For the case of equal
masses and arbitrary charges the explicit exact solutions to the canonical
equations are given in Sec.IV. By using these exact solutions we analyze in
Sec.V the motion for $\Lambda =0$ separately for four cases: attractive
charges, small repulsive charges, large repulsive charges and $H<2m$. We
analyze the motion of equal masses for $\Lambda \neq 0$ in Sec.VI for four
possible combinations of the signs of $\Lambda $ and the charges, where we
also develop the plots of phase space trajectories and the analysis of the
explicit solutions in terms of the proper time, based on a classification
given by a charge-momentum diagram. We treat the unequal mass case in
Sec.VII. In Sec.VIII we investigate the static balance problem by using both
the canonical equation and the determining equation. Sec.IX contains
concluding remarks and directions for further work. The solution of the
metric tensor, a linear approximation of the exact solutions and the causal
relationships between particles in unbounded motion are given in Appendices.

\section{CANONICALLY REDUCED HAMILTONIAN OF $N$-CHARGED PARTICLES}

Our derviation of the canonically reduced Hamiltonian for charged particles
is parallel to that given in the uncharged case \cite{2bd,2bdcoslo}. Here we
briefly review this work, highlighting those aspects that are peculiar to
the charged case.

The action integral for gravitational and electromagnetic fields coupled
with $N$ charged point masses is 
\begin{eqnarray}
I &=&\int d^{2}x\left[ \frac{1}{2\kappa }\sqrt{-g}g^{\mu \nu }\left\{ \Psi
R_{\mu \nu }+\frac{1}{2}\nabla _{\mu }\Psi \nabla _{\nu }\Psi +\frac{1}{2}%
g_{\mu \nu }\Lambda \right\} \right.   \nonumber  \label{act1} \\
&&\makebox[2em]{}\left. -A_{\mu }{\cal F}_{\;\;,\nu }^{\mu \nu }+\frac{1}{4%
\sqrt{-g}}{\cal F}^{\mu \nu }{\cal F}^{\alpha \beta }g_{\mu \alpha }g_{\nu
\beta }\right.   \nonumber \\
&&\makebox[2em]{}\left. +\sum_{a}\int d\tau _{a}\left\{ -m_{a}\left( -g_{\mu
\nu }(x)\frac{dz_{a}^{\mu }}{d\tau _{a}}\frac{dz_{a}^{\nu }}{d\tau _{a}}%
\right) ^{1/2}+e_{a}\frac{dz_{a}^{\mu }}{d\tau _{a}}A_{\mu }(x)\right\}
\delta ^{2}(x-z_{a}(\tau _{a}))\right] \;,
\end{eqnarray}
where $\Psi $ is the dilaton field, $A_{\mu }$ and ${\cal F}^{\mu \nu }$ are
the vector potential and the field strength, $g_{\mu \nu }$ and $g$ are the
metric and its determinant, $R$ is the Ricci scalar, and $e_{a}$ and $\tau
_{a}$ are the charge and the proper time of $a$-th particle, respectively,
with $\kappa =8\pi G/c^{4}$. The symbol $\nabla _{\mu }$ denotes the
covariant derivative associated with $g_{\mu \nu }$.

\bigskip The field equations derived from the action (\ref{act1}) are 
\begin{eqnarray}
&&R-g^{\mu \nu }\nabla _{\mu }\nabla _{\nu }\Psi =0\;,  \label{eq-R} \\
&&\frac{1}{2}\nabla _{\mu }\Psi \nabla _{\nu }\Psi -\frac{1}{4}g_{\mu \nu
}\nabla ^{\lambda }\Psi \nabla _{\lambda }\Psi +g_{\mu \nu }\nabla ^{\lambda
}\nabla _{\lambda }\Psi -\nabla _{\mu }\nabla _{\nu }\Psi =\kappa T_{\mu \nu
}+\frac{1}{2}g_{\mu \nu }\Lambda \;,  \label{eq-Psi} \\
&&{\cal F}_{\;\;,\nu }^{\mu \nu }=\sum_{a}e_{a}\int d\tau _{a}\frac{%
dz_{a}^{\mu }}{d\tau _{a}}\delta ^{2}(x-z_{a}(\tau _{a}))\;,  \label{eq-f} \\
&&\frac{1}{\sqrt{-g}}{\cal F}_{\mu \nu }=\partial _{\mu }A_{\nu }-\partial
_{\nu }A_{\mu }\;,  \label{eq-a} \\
&&m_{a}\left[ \frac{d}{d\tau _{a}}\left\{ g_{\mu \nu }(z_{a})\frac{%
dz_{a}^{\nu }}{d\tau _{a}}\right\} -\frac{1}{2}g_{\nu \lambda ,\mu }(z_{a})%
\frac{dz_{a}^{\nu }}{d\tau _{a}}\frac{dz_{a}^{\lambda }}{d\tau _{a}}\right]
=e_{a}\frac{dz_{a}^{\nu }}{d\tau _{a}}\left\{ A_{\nu ,\mu }(z_{a})-A_{\mu
,\nu }(z_{a})\right\} \;,  \label{eq-z}
\end{eqnarray}
where the stress-energy due to the point masses and the electric field is 
\begin{equation}
T_{\mu \nu }=\sum_{a}m_{a}\int d\tau _{a}\frac{1}{\sqrt{-g}}g_{\mu \sigma
}g_{\nu \rho }\frac{dz_{a}^{\sigma }}{d\tau _{a}}\frac{dz_{a}^{\rho }}{d\tau
_{a}}\delta ^{2}(x-z_{a}(\tau _{a}))+\frac{1}{(-g)}\left\{ {\cal F}_{\mu
\alpha }{\cal F}_{\nu \beta }g^{\alpha \beta }-\frac{1}{4}g_{\mu \nu }{\cal F%
}_{\alpha \beta }{\cal F}^{\alpha \beta }\right\} \;,
\end{equation}
recalling that in (1+1) dimensions no magnetic component of the field
exists. Eq.(\ref{eq-Psi}) guarantees the conservation of $T_{\mu \nu }$.
Inserting the trace of Eq.(\ref{eq-Psi}) into Eq.(\ref{eq-R}) yields 
\begin{equation}
R-\Lambda =\kappa T_{\;\;\mu }^{\mu }\;.  \label{RT}
\end{equation}
Eqs.(\ref{eq-f}), (\ref{eq-a}), (\ref{eq-z}) and (\ref{RT}) form a closed
sytem of equations for gravity, electromagnetism, and matter.

On transforming the action (\ref{act1}) to the canonical form we first
rewrite the electromagnetic and the particle sectors in a form appropriate
to the first-order formalism \cite{admV}. 
\begin{eqnarray}
I_{E+P} &=&\int dx^{2}\left[ -A_{\mu }{\cal F}_{\;\;,\nu }^{\mu \nu }+\frac{1%
}{4\sqrt{-g}}{\cal F}^{\mu \nu }{\cal F}^{\alpha \beta }g_{\mu \alpha
}g_{\nu \beta }\right.   \nonumber  \label{actEP} \\
&&\left. +\sum_{a}\int d\tau _{a}\left\{ \pi _{a\mu }\frac{dz_{a}^{\mu }}{%
d\tau _{a}}-\frac{1}{2}\lambda _{a}^{\prime }(\tau _{a})(\pi _{a\mu }\pi
_{a\nu }g^{\mu \nu }+m_{a}^{2})+e_{a}\frac{dz_{a}^{\mu }}{d\tau _{a}}A_{\mu
}(x)\right\} \delta ^{2}(x-z_{a}(\tau _{a}))\right] \;,
\end{eqnarray}
where $\pi _{a}^{\mu }$ is essentially the four velocity of the particle and 
$\lambda _{a}^{\prime }$ is a Lagrange multiplier. The variations with
respect to $A_{\mu }$ and ${\cal F}^{\mu \nu }$ lead to Eqs.(\ref{eq-f}) and
(\ref{eq-a}), respectively, and those with respect to $\pi _{a\mu
},z_{a}^{\mu }$ and $\lambda _{a}^{\prime }$ lead to 
\begin{eqnarray}
&&\frac{dz_{a}^{\mu }}{d\tau _{a}}-\lambda _{a}^{\prime }\pi _{a}^{\mu }=0\;,
\label{eq-ep1} \\
&&\frac{d\pi _{a\mu }}{d\tau _{a}}+\frac{1}{2}\lambda _{a}^{\prime }\pi
_{a\lambda }\pi _{a\nu }g_{\;\;,\mu }^{\lambda \nu }(z_{a})=e_{a}\frac{%
dz_{a}^{\nu }}{d\tau _{a}}\left\{ A_{\nu ,\mu }(z_{a})-A_{\mu ,\nu
}(z_{a})\right\} \;,  \label{eq-ep2} \\
&&\pi _{a\mu }\pi _{a\nu }g^{\mu \nu }+m_{a}^{2}=0\;.  \label{eq-ep3}
\end{eqnarray}
This set of three equations is equivalent to Eq.(\ref{eq-z}) when $\lambda
_{a}^{\prime }=1/m_{a}$ is chosen. \newline
By performing the integration over the parameter $\tau _{a}$ and setting 
\begin{eqnarray}
z_{a}^{\mu } &=&(t,z_{a})\;,\qquad \pi _{a\mu }=(\pi _{a0},\pi _{a})\;, 
\nonumber \\
A_{\mu } &=&(-\varphi ,A)\;,\qquad E={\cal F}^{01}\;, \\
\lambda _{a} &=&\lambda _{a}^{\prime }\frac{d\tau _{a}}{dz_{a}^{0}}%
|_{z_{a}^{0}(\tau _{a})=t}\;.
\end{eqnarray}
the action (\ref{actEP}) becomes 
\begin{eqnarray}
I_{E+P} &=&\int d^{2}x\left[ A\frac{\partial E}{\partial t}-\frac{1}{2}\sqrt{%
-g}E^{2}+\varphi \left\{ \frac{\partial E}{\partial x}-\sum_{a}e_{a}\delta
(x-z_{a}(t))\right\} \right.   \label{actEP2} \\
&&\left. +\sum_{a}\frac{dz_{a}}{dt}(\pi _{a}+e_{a}A)\delta
(x-z_{a}(t))+\sum_{a}\left\{ \pi _{a0}-\frac{1}{2}\lambda (\pi _{a\mu }\pi
_{a\nu }g^{\mu \nu }+m_{a}^{2})\right\} \delta (x-z_{a}(t))\right] \;. 
\nonumber
\end{eqnarray}
Varying the Lagrange multipliers $\varphi $ and $\lambda _{a}$ yields the
constraints 
\begin{eqnarray}
&&\frac{\partial E}{\partial x}=\sum_{a}e_{a}\delta (x-z_{a}(t))\;,
\label{eq-E} \\
&&\pi _{a\mu }\pi _{a\nu }g^{\mu \nu }+m_{a}^{2}=0\;.  \label{eq-lambda}
\end{eqnarray}
The solution to (\ref{eq-E}) is 
\begin{equation}
E=\frac{1}{2}\partial _{x}\sum_{a}e_{a}|x-z_{a}(t)|\;.  \label{sol-E}
\end{equation}
In (1+1) dimensions the electric field has no independent degrees of
freedom. When the charged particles move within a finite region $(|x|<L)$,
the electric field in an outer region $(|x|>L)$ is $E=\pm \frac{1}{2}%
\sum_{a}e_{a}$. For a system of zero total charge the electric field
vanishes in the outer region. As the solution to (\ref{eq-lambda}) we choose 
\begin{equation}
\pi _{a0}=\frac{1}{g^{00}}\left\{ -g^{01}\pi _{a}+\sqrt{(g^{01}\pi
_{a})^{2}-g^{00}(g^{11}\pi _{a}^{2}+m_{a}^{2})}\right\} \;.
\label{sol-eqlambda}
\end{equation}
After the constraints are eliminated, the action (\ref{actEP2}) is 
\begin{eqnarray}
I_{E+P} &=&\int d^{2}x\left[ \sum_{a}\pi _{a}\frac{dz_{a}}{dt}\delta
(x-z_{a}(t))\right.   \label{actEP3} \\
&&\left. +\sum_{a}\frac{1}{g^{00}}\left\{ -g^{01}\pi _{a}+\sqrt{(g^{01}\pi
_{a})^{2}-g^{00}(g^{11}\pi _{a}^{2}+m_{a}^{2})}\right\} \delta (x-z_{a}(t))-%
\frac{1}{2}\sqrt{-g}E^{2}(t,x)\right] \;.  \nonumber
\end{eqnarray}
{} From this expression we know that $\pi _{a}$ is the conjugate momentum to 
$z_{a}$ and hereafter we use the notation 
\[
p_{a}\equiv \pi _{a}\;.
\]

Note that the variations of the action (\ref{actEP2}) with respect to $A$
and $E$ lead to 
\begin{eqnarray}
&&\makebox[3em]{}\frac{\partial E}{\partial t}+\sum_{a}e_{a}\frac{dz_{a}}{dt}%
\delta (x-z_{a}(t))=0\;,  \label{eq-Et} \\
&&\makebox[3em]{}\sqrt{-g}E=-\frac{\partial \varphi }{\partial x}-\frac{%
\partial A}{\partial t}\;.  \label{eq-E2}
\end{eqnarray}
Eq.(\ref{eq-Et}) is automatically satisfied for the solution (\ref{sol-E})
to the constraint equation, in contrast to the (3+1) dimensional case where
the corresponding equation is a true dynamical equation. Upon inserting the
solution (\ref{sol-E}) into the action (\ref{actEP2}) all terms related to
the components of $A_{\mu }$ cancel. Hence we need no longer consider $%
A_{\mu }$. Its solution, if desired, is straightforwardly obtained by
solving (\ref{eq-E}) after fixing the gauge and obtaining the full solution
of the metric.

Consider next the transformation of the gravity sector to canonical form. We
decompose the scalar curvature in terms of the extrinsic curvature $K$ via 
\begin{equation}
\sqrt{-g}R=-2\partial _{0}(\sqrt{\gamma }K)+2\partial _{1}[(N_{1}K-\partial
_{1}N_{0})/\sqrt{\gamma }]\;,  \label{extK}
\end{equation}
where the metric is 
\begin{equation}
ds^{2}=-N_{0}^{2}dt^{2}+\gamma \left( dx+\frac{N_{1}}{\gamma }dt\right)
^{2}\;,  \label{lineel}
\end{equation}
with $K=(2N_{0}\gamma )^{-1}(2\partial _{1}N_{1}-\gamma ^{-1}N_{1}\partial
_{1}\gamma -\partial _{0}\gamma )$, so that $\gamma
=g_{11},N_{0}=(-g^{00})^{-1/2}$ and $N_{1}=g_{10}$. We then rewrite the
gravity sector in first-order form . We find that the action (\ref{act1})
becomes 
\begin{equation}
I=\int d^{2}x\left\{ \sum_{a}p_{a}\dot{z}_{a}\delta (x-z_{a}(t))+\pi \dot{%
\gamma}+\Pi \dot{\Psi}+N_{0}R^{0}+N_{1}R^{1}\right\}\;,  \label{act2}
\end{equation}
where $\pi $ and $\Pi $ are conjugate momenta to $\gamma $ and $\Psi $,
respectively, and 
\begin{eqnarray}
R^{0} &=&-\kappa \sqrt{\gamma }\gamma \pi ^{2}+2\kappa \sqrt{\gamma }\pi \Pi
+\frac{1}{4\kappa \sqrt{\gamma }}(\Psi ^{\prime })^{2}-\frac{1}{\kappa }%
\left( \frac{\Psi ^{\prime }}{\sqrt{\gamma }}\right) ^{\prime }-\frac{1}{2}%
\sqrt{\gamma }(E^{2}-\frac{\Lambda }{\kappa })  \nonumber \\
&&-\sum_{a}\sqrt{\frac{p_{a}^{2}}{\gamma }+m_{a}^{2}}\;\delta (x-z_{a}(t))\;,
\nonumber  \label{R0} \\
R^{1} &=&\frac{\gamma ^{\prime }}{\gamma }\pi -\frac{1}{\gamma }\Pi \Psi
^{\prime }+2\pi ^{\prime }+\sum_{a}\frac{p_{a}}{\gamma }\delta
(x-z_{a}(t))\;,  \label{R1}
\end{eqnarray}
with the symbols $(\;\dot{}\;)$ and $(\;^{\prime }\;)$ denoting $\partial
_{0}$ and $\partial _{1}$, respectively.

{}From the action (\ref{act2}) we obtain the set of equations 
\begin{eqnarray}
\dot{\pi} &+&N_{0}\left\{ \frac{3\kappa }{2}\sqrt{\gamma }\pi ^{2}-\frac{%
\kappa }{\sqrt{\gamma }}\pi \Pi +\frac{1}{8\kappa \sqrt{\gamma }\gamma }%
(\Psi ^{\prime })^{2}+\frac{1}{4\sqrt{\gamma }}(E^{2}-\frac{\Lambda }{\kappa 
})\right.  \nonumber  \label{e-pi} \\
&&\makebox[10em]{}\left. -\sum_{a}\frac{p_{a}^{2}}{2\gamma ^{2}\sqrt{\frac{%
p_{a}^{2}}{\gamma }+m_{a}^{2}}}\;\delta (x-z_{a}(t))\right\}  \nonumber \\
&+&N_{1}\left\{ -\frac{1}{\gamma ^{2}}\Pi \Psi ^{\prime }+\frac{\pi ^{\prime
}}{\gamma }+\sum_{a}\frac{p_{a}}{\gamma ^{2}}\;\delta (x-z_{a}(t))\right\}
+N_{0}^{\prime }\frac{1}{2\kappa \sqrt{\gamma }\gamma }\Psi ^{\prime
}+N_{1}^{\prime }\frac{\pi }{\gamma }=0\;,
\end{eqnarray}
\begin{eqnarray}
&&\dot{\gamma}-N_{0}(2\kappa \sqrt{\gamma }\gamma \pi -2\kappa \sqrt{\gamma }%
\Pi )+N_{1}\frac{\gamma ^{\prime }}{\gamma }-2N_{1}^{\prime }=0\;,
\label{e-gamma} \\
&&R^{0}=0\;,  \label{e-R0} \\
&&R^{1}=0\;,  \label{e-R1} \\
&&\dot{\Pi}+\partial _{1}(-\frac{1}{\gamma }N_{1}\Pi +\frac{1}{2\kappa \sqrt{%
\gamma }}N_{0}\Psi ^{\prime }+\frac{1}{\kappa \sqrt{\gamma }}N_{0}^{\prime
})=0\;,  \label{e-Pi} \\
&&\dot{\Psi}+N_{0}(2\kappa \sqrt{\gamma }\pi )-N_{1}(\frac{1}{\gamma }\Psi
^{\prime })=0\;,  \label{e-Psi} \\
&&\dot{p}_{a}+\frac{\partial N_{0}}{\partial z_{a}}\sqrt{\frac{p_{a}^{2}}{%
\gamma }+m_{a}^{2}}-\frac{N_{0}}{2\sqrt{\frac{p_{a}^{2}}{\gamma }+m_{a}^{2}}}%
\frac{p_{a}^{2}}{\gamma ^{2}}\frac{\partial \gamma }{\partial z_{a}}-\frac{%
\partial N_{1}}{\partial z_{a}}\frac{p_{a}}{\gamma }  \nonumber \\
&&\makebox[2em]{}+N_{1}\frac{p_{a}}{\gamma ^{2}}\frac{\partial \gamma }{%
\partial z_{a}}+\int dxN_{0}\sqrt{\gamma }E\frac{\partial E}{\partial z_{a}}%
=0\;,  \label{e-p} \\
&&\dot{z_{a}}-N_{0}\frac{\frac{p_{a}}{\gamma }}{\sqrt{\frac{p_{a}^{2}}{%
\gamma }+m_{a}^{2}}}+\frac{N_{1}}{\gamma }=0\;.  \label{e-z}
\end{eqnarray}
In the equations (\ref{e-p}) and (\ref{e-z}), all metric components ($N_{0}$%
, $N_{1}$, $\gamma $) are evaluated at the point $x=z_{a}$ and 
\[
\frac{\partial f}{\partial z_{a}}\equiv \left. \frac{\partial f(x)}{\partial
x}\right| _{x=z_{a}}\;. 
\]
The quantities $N_{0}$ and $N_{1}$ are Lagrange multipliers which yield the
constraint equations (\ref{e-R0}) and (\ref{e-R1}). The above set of
equations can be proved to be equivalent to the set of equations (\ref{eq-R}%
), (\ref{eq-Psi}) and (\ref{eq-z}).

To proceed to the canonical reduction of the action (\ref{act2}) we have to
eliminate the redundant variables by utilizing the constrant equations to
fix the coordinate conditions. Noticing that the only linear terms in the
constraint equations (\ref{e-R0}) and (\ref{e-R1}) are $\left( \Psi ^{\prime
}/\sqrt{\gamma }\right) ^{\prime }$ and $\pi ^{\prime }$, respectively, and
the equations may be solved for these quantities, we can transform the total
generator obtained from the end point variation into an approriate form to
fix the coordinate conditions. Generalizing the procedure described in our
previous papers \cite{OR,2bd} for the case of $\Lambda =e_{a}=0$, we find
that we can consistently choose the coordinate conditions{\bf \ } 
\begin{equation}
\gamma =1\makebox[2em]{}\mbox{and}\makebox[2em]{}\Pi =0\;.  \label{cc}
\end{equation}
Eliminating the constraints, the action (\ref{act2}) reduces to 
\begin{equation}
I=\int dx^{2}\left\{ \sum_{a}p_{a}\dot{z}_{a}\delta (x-z_{a})-{\cal H}%
\right\} \;,
\end{equation}
where the reduced Hamiltonian for the system of particles is defined by 
\begin{equation}
H=\int dx{\cal H=-}\frac{1}{\kappa }\int {\cal \triangle }\Psi \;.
\label{ham1}
\end{equation}
Here $\Psi $ is a function of $z_{a}$ and $p_{a}$ and is determined by
solving the constraints which under the coordinate conditions (\ref{cc})
become 
\begin{equation}
\triangle \Psi -\frac{1}{4}(\Psi ^{\prime })^{2}+\kappa ^{2}\pi ^{2}+\frac{1%
}{2}(\kappa E^{2}-\Lambda )+\kappa \sum_{a}\sqrt{p_{a}^{2}+m_{a}^{2}}\delta
(x-z_{a})=0\;,  \label{Psi}
\end{equation}
\begin{equation}
2\pi ^{\prime }+\sum_{a}p_{a}\delta (x-z_{a})=0\;.  \label{pi}
\end{equation}

The consistency of this canonical reduction is proved in an analogous way to
the case of $\Lambda=e_{a}=0$ : namely the canonical equations of motion
derived from the reduced Hamiltonian (\ref{ham1}) are identical with the
equations (\ref{e-p}) and (\ref{e-z}) \cite{2bd,2bdcoslo}.

\section{SOLUTION TO THE CONSTRAINT EQUATIONS AND THE HAMILTONIAN FOR A
SYSTEM OF TWO PARTICLES}

The standard approach for investigating the dynamics of particles is to get
first an explicit expression of the Hamiltonian and to derive the equations
of motion, from which the solution of trajectories are obtained. In this
section we show how to derive the Hamiltonian from the solution to the
constraint equations (\ref{Psi}) and (\ref{pi}) and get the explicit
Hamiltonian for a system of two charged particles. Since the electric field
appears in the combination $(E^{2}-\Lambda /\kappa )$ in all equations we
set 
\begin{equation}
V(x)\equiv E^{2}-C\quad \mbox{and}\quad \Lambda _{e}\equiv \Lambda -\kappa C
\end{equation}
with $C\equiv \frac{1}{4}(\sum_{a}e_{a})^{2}$. Thus $V(x)$ vanishes in the
outer region and $\Lambda _{e}$ is an effective cosmological constant, which
includes the contribution from the electric field. This latter situation
arises from the well-known fact that in (1+1) dimensions the electromagnetic
field strength is a 2-form, and so in compact spatial regions it contributes
to the stress-energy tensor in the same manner as a cosmological constant,
analogous to the way in which a 4-form behaves in 3+1 dimensions \cite
{aurilia}. We shall later see that when $\Lambda _{e}$ vanishes we get the
Hamiltonian which leads, in the limit $\kappa \rightarrow 0$, to the correct
special-relativistic electrodynamics in (1+1) flat space-time.

We express equations (\ref{Psi}) and (\ref{pi}) as 
\begin{equation}
\triangle \Psi =\frac{1}{4}\left( \Psi ^{\prime }\right) ^{2}-\kappa
^{2}\left( \chi ^{\prime }\right) ^{2}-\frac{1}{2}(\kappa V-\Lambda
_{e})-\kappa \sum_{a}\sqrt{p_{a}^{2}+m_{a}^{2}}\delta (x-z_{a})\;,
\label{Psi1}
\end{equation}
\begin{equation}
\triangle \chi =-\frac{1}{2}\sum_{a}p_{a}\delta (x-z_{a})\;,  \label{chi1}
\end{equation}
where we set $\chi ^{\prime }=\pi $. Rewriting (\ref{Psi1}) as 
\begin{equation}
(1+\frac{\Psi }{4})\triangle \Psi =\frac{1}{8}\triangle (\Psi ^{2}-4\kappa
^{2}\chi ^{2})-\frac{1}{2}(\kappa V-\Lambda _{e})+\kappa ^{2}\chi \triangle
\chi -\kappa \sum_{a}\sqrt{p_{a}^{2}+m_{a}^{2}}\delta (x-z_{a})\;,
\label{Psi1a}
\end{equation}
yields, upon insertion into the Right-hand side (RHS) of (\ref{ham1}) 
\begin{eqnarray}
H &=&\sum_{a}\frac{\sqrt{p_{a}^{2}+m_{a}^{2}}}{1+\frac{1}{4}\Psi (z_{a})}+%
\frac{\kappa }{2}\sum_{a}\frac{p_{a}\chi (z_{a})}{1+\frac{1}{4}\Psi (z_{a})}+%
\frac{1}{2}\int dx\frac{V(x)}{1+\frac{1}{4}\Psi (x)}  \nonumber  \label{ham2}
\\
&&-\frac{1}{8\kappa }\int dx\frac{1}{1+\frac{1}{4}\Psi (x)}\triangle \left(
\Psi ^{2}-4\kappa ^{2}\chi ^{2}+2\Lambda _{e}x^{2}\right) \;.
\end{eqnarray}
an expression which can also be obtained repeated iteration of the insertion
of the RHS of (\ref{Psi1}) into the RHS of (\ref{ham1}).

Defining $\phi $ by 
\begin{equation}
\Psi =-4\ln |\phi |\;,
\end{equation}
the constraints (\ref{Psi1}) and (\ref{chi1}) for a two-particle system
become 
\begin{eqnarray}
\triangle \phi -\frac{1}{4}\left\{ \kappa ^{2}\left( \chi ^{\prime }\right)
^{2}+\frac{\kappa }{2}V-\frac{1}{2}\Lambda _{e}\right\} \phi &=&\frac{\kappa 
}{4}\left\{ \sqrt{p_{1}^{2}+m_{1}^{2}}\;\phi (z_{1})\delta (x-z_{1})\right. 
\nonumber \\
&&\left. \quad +\sqrt{p_{2}^{2}+m_{2}^{2}}\;\phi (z_{2})\delta
(x-z_{2})\right\} \;,  \label{phi-eq}
\end{eqnarray}
\begin{equation}
\triangle \chi =-\frac{1}{2}\left\{ p_{1}\delta (x-z_{1})+p_{2}\delta
(x-z_{2})\right\} \;.  \label{chi-eq}
\end{equation}
The general solution to (\ref{chi-eq}) is 
\begin{equation}
\chi =-\frac{1}{4}\left\{ p_{1}\mid x-z_{1}\mid +p_{2}\mid x-z_{2}\mid
\right\} -\epsilon Xx+\epsilon C_{\chi }\;\;.  \label{chi-sol}
\end{equation}
The factor $\epsilon $ ($\epsilon ^{2}=1$) has been introduced in the
constants $X$ and $C_{\chi }$ so that the T-inversion (time reversal)
properties of $\chi $ are explicitly manifest. By definition, $\epsilon $
changes sign under time reversal and so, therefore, does $\chi $.

Consider first the case $z_{2}<z_{1}$, for which we may divide space into
three regions: $z_{1}<x$ ((+) region), $z_{2}<x<z_{1}$ ((0) region) and $%
x<z_{2}$ ((-) region). In each region, $V$ and $\chi^{\prime}$ are constant: 
\begin{equation}
V=\left\{ 
\begin{array}{ll}
0 & \qquad \mbox{(+) region}, \\ 
-\frac{1}{2}e_{1}e_{2} & \qquad \mbox{(0) region}, \\ 
0 & \qquad \mbox{(-) region},
\end{array}
\right.
\end{equation}
\begin{equation}
\chi^{\prime}=\left\{ 
\begin{array}{ll}
-\epsilon X-\frac{1}{4}(p_{1}+p_{2}) & \makebox[3em]{}\mbox{(+) region}, \\ 
-\epsilon X+\frac{1}{4}(p_{1}-p_{2}) & \makebox[3em]{}\mbox{(0) region}, \\ 
-\epsilon X+\frac{1}{4}(p_{1}+p_{2}) & \makebox[3em]{}\mbox{(-) region}\;.
\end{array}
\right.
\end{equation}
General solutions to the homogeneous equation $\triangle\phi-\frac{1}{4}%
\left\{\kappa^{2}\left(\chi^{\prime}\right)^{2} +\frac{\kappa}{2}V-\frac{1}{2%
}\Lambda_{e}\right\}\phi=0$ in each region are 
\begin{equation}  \label{phi2}
\left\{ 
\begin{array}{l}
\phi_{+}(x)=A_{+}e^{\frac{1}{2}K_{+}x}+B_{+}e^{-\frac{1}{2}K_{+}x}\;, \\ 
\phi_{0}(x)=A_{0}e^{\frac{1}{2}K_{0}x}+B_{0}e^{-\frac{1}{2}K_{0}x}\;, \\ 
\phi_{-}(x)=A_{-}e^{\frac{1}{2}K_{-}x}+B_{-}e^{-\frac{1}{2}K_{-}x}\;,
\end{array}
\right.
\end{equation}
where 
\begin{equation}
\left\{ 
\begin{array}{lll}
K_{+} & = \sqrt{\kappa^{2}\left(X+\frac{\epsilon}{4}(p_{1}+
p_{2})\right)^{2}-\frac{1}{2}\Lambda_{e}} & \qquad \mbox{(+) region}\;, \\ 
K_{0} & = \sqrt{\kappa^{2}\left(X-\frac{\epsilon}{4}(p_{1}
-p_{2})\right)^{2}-\frac{\kappa}{2}e_{1}e_{2}-\frac{1}{2}\Lambda_{e}} & 
\qquad \mbox{(0) region}\label{K}\;, \\ 
K_{-} & = \sqrt{\kappa^{2}\left(X-\frac{\epsilon}{4}(p_{1}+
p_{2})\right)^{2}-\frac{1}{2}\Lambda_{e}} & \qquad \mbox{(-) region}\;.
\end{array}
\right.
\end{equation}
For these solutions to be the actual solutions to Eq.(\ref{phi-eq}) with
delta function source terms, they must satisfy the following matching
conditions at $x=z_{1}, z_{2}$: {\ \setcounter{enumi}{\value{equation}} %
\addtocounter{enumi}{1} \setcounter{equation}{0} \renewcommand{%
\theequation}{\theenumi\alph{equation}} 
\begin{eqnarray}
&&\phi_{+}(z_{1})=\phi_{0}(z_{1})=\phi(z_{1})\;,  \label{match1} \\
&&\phi_{-}(z_{2})=\phi_{0}(z_{2})=\phi(z_{2})\;,  \label{match2} \\
&&\phi^{\prime}_{+}(z_{1})-\phi^{\prime}_{0}(z_{1}) =\frac{\kappa}{4}\sqrt{%
p^{2}_{1}+m^{2}_{1}}\phi(z_{1})\;,  \label{match3} \\
&&\phi^{\prime}_{0}(z_{2})-\phi^{\prime}_{-}(z_{2}) =\frac{\kappa}{4}\sqrt{%
p^{2}_{2}+m^{2}_{2}}\phi(z_{2}) \;.  \label{match4}
\end{eqnarray}
\setcounter{equation}{\value{enumi}} }

Since the magnitudes of both $\Psi $ and $\chi $ increase with increasing $%
|x|$, it is necessary to impose a boundary condition which guarantees that
the surface terms which arise in transforming the action vanish and
simultaneously preserves the finiteness of the Hamiltonian. From the
iterative expression (\ref{ham2}) we know that we may choose the boundary
condition 
\begin{equation}
\Psi ^{2}-4\kappa ^{2}\chi ^{2}+2\Lambda_{e} x^{2}=C_{\pm }x\qquad 
\mbox{for (+)
and (-) regions}  \label{bound}
\end{equation}
with $C_{\pm }$ being constants to be determined.

The above matching conditions accompanied by the boundary condition (\ref
{bound}) determine the solution $\phi $ ( and also all coefficients)
completely. The process of the calculation is quite analogous to the
procedure in the previous paper \cite{2bdcoslo}, and we shall omit the
details here. The compact expression of the $\phi $ solution is 
\begin{eqnarray}
\phi _{+} &=&\left( \frac{K_{1}}{{\cal M}_{1}}\right) ^{\frac{1}{2}}\;e^{-%
\frac{1}{4}(K_{01}z_{1}-K_{02}z_{2})+\frac{1}{2}K_{+}(x-z_{1})}\;,  \nonumber
\label{phi-sol2} \\
\phi _{0} &=&\frac{1}{4K_{0}}\;e^{-\frac{1}{4}(K_{01}z_{1}-K_{02}z_{2})}%
\left\{ (K_{1}{\cal M}_{1})^{1/2}e^{-\frac{1}{2}K_{0}(x-z_{1})}+(K_{2}{\cal M%
}_{2})^{1/2}e^{\frac{1}{2}K_{0}(x-z_{2})}\right\} \;, \\
\phi _{-} &=&\left( \frac{K_{2}}{{\cal M}_{2}}\right) ^{\frac{1}{2}}\;e^{-%
\frac{1}{4}(K_{01}z_{1}-K_{02}z_{2})-\frac{1}{2}K_{-}(x-z_{2})}\;,  \nonumber
\end{eqnarray}
where 
\begin{eqnarray}
K_{1} &\equiv &2K_{0}+2K_{-}-\kappa \sqrt{p_{2}^{2}+m_{2}^{2}}\;,  \nonumber
\label{nota} \\
K_{2} &\equiv &2K_{0}+2K_{+}-\kappa \sqrt{p_{1}^{2}+m_{1}^{2}}\;,  \nonumber
\\
K_{01} &\equiv &K_{0}-K_{+}+\frac{\kappa \epsilon }{2}p_{1}\;, \\
K_{02} &\equiv &K_{0}-K_{-}-\frac{\kappa \epsilon }{2}p_{2}\;,  \nonumber \\
{\cal M}_{1} &\equiv &\kappa \sqrt{p_{1}^{2}+m_{1}^{2}}+2K_{0}-2K_{+}\;, 
\nonumber \\
{\cal M}_{2} &\equiv &\kappa \sqrt{p_{2}^{2}+m_{2}^{2}}+2K_{0}-2K_{-}\;, 
\nonumber
\end{eqnarray}
and among these quantites there exists one relation 
\begin{equation}
K_{1}K_{2}={\cal M}_{1}{\cal M}_{2}\;e^{K_{0}(z_{1}-z_{2})}\;,  \label{H0}
\end{equation}
which we refer to as the determining equation of $X$.

The Hamiltonian (\ref{ham1}) becomes 
\begin{eqnarray}
H &=&-\frac{1}{\kappa }\int dx\triangle \Psi =\frac{4}{\kappa }\left[ \frac{%
\phi ^{\prime }}{\phi }\right] _{-\infty }^{\infty }  \nonumber
\label{ham2a} \\
&=&\frac{2(K_{+}+K_{-})}{\kappa }\;\;.
\end{eqnarray}
Once the solution $X$ to (\ref{H0}) is obtained, the Hamiltonian is
explicitly determined. Consequently (\ref{H0}) is the determining equation
of the Hamiltonian.

Repeating the analysis for $z_{1}<z_{2}$ yields a similar solution with $%
p_{i} \rightarrow -p_{i}$ and so the full solution is obtained by replacing $%
p_{i}$ and $z_{1}-z_{2}$ by $\tilde{p}_{i}=p_{i}\;\makebox{sgn}(z_{1}-z_{2})$
and $|z_{1}-z_{2}|$, respectively. The determining equation (\ref{H0}) of
the Hamiltonian becomes 
\begin{equation}  \label{H1}
K_{1}K_{2}={\cal M}_{1}{\cal M}_{2}\;e^{K_{0}|z_{1}-z_{2}|}\;\;.
\end{equation}
or more explicitly 
\begin{eqnarray}  \label{H2}
&&\left(4K_{0}^{2}+[\kappa\sqrt{p_{1}^{2}+m^{2}_{1}}-2K_{+}] [\kappa\sqrt{%
p_{2}^{2}+m^{2}_{2}}-2K_{-}]\right) \mbox{tanh}\left(\frac{1}{2}%
K_{0}|z_{1}-z_{2}|\right)  \nonumber \\
&&\makebox[5em]{}=-2K_{0} \left([\kappa\sqrt{p_{1}^{2}+m^{2}_{1}}-2K_{+}]
+[\kappa\sqrt{p_{2}^{2}+m^{2}_{2}}-2K_{-}]\right)
\end{eqnarray}
where the momentum $p_{i}$ is replaced by $\tilde{p}_{i}$.

For the expression (\ref{ham2a}) to have a definite meaning as the
Hamiltonian, $K_{\pm }$ must both be real, with positive sum. This imposes a
restriction on $X$ corresponding to a value of the cosmological constant $%
\Lambda _{e}$. However $K_{0}$ is not necessarily real. If $e_{1}e_{2}$
takes a large positive value (strong electromagnetic repulsion), $K_{0}$ may
be imaginary. In this case we need to reconsider the above analysis, because
in the (0) region the soluton to the $\phi $ equation (\ref{phi-eq}) becomes 
\begin{equation}
\phi _{0}(x)=A_{s}\;\mbox{sin}\frac{1}{2}\tilde{K}_{0}x+A_{c}\;\mbox{cos}%
\frac{1}{2}\tilde{K}_{0}x \;,  \label{phi-0}
\end{equation}
where 
\begin{eqnarray}
\tilde{K}_{0} &=&-iK_{0}  \nonumber \\
&=&\sqrt{\frac{\kappa }{2}e_{1}e_{2}+\frac{1}{2}\Lambda _{e}-\kappa
^{2}\left( X-\frac{\epsilon }{4}(\tilde{p}_{1}-\tilde{p}_{2})\right) ^{2}}%
\;\;.
\end{eqnarray}
Under the same matching conditions (\ref{match1}-\ref{match4}) and the
boundary condition (\ref{bound}) we get, instead of (\ref{H2}), a new
determining equation of the Hamiltonian 
\begin{eqnarray}
&&\left( 4\tilde{K}_{0}^{2}-[\kappa \sqrt{p_{1}^{2}+m_{1}^{2}}%
-2K_{+}][\kappa \sqrt{p_{2}^{2}+m_{2}^{2}}-2K_{-}]\right) \mbox{tan}\left( 
\frac{1}{2}\tilde{K}_{0}|z_{1}-z_{2}|\right)  \nonumber  \label{H2new} \\
&&\makebox[5em]{}=2\tilde{K}_{0}\left( [\kappa \sqrt{p_{1}^{2}+m_{1}^{2}}%
-2K_{+}]+[\kappa \sqrt{p_{2}^{2}+m_{2}^{2}}-2K_{-}]\right) \;\;.
\end{eqnarray}
Actually, this is just the equation derived from (\ref{H2}) by formally
replacing $K_{0}$ with $i\tilde{K}_{0}$. The solution of $\phi $ for
imaginary $K_{0}$ is also identical with that obtained from (\ref{phi-sol2})
by the same replacement. We can therefore use equation (\ref{H1}) for all
values of $K_{0}$; it is a transcendental equation which determines $H$ in
terms of the momenta and positions of the particles. We have previously
shown that in the case of zero cosmological constant and no charges the
solution for $H$ is expressed in terms of the Lambert $W$ function. More
generally, with $\Lambda _{e}$ and $e_{a}$ all nonzero, a solution to (\ref
{H1}) for $H$ cannot be explicitly expressed in terms of known fuctions.
However it can be obtained in successive approximations in the parameters $%
c^{-1}$, $\kappa $, etc. The examples will be shown later in the sections VI
and VIII.

Though in the general case the Hamiltonian cannot be expressed explicitly in
terms of known functions, the canonical equations of motion can be exactly
derived from the determining equation (\ref{H1}) by differentiating it with
respect to the variables $z_{i}$ and $p_{i}$ . For the variables $p_{1}$ and 
$z_{1}$ we have 
\begin{eqnarray}
\dot{p}_{1} &=&-\frac{\partial H}{\partial z_{1}}=-\frac{2}{\kappa }\left( 
\frac{\partial K_{+}}{\partial z_{1}}+\frac{\partial K_{-}}{\partial z_{1}}%
\right) =-2\left( \frac{Y_{+}}{K_{+}}+\frac{Y_{-}}{K_{-}}\right) \frac{%
\partial X}{\partial z_{1}}  \nonumber  \label{p1} \\
&=&-\frac{2}{\kappa }\left( \frac{Y_{+}}{K_{+}}+\frac{Y_{-}}{K_{-}}\right) 
\frac{K_{0}K_{1}K_{2}}{J}\;\;,
\end{eqnarray}
\begin{eqnarray}
\dot{z}_{1} &=&\frac{\partial H}{\partial p_{1}}=\frac{2}{\kappa }\left( 
\frac{\partial K_{+}}{\partial p_{1}}+\frac{\partial K_{-}}{\partial p_{1}}%
\right)  \nonumber  \label{z1} \\
&=&\frac{\epsilon }{2}\left( \frac{Y_{+}}{K_{+}}-\frac{Y_{-}}{K_{-}}\right)
+2\left( \frac{Y_{+}}{K_{+}}+\frac{Y_{-}}{K_{-}}\right) \frac{\partial X}{%
\partial p_{1}}  \nonumber \\
&=&\epsilon \frac{Y_{+}}{K_{+}}+\frac{8}{J}\left( \frac{Y_{+}}{K_{+}}+\frac{%
Y_{-}}{K_{-}}\right) \frac{K_{0}K_{1}}{{\cal M}_{1}}\left\{ \frac{p_{1}}{%
\sqrt{p_{1}^{2}+m_{1}^{2}}}-\epsilon \frac{Y_{+}}{K_{+}}\right\}\;\;,
\end{eqnarray}
where 
\begin{eqnarray}
Y_{+} &\equiv &\kappa \left[ X+\frac{\epsilon }{4}(p_{1}+p_{2})\right] \;, 
\nonumber  \label{notaY} \\
Y_{0} &\equiv &\kappa \left[ X-\frac{\epsilon }{4}(p_{1}-p_{2})\right] \;, \\
Y_{-} &\equiv &\kappa \left[ X-\frac{\epsilon }{4}(p_{1}+p_{2})\right] \;, 
\nonumber
\end{eqnarray}
and 
\begin{eqnarray}
J &=&2\left\{ \left( \frac{Y_{0}}{K_{0}}+\frac{Y_{+}}{K_{+}}\right)
K_{1}+\left( \frac{Y_{0}}{K_{0}}+\frac{Y_{-}}{K_{-}}\right) K_{2}\right\} 
\nonumber  \label{J} \\
&&-2\left\{ \left( \frac{Y_{0}}{K_{0}}-\frac{Y_{+}}{K_{+}}\right) \frac{1}{%
{\cal M}_{1}}+\left( \frac{Y_{0}}{K_{0}}-\frac{Y_{-}}{K_{-}}\right) \frac{1}{%
{\cal M}_{2}}\right\} K_{1}K_{2}-\frac{Y_{0}}{K_{0}}K_{1}K_{2}(z_{1}-z_{2})%
\;.
\end{eqnarray}

Similarly for particle 2 the equations are 
\begin{eqnarray}
\dot{p}_{2} &=&\frac{2}{\kappa }\left( \frac{Y_{+}}{K_{+}}+\frac{Y_{-}}{K_{-}%
}\right) \frac{K_{0}K_{1}K_{2}}{J}\;,  \label{p2} \\
\dot{z}_{2} &=&-\epsilon \frac{Y_{-}}{K_{-}}+\frac{8}{J}\left( \frac{Y_{+}}{%
K_{+}}+\frac{Y_{-}}{K_{-}}\right) \frac{K_{0}K_{2}}{{\cal M}_{2}}\left\{ 
\frac{p_{2}}{\sqrt{p_{2}^{2}+m_{2}^{2}}}+\epsilon \frac{Y_{-}}{K_{-}}%
\right\} \;\;.  \label{z2}
\end{eqnarray}
These canonical equations guarantee the conservation of the Hamiltonian and
the total momentum $p_{1}+p_{2}$.

On the other hand the equations of motion (\ref{e-p}) and (\ref{e-z})
derived from the action (\ref{act2}) become under the coordinate conditions (%
\ref{cc}) 
\begin{eqnarray}
\dot{p}_{a}&=&-\frac{\partial N_{0}}{\partial z_{a}}\sqrt{p^{2}_{a}
+m^{2}_{a}}+\frac{\partial N_{1}}{\partial z_{a}}p_{a} +\frac{1}{2}%
\sum_{b}e_{a}e_{b}N_{0}\frac{\partial}{\partial z_{a}}|z_{a}-z_{b}| \;,
\label{pa} \\
\dot{z_{a}}&=&N_{0}\frac{p_{a}}{\sqrt{p^{2}_{a}+m^{2}_{a}}}-N_{1}\;\;.
\label{za}
\end{eqnarray}

It is straightforward to verify that insertion of the solutions of the
metric components given in Appendix A into (\ref{pa}) and (\ref{za})
reproduces the canonical equations of motion (\ref{p1}), (\ref{z1}), (\ref
{p2}) and (\ref{z2}) where the partial derivatives at $z{1},z{2}$ are
defined by 
\begin{equation}
\frac{\partial N_{0,1}}{\partial z_{i}}\equiv \frac{1}{2}\left\{ \left. 
\frac{\partial N_{0,1}}{\partial x}\right| _{x=z_{i}+0}+\left. \frac{%
\partial N_{0,1}}{\partial x}\right| _{x=z_{i}-0}\right\} \;\;.
\end{equation}
Thus consistency between the geodesic equations and the canonical equations
of motion is explicitly assured, while formal proof of consistency in the
case of $\Lambda _{e}=0$ and $e_{a}=0$ can be easily generalized \cite{OR}.

The components of the metric are determined from the equations (\ref{e-pi}),
(\ref{e-gamma}), (\ref{e-Pi}) and (\ref{e-Psi}) under the coordinate
conditions (\ref{cc}). The derivation and the explicit solutions of the
metric are given in Appendix A. With these solutions we can trace how the
structure of space-time changes due to the motion of the two bodies.

\section{EXACT SOLUTIONS OF THE TRAJECTORIES FOR EQUAL MASSES AND ARBITRARY
CHARGES}

In this section we consider a system of two particles with equal mass. Since
the total momentum is conserved we can always choose the center of inertia
(C.I.) system $p_{1}=-p_{2}=p$. Corresponding to the sign of $(\sqrt{%
H^{2}+8\Lambda _{e}/\kappa ^{2}}-2\epsilon \tilde{p})^{2}-8e_{1}e_{2}/\kappa
-8\Lambda _{e}/\kappa ^{2}$ the determining equations (\ref{H2}) and (\ref
{H2new}) become 
\begin{equation}
({\cal J}_{\Lambda }^{\;2}+B^{2})\;\mbox{tanh}\left( \frac{\kappa }{8}{\cal J%
}_{\Lambda }\;|r|\right) =2{\cal J}_{\Lambda }B \;,  \label{deteq-1}
\end{equation}
and 
\begin{equation}
(\tilde{{\cal J}}_{\Lambda }^{\;2}-B^{2})\;\mbox{tan}\left( \frac{\kappa }{8}%
\tilde{{\cal J}}_{\Lambda }\;|r|\right) =-2\tilde{{\cal J}}_{\Lambda }B\;,
\label{deteq-2}
\end{equation}
respectively, where 
\begin{eqnarray}
&&{\cal J}_{\Lambda }=\sqrt{\left( \sqrt{H^{2}+\frac{8\Lambda _{e}}{\kappa
^{2}}}-2\epsilon \tilde{p}\right) ^{2}-\frac{8e_{1}e_{2}}{\kappa }-\frac{%
8\Lambda _{e}}{\kappa ^{2}}}\;,  \nonumber \\
&&\tilde{{\cal J}}_{\Lambda }=\sqrt{\frac{8e_{1}e_{2}}{\kappa }+\frac{%
8\Lambda _{e}}{\kappa ^{2}}-\left( \sqrt{H^{2}+\frac{8\Lambda _{e}}{\kappa
^{2}}}-2\epsilon \tilde{p}\right) ^{2}}\;,  \nonumber \\
&&B=H-2\sqrt{p^{2}+m^{2}}\;\;.
\end{eqnarray}
The equation (\ref{deteq-1}) is further divided into two types: 
\begin{equation}
\mbox{tanh}\left( \frac{\kappa }{16}{\cal J}_{\Lambda }|r|\right) =\frac{B}{%
{\cal J}_{\Lambda }}\;,\qquad \mbox{(tanh-type A)}  \label{TanhA}
\end{equation}
or 
\begin{equation}
\mbox{tanh}\left( \frac{\kappa }{16}{\cal J}_{\Lambda }|r|\right) =\frac{%
{\cal J}_{\Lambda }}{B}\;,\qquad \mbox{(tanh-type B)}\;\;.  \label{TanhB}
\end{equation}
In the case of $\Lambda _{e}=0$ and $e_{a}=0$ the tanh-type B equation is
excluded, because ${\cal J}_{\Lambda }/B$ exceeds unity. When a cosmological
constant and/or charge are introduced, this equation may have a solution in
some restricted range of the parameters. \newline
\noindent Eq.(\ref{deteq-2}) is also divided into 
\begin{equation}
\mbox{tan}\left( \frac{\kappa }{16}\tilde{{\cal J}}_{\Lambda }|r|\right) =-\;%
\frac{B}{\tilde{{\cal J}}_{\Lambda }}\;,\qquad \mbox{(tan-type A)}
\label{TanA}
\end{equation}
or 
\begin{equation}
\mbox{tan}\left( \frac{\kappa }{16}\tilde{{\cal J}}_{\Lambda }|r|\right) =%
\frac{\tilde{{\cal J}}_{\Lambda }}{B}\;,\qquad \mbox{(tan-type B)}\;\;.
\label{TanB}
\end{equation}

For all four types of the determining equations the canonical equations of
motion are identical: 
\begin{eqnarray}
\dot{p}&=&-\;\frac{\kappa{\cal J}_{\Lambda}^{2}({\cal J}_{\Lambda}^{2}-B^2)%
} {16C}\;\mbox{sgn(r)}\;,  \label{can-p} \\
\dot{r}&=&2\epsilon \;\sqrt{1+\frac{8\Lambda_{e}}{\kappa^2 H^2}} \left(1-%
\frac{{\cal J}_{\Lambda}^{2}}{C}\right)\mbox{sgn(r)} +\frac{2{\cal J}%
_{\Lambda}^{2}}{C}\frac{p}{\sqrt{p^{2}+m^{2}}}\;.  \label{can-z}
\end{eqnarray}
where 
\begin{equation}  \label{C}
C=\frac{1}{\sqrt{1+\frac{8\Lambda_{e}}{\kappa^2 H^2}}} \left\{\sqrt{1+\frac{%
8\Lambda_{e}}{\kappa^2 H^2}}\;{\cal J}_{\Lambda}^{2} -\left(\sqrt{H^2+\frac{%
8\Lambda_{e}}{\kappa^2}}-2\epsilon\tilde{p}\right) \left(B+\frac{\kappa}{16}(%
{\cal J}_{\Lambda}^{2}-B^2)\;r \right)\right\}\;.
\end{equation}

For given values of $\Lambda _{e}$ and $e_{a}$, the equations (\ref{deteq-1}%
) or (\ref{deteq-2}) describe the surface in $(r,p,H)$ space of all allowed
phase-space trajectories. Since $H$ is a constant of motion, we can draw a
phase space trajectory in $(r,p)$ space by setting $H=H_{0}$ in (\ref
{deteq-1}) or (\ref{deteq-2}). This same trajectory can also be obtained
directly from the solution $r(t),p(t)$ to the canonical equations (\ref
{can-p}) and (\ref{can-z}) by eliminating the time variable $t$. Numerical
solutions to (\ref{can-p}) and (\ref{can-z}) confirm this, but in $r(t)$ and 
$p(t)$ superficial singularities appear due to the zero points of $C$.

It is therefore preferable to describe the particles' trajectories in terms
of some invariant parameter. The common proper time $\tau _{a}$ of each
particle is the best candidate as seen in the starting action (\ref{act1}).
{}From the metric components given in the Appendix A and the canonical
equations (\ref{za}), the proper time is 
\begin{eqnarray}
d\tau _{a}^{2} &=&dt^{2}\left\{ N_{0}(z_{a})^{2}-(N_{1}(z_{a})+\dot{z}%
_{a})^{2}\right\} \;,  \nonumber \\
&=&dt^{2}N_{0}(z_{a})^{2}\frac{m_{a}^{2}}{p_{a}^{2}+m_{a}^{2}}\qquad \qquad
(a=1,2)\;\;.  \label{proper}
\end{eqnarray}
For the equal mass case it is identical for particles 1 and 2 
\begin{equation}
d\tau =d\tau _{1}=d\tau _{2}=\frac{m}{\sqrt{p^{2}+m^{2}}}\frac{{\cal J}%
_{\Lambda }^{2}}{C}dt\;,  \label{Tau}
\end{equation}
from which the canonical equations (\ref{can-p}) and (\ref{can-z}) may be
expressed in the form

\begin{eqnarray}
&&\frac{dp}{d\tau}=-\;\frac{\kappa\sqrt{p^2+m^2}({\cal J}_{\Lambda}^2-B^2)} {%
16m}\;\mbox{sgn(r)}\;,  \label{p-Tau} \\
&&\frac{dr}{d\tau}=\frac{2\epsilon}{m} \left\{\frac{\sqrt{p^2+m^2}\;C}{{\cal %
J}_{\Lambda}^2} -(\sqrt{p^2+m^2}-\epsilon\tilde{p})\right\}\mbox{sgn(r)}
\;\;.  \label{r-Tau}
\end{eqnarray}

Remarkably these equations can be solved exactly. First we solve Eq.(\ref
{p-Tau}) for $p(\tau )$ and then obtain $r(\tau )$, either by directly
solving (\ref{r-Tau}) after substituting the solution for $p$ or by solving (%
\ref{TanhA})-(\ref{TanB}) for $r$. For the $r>0$ region Eq.(\ref{p-Tau})
leads to 
\begin{eqnarray}
\int_{p_{0}}^{p}\;\frac{dp}{\left\{ \sqrt{p^{2}+m^{2}}-\epsilon \sqrt{1+%
\frac{8\Lambda _{e}}{\kappa ^{2}H^{2}}}\;p-\frac{1}{H}\left( m^{2}+\frac{%
2e_{1}e_{2}}{\kappa }\right) \right\} \sqrt{p^{2}+m^{2}}} &=&-\;\frac{\kappa
H}{4m}\int_{\tau _{0}}^{\tau }\;d\tau  \nonumber \\
&=&-\;\frac{\kappa H}{4m}(\tau -\tau _{0})\;\;.  \label{pintegrate}
\end{eqnarray}
This expression implies the condition 
\begin{equation}
1+\frac{8\Lambda _{e}}{\kappa ^{2}H^{2}}\geq 0\;\;
\end{equation}
which is satisfied for all $\Lambda _{e}>0$. For negative $\Lambda _{e}$ the
motion is allowed provided $H$ satisfies 
\begin{equation}
H\geq \sqrt{-\;\frac{8\Lambda _{e}}{\kappa ^{2}}}\;\;.  \label{cosmolimit}
\end{equation}
We perfrom the integration of the Left-hand side (LHS) of (\ref{pintegrate})
in three cases, separately, depending on the value of $\Lambda _{e}$. The
solution $p(\tau )$ is 
\begin{equation}
p(\tau )=\frac{\epsilon m}{2}\left( f(\tau )-\frac{1}{f(\tau )}\right)
\label{p-exact1}
\end{equation}
with 
\begin{equation}
f(\tau )=\left\{ 
\begin{array}{ll}
\frac{\frac{H}{m}\left( 1+\sqrt{\gamma _{H}}\right) \left\{ 1-\eta \;e^{%
\frac{\epsilon \kappa m}{4}\sqrt{\gamma _{m}}(\tau -\tau _{0})}\right\} }{%
\gamma _{e}+\sqrt{\gamma _{m}}+\left( \sqrt{\gamma _{m}}-\gamma _{e}\right)
\;\eta \;e^{\frac{\epsilon \kappa m}{4}\sqrt{\gamma _{m}}(\tau -\tau _{0})}}
& \qquad \gamma _{m}>0\;, \\ 
&  \\ 
\frac{1+\sqrt{\gamma _{H}}}{\frac{m}{H}\gamma _{e}+\frac{\sigma }{m-\sigma 
\frac{\epsilon \kappa H}{8}(\tau -\tau _{0})}} & \qquad \gamma _{m}=0\;, \\ 
&  \\ 
\frac{\frac{H}{m}(1+\sqrt{\gamma _{H}})}{\gamma _{e}+\sqrt{-\gamma _{m}}%
\frac{\sigma +\frac{m^{2}}{H}\sqrt{-\gamma _{m}}\tan \left[ \frac{\epsilon
\kappa m}{8}\sqrt{-\gamma _{m}}(\tau -\tau _{0})\right] }{\frac{m^{2}}{H}%
\sqrt{-\gamma _{m}}-\sigma \tan \left[ \frac{\epsilon \kappa m}{8}\sqrt{%
-\gamma _{m}}(\tau -\tau _{0})\right] }} & \qquad \gamma _{m}<0\;,
\end{array}
\right.
\end{equation}
where 
\begin{equation}
\begin{array}{ll}
\gamma _{H}=1+\frac{8\Lambda _{e}}{\kappa ^{2}H^{2}}\;, & \qquad \gamma
_{e}=1+\frac{2e_{1}e_{2}}{\kappa m^{2}}\;, \\ 
\gamma _{m}=\gamma _{e}^{2}+\frac{8\Lambda _{e}}{\kappa ^{2}m^{2}}\;, & 
\qquad \sigma =(1+\sqrt{\gamma _{H}})(\sqrt{p_{0}^{2}+m^{2}}-\epsilon p_{0})-%
\frac{m^{2}}{H}\gamma _{e}\;, \\ 
\eta =\frac{\sigma -\frac{m^{2}}{H}\sqrt{\gamma _{m}}}{\sigma +\frac{m^{2}}{H%
}\sqrt{\gamma _{m}}}\;, & 
\end{array}
\end{equation}
with $p_{0}$ being the initial momentum at $\tau =\tau _{0}$.

Similarly the solution in $r<0$ region is 
\begin{equation}  \label{p-exact2}
p(\tau)=-\frac{\epsilon m}{2}\left(\bar{f}(\tau)-\frac{1}{\bar{f}(\tau)}
\right)
\end{equation}
with 
\begin{equation}
\bar{f}(\tau) = \left\{ 
\begin{array}{ll}
\frac{\frac{H}{m}\left(1+\sqrt{\gamma_H}\right) \left\{1-\bar{\eta}\;e^{%
\frac{\epsilon\kappa m}{4}\sqrt{\gamma_m}(\tau-\tau_{0})}\right\}} {%
\gamma_{e}+\sqrt{\gamma_m} +\left(\sqrt{\gamma_m}-\gamma_{e}\right) \;\bar{%
\eta}\;e^{\frac{\epsilon\kappa m}{4}\sqrt{\gamma_m}(\tau-\tau_{0})}} & 
\qquad \gamma_{m}> 0 \;, \\ 
&  \\ 
\frac{1+\sqrt{\gamma_H}} {\frac{m}{H}\gamma_{e}+\frac{\bar{\sigma}} {m-\bar{%
\sigma}\frac{\epsilon\kappa H}{8}(\tau-\tau_{0})}} & \qquad \gamma_{m}=0 \;,
\\ 
&  \\ 
\frac{\frac{H}{m}(1+ \sqrt{\gamma_H})} {\gamma_{e}+\sqrt{-\gamma_m} \frac{%
\bar{\sigma}+\frac{m^2}{H}\sqrt{-\gamma_m} \tan\left[\frac{\epsilon\kappa m}{%
8}\sqrt{-\gamma_m}(\tau-\tau_{0})\right]} {\frac{m^2}{H}\sqrt{-\gamma_m} -%
\bar{\sigma} \tan\left[\frac{\epsilon\kappa m}{8}\sqrt{-\gamma_m}%
(\tau-\tau_{0})\right]}} & \qquad \gamma_{m} < 0 \;,
\end{array}
\right.
\end{equation}
where 
\begin{equation}
\bar{\sigma} = (1+\sqrt{\gamma_H}) (\sqrt{p_{0}^2+m^2}+\epsilon p_{0})-\frac{%
m^2}{H}\gamma_{e}\;, \qquad \bar{\eta}=\frac{\bar{\sigma} -\frac{m^2}{H}%
\sqrt{\gamma_m}} {\bar{\sigma} + \frac{m^2}{H}\sqrt{\gamma_m}}\;\;.
\end{equation}

Corresponding to each type of the determining equations (\ref{TanhA})- (\ref
{TanB}) the solution for $r(\tau)$ is obtained as follows,

\noindent tanh-type A: 
\begin{equation}  \label{sol-tanhA}
r(\tau)=\left\{ 
\begin{array}{ll}
\frac{16\;\; \mbox{tanh}^{-1}\left[\frac{\kappa\left(H- m\left|f(\tau)+\frac{%
1}{f(\tau)} \right|\right)} {\sqrt{\left(\sqrt{\kappa^2 H^2+8\Lambda_{e}}
-m\kappa(f(\tau)-\frac{1}{f(\tau)})\right)^{2}-8\kappa e_{1}e_{2}
-8\Lambda_{e}}}\right]} {\sqrt{\left(\sqrt{\kappa^2 H^2+8\Lambda_{e}}
-m\kappa(f(\tau)-\frac{1}{f(\tau)})\right)^{2}-8\kappa e_{1}e_{2}
-8\Lambda_{e} }} & \qquad r>0\;, \\ 
&  \\ 
\frac{-16\;\; \mbox{tanh}^{-1}\left[\frac{\kappa\left(H- m\left|{\bar f}%
(\tau)+\frac{1} {{\bar f}(\tau)}\right|\right)} {\sqrt{\left(\sqrt{\kappa^2
H^2+8\Lambda_{e}} -m\kappa({\bar f}(\tau)-\frac{1}{{\bar f}(\tau)}%
)\right)^{2} -8\kappa e_{1}e_{2}-8\Lambda_{e} }}\right]} {\sqrt{\left(\sqrt{%
\kappa^2 H^2+8\Lambda_{e}} -m\kappa({\bar f}(\tau)-\frac{1}{{\bar f}(\tau)}%
)\right)^{2} -8\kappa e_{1}e_{2}-8\Lambda_{e} }} & \qquad r<0 \;,
\end{array}
\right.
\end{equation}
tanh-type B: 
\begin{equation}  \label{sol-tanhB}
r(\tau)=\left\{ 
\begin{array}{ll}
\frac{16\;\; \mbox{tanh}^{-1}\left[ \frac{\sqrt{\left(\sqrt{\kappa^2
H^2+8\Lambda_{e}} -m\kappa(f(\tau)-\frac{1}{f(\tau)})\right)^{2}-8\kappa
e_{1}e_{2} -8\Lambda_{e}}} {\kappa\left(H- m\left|f(\tau)+\frac{1}{f(\tau)}%
\right|\right)} \right] } {\sqrt{\left(\sqrt{\kappa^2 H^2+8\Lambda_{e}}
-m\kappa(f(\tau)-\frac{1}{f(\tau)})\right)^{2}-8\kappa e_{1}e_{2}
-8\Lambda_{e} }} & \qquad r>0\;, \\ 
&  \\ 
\frac{-16\;\; \mbox{tanh}^{-1}\left[ \frac{\sqrt{\left(\sqrt{\kappa^2
H^2+8\Lambda_{e}} -m\kappa({\bar f}(\tau)-\frac{1}{{\bar f}(\tau)}%
)\right)^{2} -8\kappa e_{1}e_{2}-8\Lambda_{e} }} {\kappa\left(H- m\left|{%
\bar f}(\tau)+\frac{1}{{\bar f}(\tau)}\right|\right)} \right]} {\sqrt{\left(%
\sqrt{\kappa^2 H^2+8\Lambda_{e}} -m\kappa({\bar f}(\tau)-\frac{1}{{\bar f}%
(\tau)})\right)^{2} -8\kappa e_{1}e_{2}-8\Lambda_{e} }} & \qquad r<0 \;,
\end{array}
\right.
\end{equation}
tan-type A: 
\begin{equation}  \label{sol-tanA}
r(\tau)=\left\{ 
\begin{array}{ll}
\frac{16\;\; \left(\mbox{tan}^{-1}\left[\frac{\kappa\left(m \left|f(\tau)+%
\frac{1}{f(\tau)} \right|-H\right)} {\sqrt{8\Lambda_{e} + 8\kappa e_{1}e_{2}
- \left(\sqrt{\kappa^2 H^2+8\Lambda_{e}} -m\kappa(f(\tau)-\frac{1}{f(\tau)}%
)\right)^{2} }}\right]+n\pi\right)} {\sqrt{8\Lambda_{e} + 8\kappa e_{1}e_{2}
- \left(\sqrt{\kappa^2 H^2+8\Lambda_{e}} -m\kappa(f(\tau)-\frac{1}{f(\tau)}%
)\right)^{2} }}\; & \quad r>0\;, \\ 
&  \\ 
\frac{-16\;\; \left(\mbox{tan}^{-1}\left[\frac{\kappa\left(m \left|{\bar f}%
(\tau) +\frac{1}{{\bar f}(\tau)}\right|-H\right)} {\sqrt{8\Lambda_{e} +
8\kappa e_{1}e_{2} - \left(\sqrt{\kappa^2 H^2+8\Lambda} -m\kappa({\bar f}%
(\tau)-\frac{1}{{\bar f}(\tau)})\right)^{2} }}\right]+n\pi\right)} {\sqrt{%
8\Lambda_{e} + 8\kappa e_{1}e_{2} - \left(\sqrt{\kappa^2 H^2+8\Lambda_{e}}
-m\kappa({\bar f}(\tau)-\frac{1}{{\bar f}(\tau)})\right)^{2} }}\; & \quad
r<0 \;,
\end{array}
\right.
\end{equation}
tan-type B: 
\begin{equation}  \label{sol-tanB}
r(\tau)=\left\{ 
\begin{array}{ll}
\frac{16\;\; \left(\mbox{tan}^{-1}\left[\frac{\sqrt{8\Lambda_{e} + 8\kappa
e_{1}e_{2} - \left(\sqrt{\kappa^2 H^2+8\Lambda_{e}} -m\kappa(f(\tau)-\frac{1%
}{f(\tau)})\right)^{2} }}{\kappa\left(H-m \left|f(\tau)+\frac{1}{f(\tau)}%
\right|\right)}\right]+n\pi\right)} {\sqrt{8\Lambda_{e} + 8\kappa e_{1}e_{2}
- \left(\sqrt{\kappa^2 H^2+8\Lambda_{e}} -m\kappa(f(\tau)-\frac{1}{f(\tau)}%
)\right)^{2} }}\; & \quad r>0\;, \\ 
&  \\ 
\frac{-16\;\; \left(\mbox{tan}^{-1}\left[\frac{\sqrt{8\Lambda_{e} + 8\kappa
e_{1}e_{2} - \left(\sqrt{\kappa^2 H^2+8\Lambda_{e}} -m\kappa({\bar f}(\tau)-%
\frac{1}{{\bar f}(\tau)})\right)^{2} }} {\kappa\left(H-m \left|{\bar f}%
(\tau)+\frac{1}{{\bar f}(\tau)}\right|\right)} \right]+n\pi\right)} {\sqrt{%
8\Lambda_{e} + 8\kappa e_{1}e_{2} - \left(\sqrt{\kappa^2 H^2+8\Lambda_{e}}
-m\kappa({\bar f}(\tau)-\frac{1}{{\bar f}(\tau)})\right)^{2} }}\; & \quad
r<0 \;\;.
\end{array}
\right.
\end{equation}

\section{ANALYSIS OF ELECTRODYNAMIC MOTION WITH $\Lambda _{e}=0$}

Based on the exact solutions in the previous section we can analyze the
dynamics of two-body problem. In this section we investigate electrodynamics
in a space-time with $\Lambda _{e}=0$. The solution $p(\tau )$ in this case
is given for $r>0$ by 
\begin{equation}
p(\tau )=\frac{\epsilon m}{2}\left( f_{e}(\tau )-\frac{1}{f_{e}(\tau )}%
\right)
\end{equation}
with 
\begin{eqnarray}
f_{e}(\tau ) &=&\frac{H}{m\gamma _{e}}\left\{ 1-\eta _{e}\;e^{\frac{\epsilon
\kappa m}{4}\gamma _{e}(\tau -\tau _{0})}\right\} \;,  \nonumber \\
&& \\
\eta _{e} &=&\frac{\sqrt{p_{0}^{2}+m^{2}}-\epsilon p_{0}-\frac{m^{2}}{H}%
\gamma _{e}}{\sqrt{p_{0}^{2}+m^{2}}-\epsilon p_{0}}\;,  \nonumber
\end{eqnarray}
and for $r<0$ by 
\begin{equation}
p(\tau )=-\frac{\epsilon m}{2}\left( \bar{f}_{e}(\tau )-\frac{1}{\bar{f}%
_{e}(\tau )}\right)
\end{equation}
with 
\begin{eqnarray}
\bar{f}_{e}(\tau ) &=&\frac{H}{m\gamma _{e}}\left\{ 1-\bar{\eta}_{e}\;e^{%
\frac{\epsilon \kappa m}{4}\gamma _{e}(\tau -\tau _{0})}\right\} \;, 
\nonumber \\
&& \\
\bar{\eta}_{e} &=&\frac{\sqrt{p_{0}^{2}+m^{2}}+\epsilon p_{0}-\frac{m^{2}}{H}%
\gamma _{e}}{\sqrt{p_{0}^{2}+m^{2}}+\epsilon p_{0}}\;\;.  \nonumber
\end{eqnarray}
The relative distance $r(\tau )$ is obtained from (\ref{TanhA}) - (\ref{TanB}%
) by simply replacing $f(\tau )$ and $\bar{f}(\tau )$ with $f_{e}(\tau )$
and $\bar{f}_{e}(\tau )$, respectively.

Relative motion of two charged particles is classified by the signs and the
magnitudes of the charges. Since the charges always appear as the product $%
e_{1}e_{2}$ in all solutions, it is sufficient to analyze the attractive
case by setting $e_{1}=-e_{2}=q$ and the repulsive case by setting $%
e_{1}=e_{2}=q$ so that the charges have equal magnitude. Throughout this
paper, in the numerical analysis, we set $\epsilon =1,\kappa =1$ and rescale
everything in units of $\ m$ (effectively setting $m=1$) for simplicity,
except when otherwise stated. It should be remarked that for every phase
space trajectory there exists a time-reversed trajectory with $\epsilon =-1$%
, which is obtained by reversing the former with respect to the $r$-axis.

\subsection{The attractive case: $e_{1}= -e_{2}=q$}

When the electric force between charges is attractive, the value of $\kappa
^{2}\left( H-mf(\tau )+\frac{m}{f(\tau )}\right) ^{2}-8\kappa e_{1}e_{2}$ is
always positive and two particles follow a bounded periodic motion described
by tanh-type solution (\ref{TanhA}). The period is determined from the
initial value of $p_{0}=\sqrt{(H/2)^{2}-m^{2}}$ at $r=0$: 
\begin{equation}
T=\frac{16}{\kappa m\gamma _{e}}\;\mbox{tanh}^{-1}\left( \frac{\gamma _{e}%
\sqrt{H^{2}-4m^{2}}}{(2-\gamma _{e})H}\right) \;.  \label{period}
\end{equation}
Although the above expression diverges when $\gamma _{e}=\frac{2H}{2p_{0}+H}$%
, this situation is never realized in the attractive case, since $\gamma
_{e}<1$ \ whereas $\frac{2H}{2p_{0}+H}>1$.

In Figs.1 and 2 we plot $r(\tau )$ and the phase space trajectories ,
respectively, for two particles initially at $r=0$ for fixed $|q|=1$ and
four different values of $H$. We see that as $H$ increases, the phase space
trajectory becomes more $S$-shaped, $r(\tau )$ becomes steep and the
amplitude increases, but the period approaches a constant value $(16/\kappa
m\gamma _{e})\;\mbox{tanh}^{-1}[\gamma _{e}/(2-\gamma _{e})]$. This $S$%
-shaped deformation of the trajectory at higher energy is caused by the $p$%
-dependence of the gravitational potential and is common to the $q=0$ case
described with the $W$ function \cite{2bd}.
\begin{figure}
\begin{center}
\epsfig{file=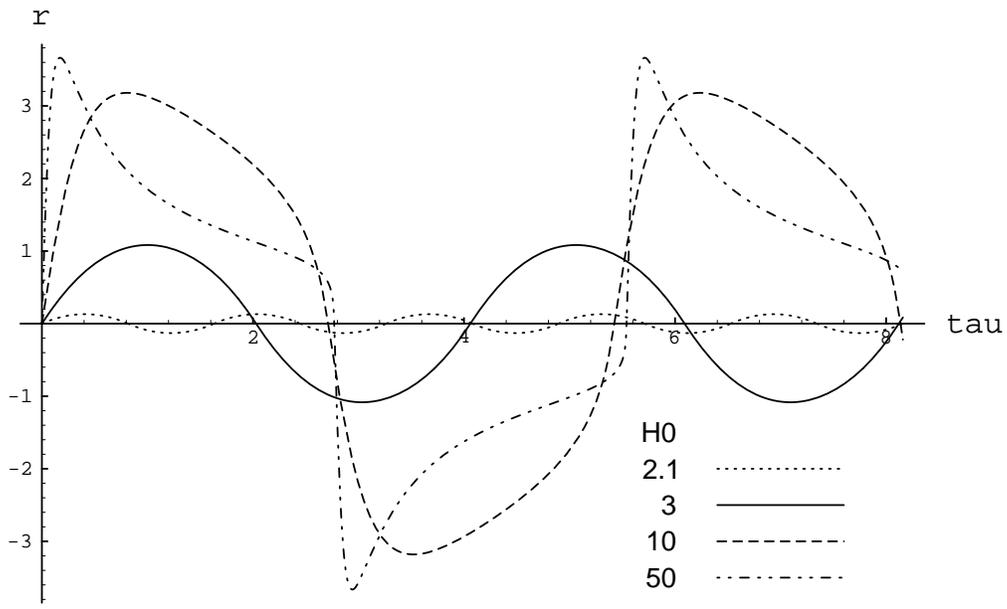,width=0.8\linewidth}
\end{center}
\caption{The exact $r(\tau)$ plots for $q=1$ and four different values of $H_0$}
\label{fig1}
\end{figure}
\begin{figure}
\begin{center}
\epsfig{file=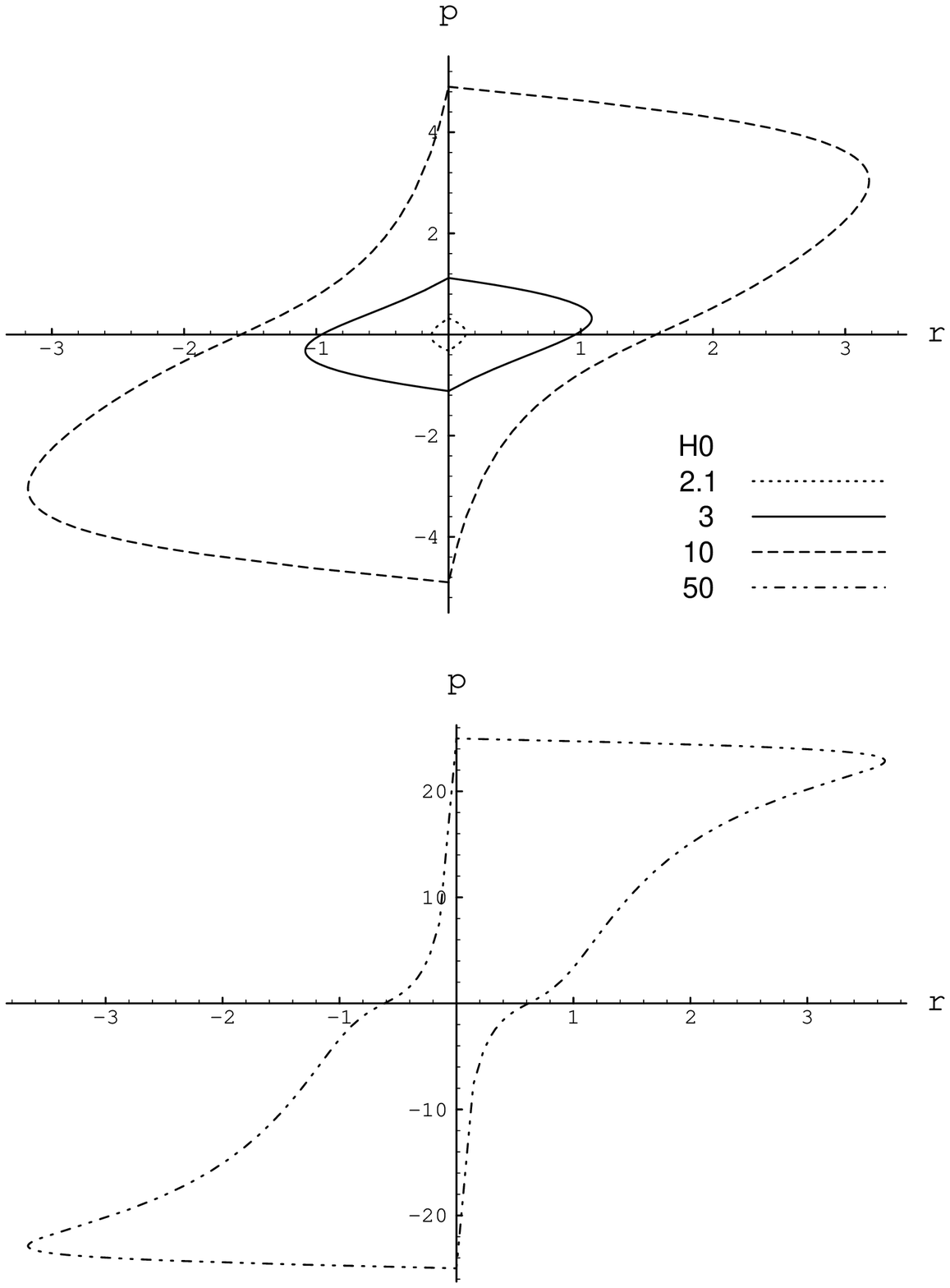,width=0.8\linewidth}
\end{center}
\caption{Phase space trajectories correponding to the $r(\tau)$ plots in Fig.1.}
\label{fig2}
\end{figure}
\noindent Figs.3 and 4 show similar plots for fixed $H=3$ and four different
values of $|q|$. For large $|q|$ ($\gamma _{e}<0$) due to the attractive
force between charges the phase space trajectory contracts toward the
origin, and the period and the amplitude of $r(\tau )$ become small. As $|q|$
becomes small, $\gamma _{e}$ passes zero at $|q|=1/\sqrt{2}$ and approaches
1 at $|q|=0$. As we can see from the figures, no peculiar changes in either $%
r(\tau )$ or the phase space trajectories occur except for a growing
amplitude and period, a natural tendency due to the weakening attractive
force.
\begin{figure}
\begin{center}
\epsfig{file=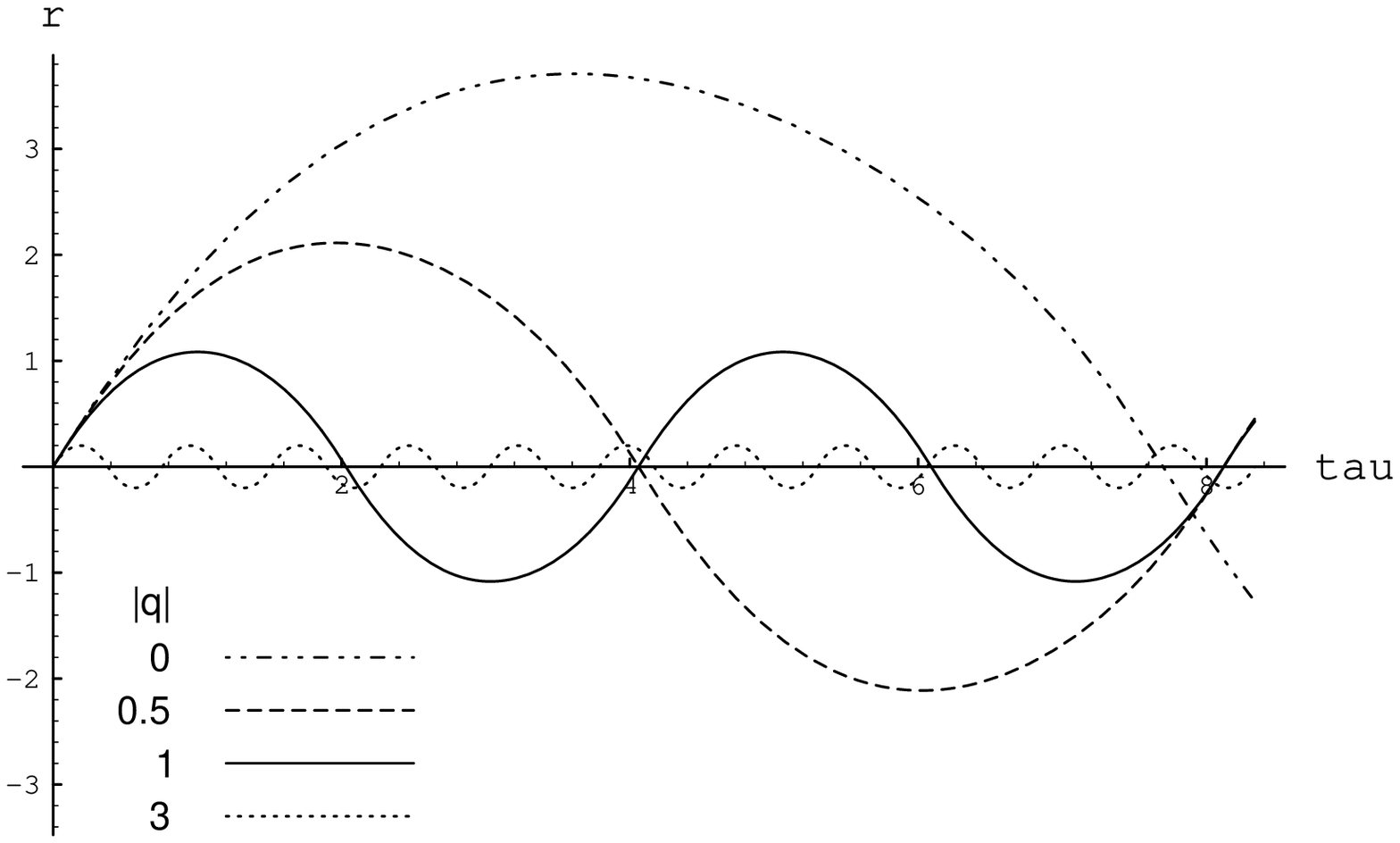,width=0.8\linewidth}
\end{center}
\caption{The $r(\tau)$ plots for $H_{0}=3$ and four different values of $|q|$. }
\label{fig3}
\end{figure}
\begin{figure}
\begin{center}
\epsfig{file=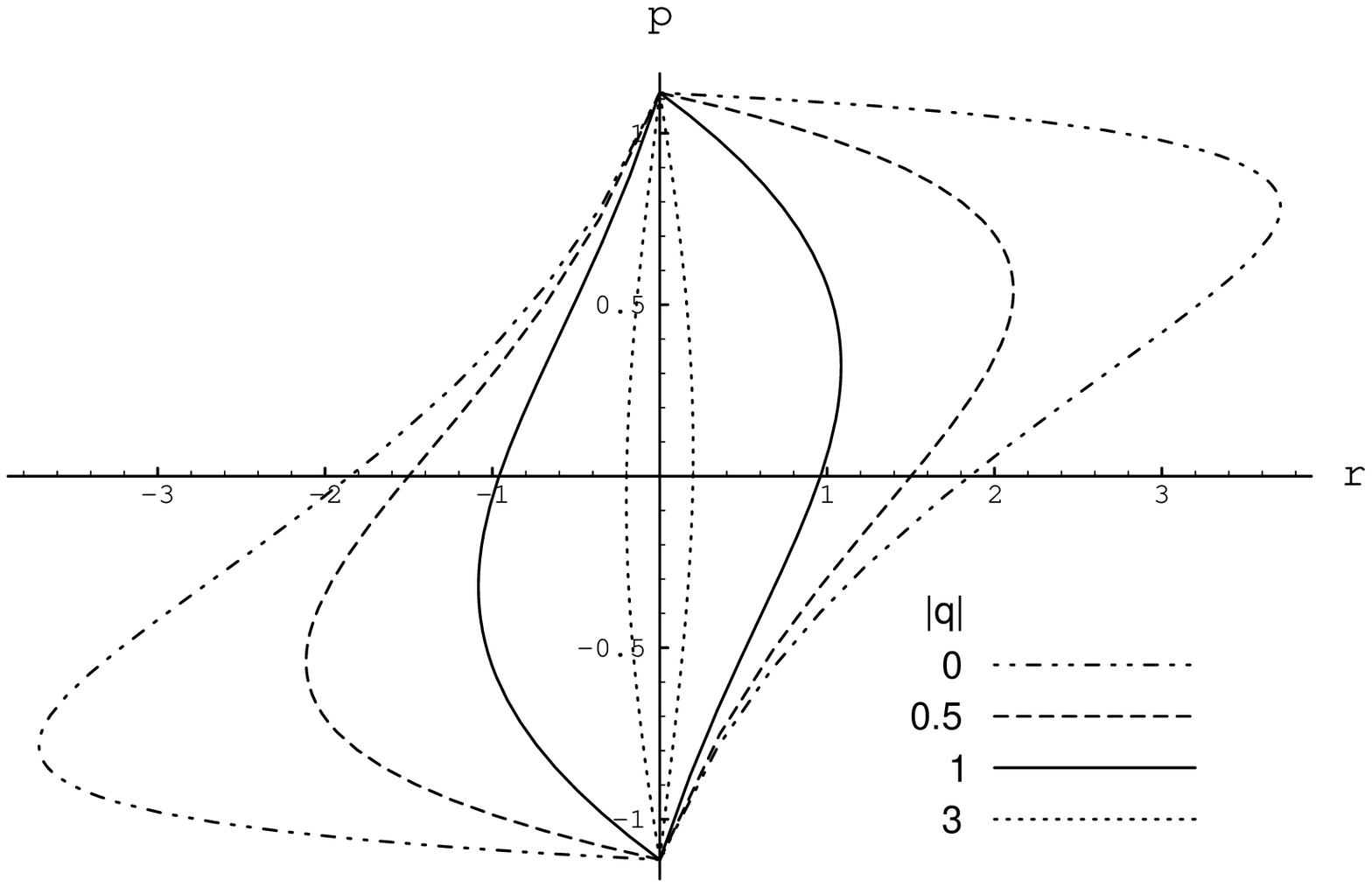,width=0.8\linewidth}
\end{center}
\caption{Phase space trajectories correponding to the plots in Fig.3.}
\label{fig4}
\end{figure}

The distinct characteristics of these relativistic motions are more easily
seen by comparing the exact trajectory with those of the motions in three
approximations : \newline
(1) the non-relativistic motion described by the Hamiltonian 
\begin{equation}
H=2m+\frac{p^{2}}{m}+\frac{q^{2}}{2}\;|r|+\frac{\kappa m^{2}}{4}\;|r|\;,
\label{Hnonrel}
\end{equation}
(2) the linear approximation in $\kappa $, its Hamiltonian being 
\begin{eqnarray}
H&=&2\sqrt{p^2 + m^2}+\frac{1}{2}q^2 |r| +\frac{\kappa}{4}\left\{(2p^2 + m^2
- 2\epsilon \tilde{p}\sqrt{p^2 + m^2})|r| \right.  \nonumber \\
&&\left.+\frac{1}{2}q^2 (\sqrt{p^2 + m^2}-\epsilon\tilde{p})r^2 + \frac{1}{24%
}q^4 |r|^3 \right\}\;,
\end{eqnarray}
\newline
\noindent (3) the $\kappa \rightarrow 0$ limit theory, namely,
special-relativistic electrodynamics in (1+1) flat space-time described by
the Hamiltonian 
\begin{equation}
H=2\sqrt{p^{2}+m^{2}}+\frac{q^{2}}{2}\;|r|\;.
\end{equation}
In Fig.5 the trajectories of the exact and the three approximate solutions
under identical total energy $H_{0}=3$ are drawn for $q=0.5,1,5$ and $10$.
For small $q=0.5$ both the exact solution (solid curve) and the linear
approximation (dashed curve) follow the $S$-shaped trajectories, while the
trajectories of the non-relativistic solution (dotted curve) and flat
electrodynamics (dot-dashed curve) have symmetrical oval forms. As $|q|$
increases all trajectories tend to coincide with the trajectory of flat
electrodynamics (since the effect of gravity becomes relatively weak) though
the initial value of $p_{0}=\sqrt{m(H-2m)}=1$ in the non-relativistic case
is slightly different from the value of others ($p_{0}=\sqrt{(H/2)^{2}-m^{2}}%
=\sqrt{5}/2$). If the motion in the non- relativistic case starts from the
same initial value $p_{0}=\sqrt{5}/2$ the oval trajectory for large $|q|$
region as shown in Fig.6 becomes larger than others, reflecting the
difference between the relativistic and non- relativistic effects.
\begin{figure}
\begin{center}
\epsfig{file=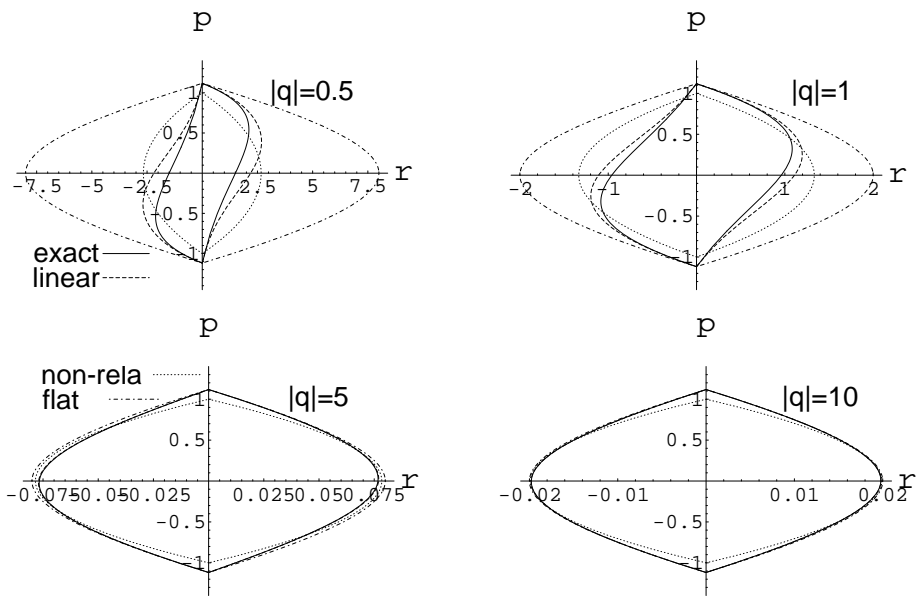,width=0.8\linewidth}
\end{center}
\caption{The phase space trajectories of $H_{0}=3$ for the exact, the linear,
the non-relativistic cases and the flat electrodynamics.  }
\label{fig5}
\end{figure}
\begin{figure}
\begin{center}
\epsfig{file=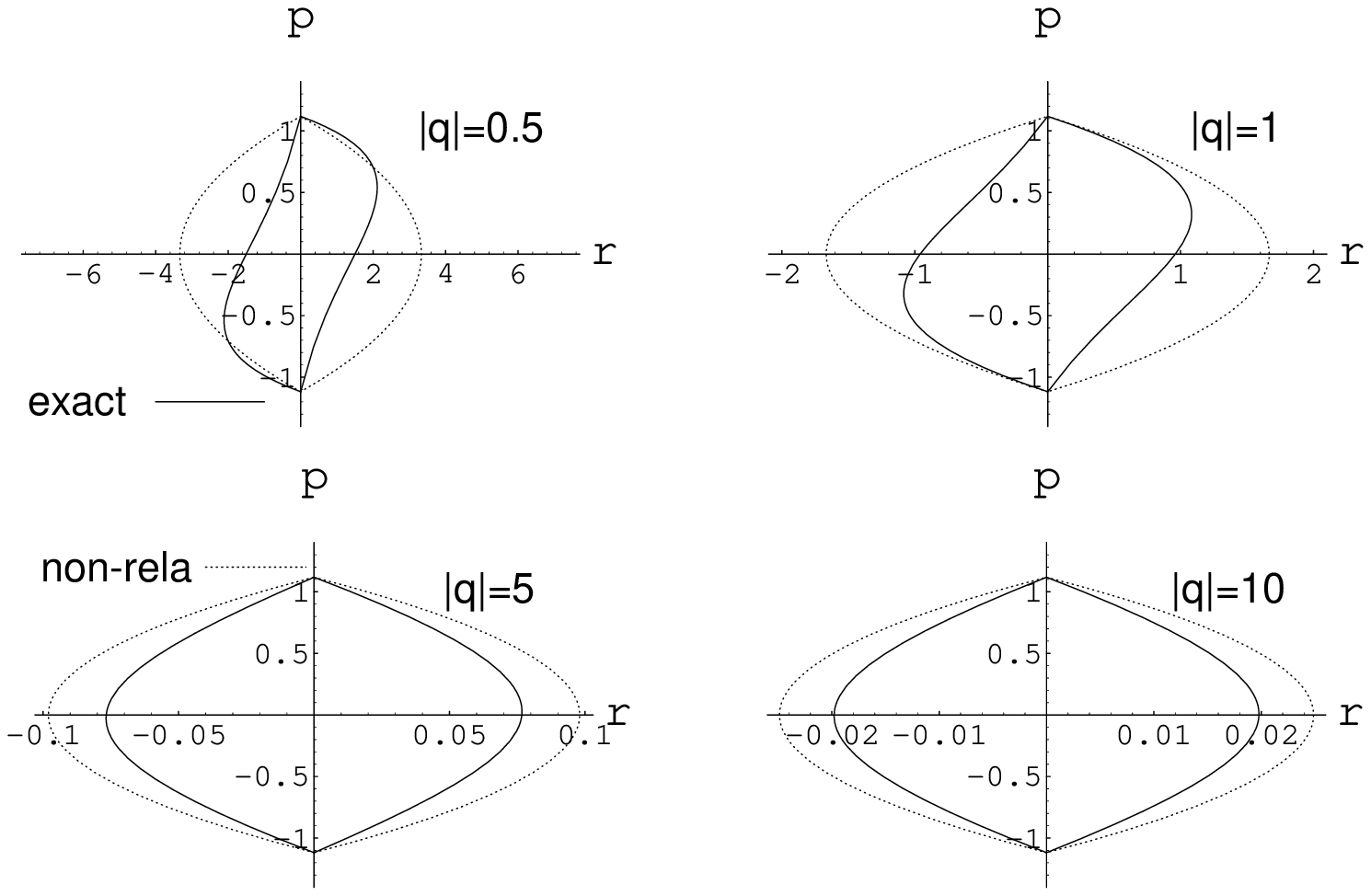,width=0.8\linewidth}
\end{center}
\caption{Phase space trajectories of the relativistic and the non-relativistic
cases under the same initial condition $p_{0}=\sqrt{5}/2$ at $r=0$. }
\label{fig6}
\end{figure}

\subsection{The repulsive case: the small $|q|$ regime ($e_{1}=e_{2}=q,|q|%
\leq q_{c}$)}

For the case where the charges are of the same sign, the electric force is
repulsive and competes with the attractive gravitational force. Depending on
the strengths of the two forces the solutions become either tanh-type and/or
tan-type. The transition from tanh-type to tan-type is given by the zeroes
of ${\cal J}=\sqrt{(H-2\epsilon \tilde{p})^{2}-8q^{2}/\kappa }$, which lead
to a critical value of the charge 
\begin{equation}  \label{qc}
q_{c}=\sqrt{\frac{\kappa }{8}}\left( H-\sqrt{H^{2}-4m^{2}}\right) \;,
\end{equation}
which separates two qualitatively different kinds of motion. In the regime
of $|q|<q_{c}$, $({\cal J})^{2}$ takes both positive and negative values and 
$r(\tau )$ obeys both tanh-type and tan-type solutions, representing bounded
and unbounded motions respectively. Alternatively, (\ref{qc})
gives\thinspace the critical value of $H$ for fixed $\kappa $ and $q$, or
the critical value of $\kappa $ for fixed $H$ and $q$, both corresponding to
the transition from bounded to unbounded motion.

In Fig.7 we show $r(\tau )$ plots for $H_{0}=3$ and five different values of 
$q$. As $|q|$ increases the period becomes large. When $|q|$ exceeds the
critical value $q_{c}=0.2700907567$, the motion becomes unbounded and the
separation of the two particles diverges at finite $\tau $. A similar
transition is found in Fig.8, where $r(\tau )$ plots are depicted for $%
|q|=0.1$ and five different values of $H_{0}$ and the transition occurs at $%
H_{0}=7.21249$.

\begin{figure}
\begin{center}
\epsfig{file=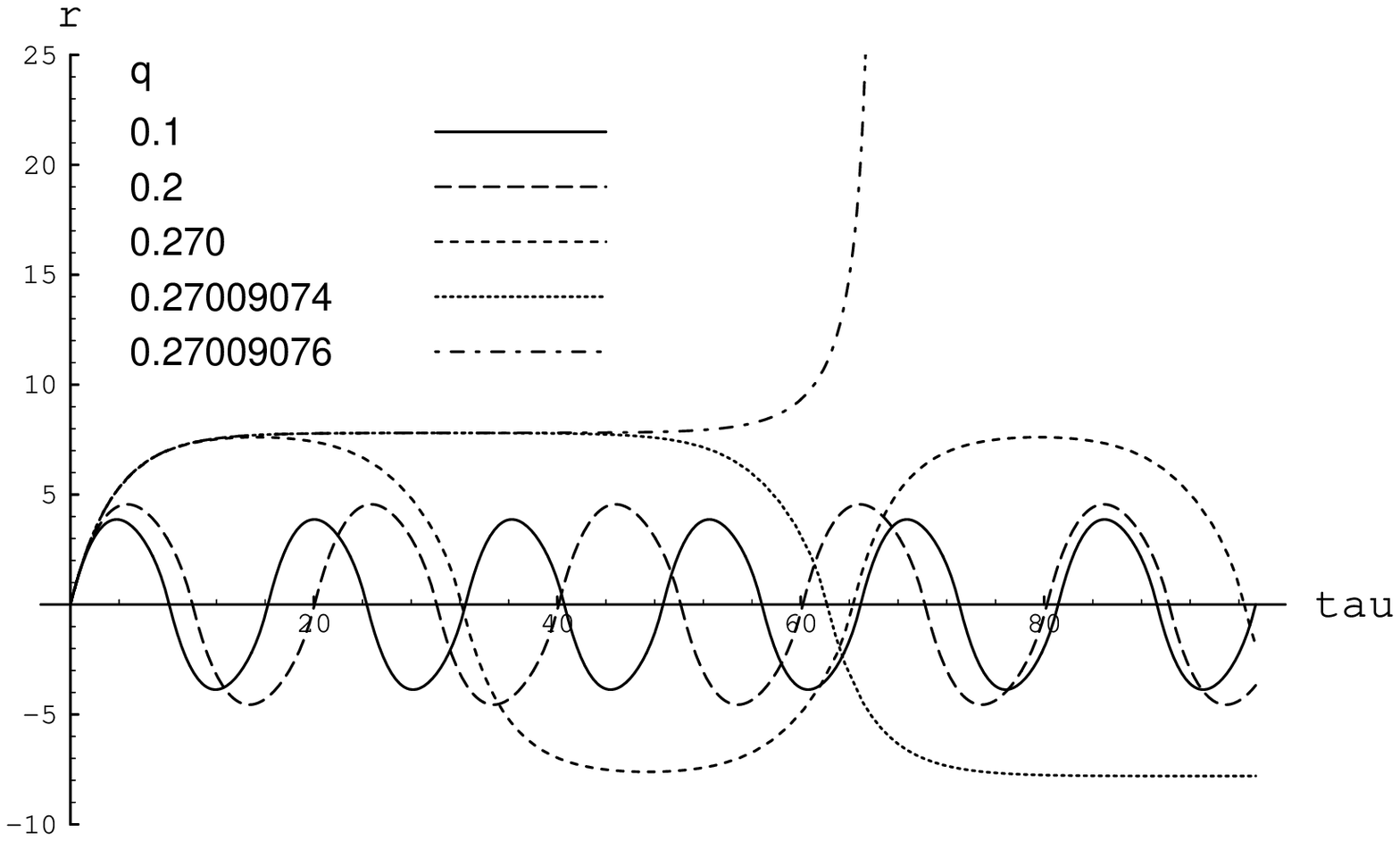,width=0.8\linewidth}
\end{center}
\caption{The $r(\tau)$ plots for $H_{0}=3$ and five different values of $|q|$ in
the repulsive case.}
\label{fig7}
\end{figure}
\begin{figure}
\begin{center}
\epsfig{file=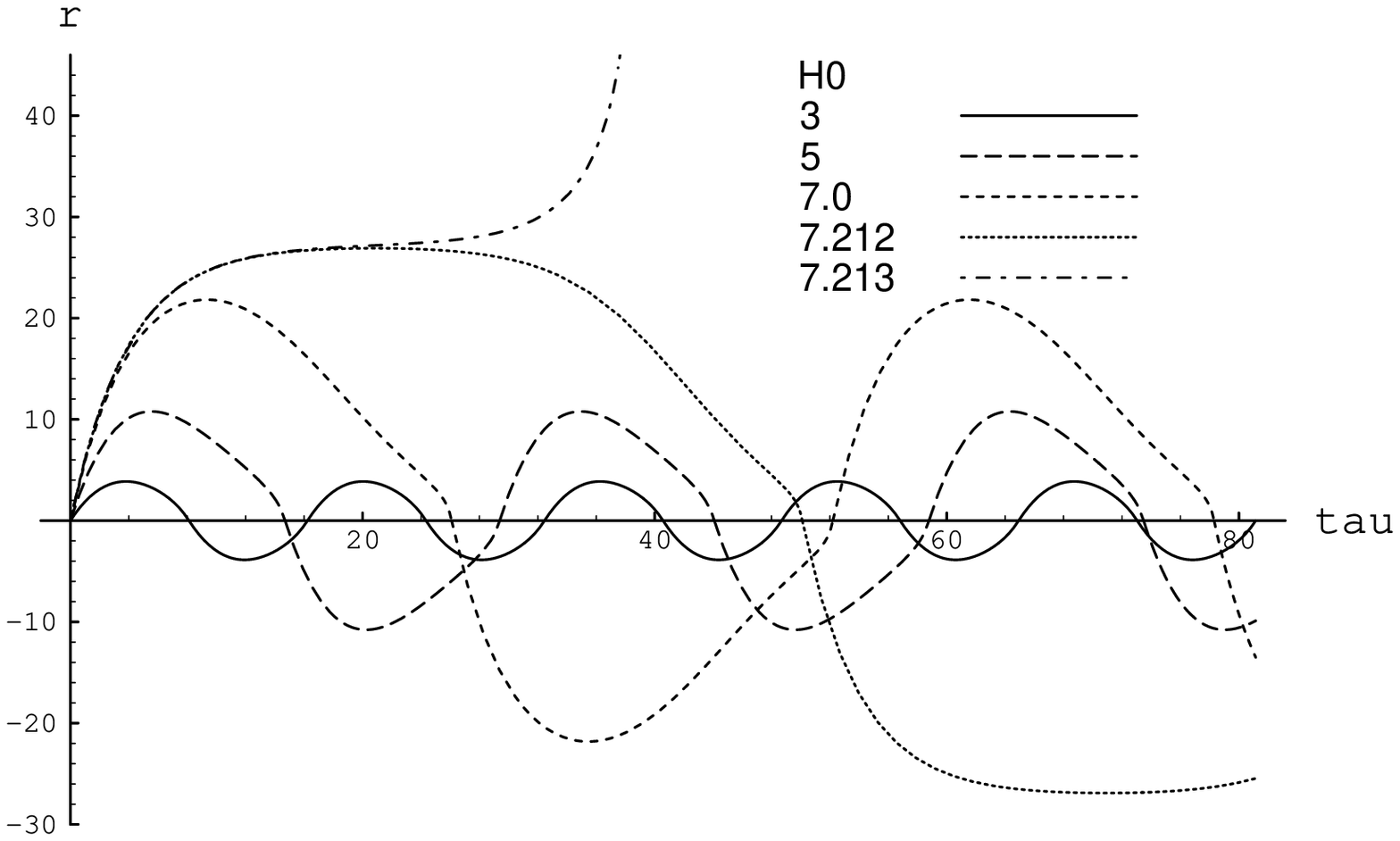,width=0.8\linewidth}
\end{center}
\caption{The $r(\tau)$ plots for $|q|=0.1$ and five different values of $H_{0}$
in the repulsive case.}
\label{fig8}
\end{figure}

Before proceeding to the analysis of the phase space trajectories, it is
instructive to compare the general structure of the determining equation for
the repulsive case with that in no-charge case. In the $q=0$ case the
determining equation (\ref{H1}) of the Hamiltonian for equal masses becomes 
\begin{equation}  \label{YZeq}
Y^2\;e^{2Y}=Z^2\;e^{2Z} \;,
\end{equation}
where $Y$ and $Z$ are defined by 
\begin{eqnarray}
H&=&\sqrt{p^2 + m^2} + \epsilon \tilde{p} - \frac{8}{\kappa |r|}Y(p,r)\;, \\
Z&=&\frac{\kappa |r|}{8}(\sqrt{p^2 + m^2} - \epsilon \tilde{p})\;.
\end{eqnarray}
Equation (\ref{YZeq}) has three formal solutions shown in Fig.9: \newline

sol 1:\hspace{5mm} line F-O ;\hspace{15mm} trivial solution $Y=Z$, \newline

sol 2:\hspace{5mm} curve A-B-O-C-D; $Y=W(-Z\;e^{Z}), \quad Z <
W^{-1}(e^{-1})=0.278$, \newline

sol 3:\hspace{5mm} curve E-F-G;\hspace{10mm} $Y=W(Z\;e^{Z}),\quad Z<0$. 
\newline
Here $W(x)$ is the Lambert $W$ function defined via 
\begin{equation}
y\cdot e^{y}=x\quad \Rightarrow \quad y=W(x)\;.
\end{equation}
Since $Z>0$ for real $p$, only sol.2 represents a physical solution. (Sol.1
holds for the special case of massless particles with no interactions.)
\begin{figure}
\begin{center}
\epsfig{file=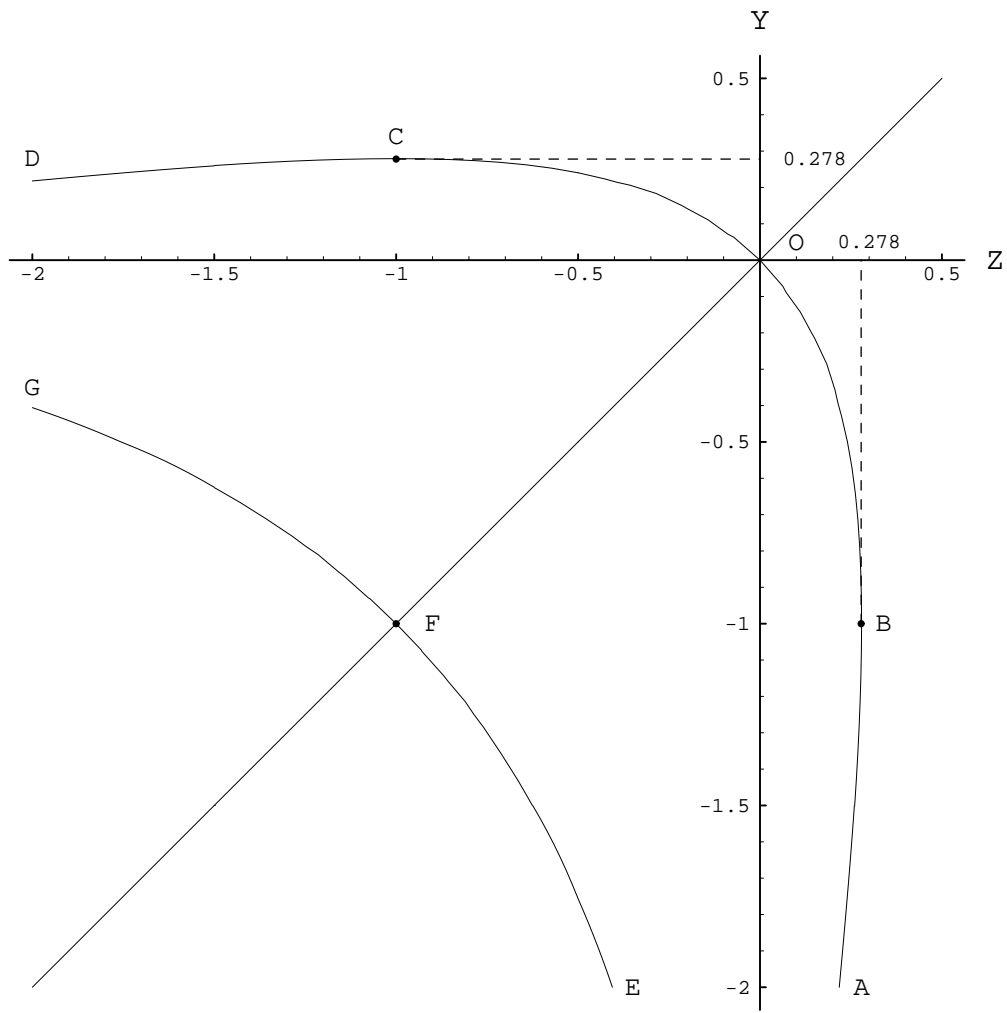,width=0.8\linewidth}
\end{center}
\caption{Solution to Eq.(\ref{YZeq}). The points $B$ and $C$ represent the
extremal $Z$ and $Y$ values of $W^{-1}(1/e)=0.278$. }
\label{fig9}
\end{figure}

Of the four types of the determining equations (\ref{TanhA})-(\ref{TanB}) in
which ${\cal J}_{\Lambda }$ is replaced by ${\cal J}$, tanh-type A is a
generalization of sol.2 and tanh-type B is that of sol.3. Tan-type A and
tan-type B are new equations appearing specifically in the repulsive case
and sol.1 is a special case of \ tan-type A and B solutions with ${\cal J}=0 
$ and $q=0$. \ Tan-type A and tan-type B trajectories yield a countably
infinite series of unbounded motions of the particles. This is in strong
contrast with the Newtonian case, in which only one trajectory exists for
fixed $H$ and $|q|$.

Fig.10 is a diagram of the physical region of $(|q|,p)$ parameter space in
the case of $H_{0}=3$. The shaded area of ${\cal J}^{2}>0$ and $B>0$ is the
region where tanh-type A and tanh-type B give the actual trajectories. The
boundary of this area is fixed by $p=\pm p_{0}=\pm \sqrt{(H_{0}/2)^{2}-m^{2}}
$ and $p=-\sqrt{2/\kappa }\;|q|+H_{0}/2$. The values of $|q|$ at the
intersections of these boundary lines are denoted as $q_{c}$ and $q_{d}$, of
which $q_{c}$ is the critical value (\ref{qc}). In this area the tanh-type B
solution is realized in a quite narrow region between $p=-\sqrt{2/\kappa }%
\;|q|+H_{0}/2$ and ${\cal J}-B=0$ (dashed curve). The motions of tan-type A,
B are realized in the area of ${\cal J}^{2}<0$ whose boundaries are $p=\pm 
\sqrt{2/\kappa }\;|q|+H_{0}/2$. Consider a $|q|=const$ line (the dotted line
in Fig.10) and define by $p_{1}$ and $p_{2}$ the momenta of the
intersections of the line with $p=\pm \sqrt{2/\kappa }\;|q|+H_{0}/2$ .
\begin{figure}
\begin{center}
\epsfig{file=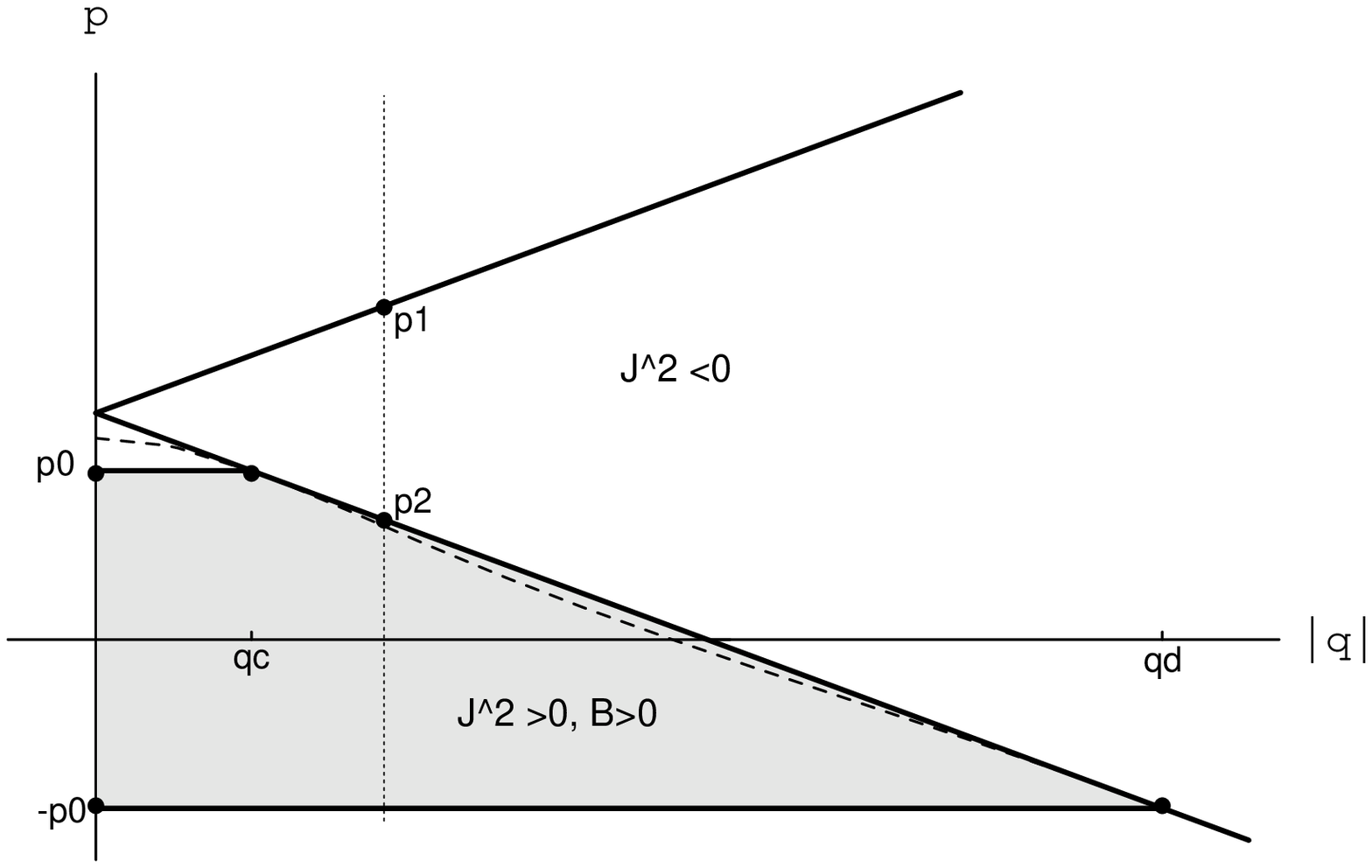,width=0.8\linewidth}
\end{center}
\caption{The diagram of the physical region of $(|q|, p)$ for $H_{0}=3$ 
in the charged repulsive case. }
\label{fig10}
\end{figure}

For the case of $0<|q|<q_{c}$ the allowed value of $p$ is divided into two
parts: $-p_{0}<p<p_{0}$ and $p_{2}<p<p_{1}$. The solution in the former
region is tanh-type A and the solutions in the latter region are both
tan-type A and B. We show in Fig.11 the phase space trajectories for $%
H_{0}=3 $ and $|q|=0.25$, in which the solid curve ($N$) denotes the bounded
motion given by tanh-type A, and the dotted ($A_{n}:n=0,1,\ldots $) and the
dashed ($B_{n}:n=1,2,\ldots $) curves represent the infinite series of
unbounded motions specified by tan-type A and tan-type B, respectively. For
the unbounded motions $p_{1}$ and $p_{2}$ are the asymptotic values of the
momentum and the two particles simply approach one another at some minimal
value of $|r|$ and then reverse direction toward infinity. In the figure we
added also the trajectory \ of flat-space electrodynamics (a dot-dashed
curve), which is the only solution of the theory for given values of $H_{0}$
and $|q|$. The motion of tan-type A, B has a specific feature that $r(\tau )$
becomes infinite at a finite proper time (but an infinite coordinate time).
In Appendix C we shall present a simple model in flat space-time that has
this feature. Fig.12 shows the $r(\tau )$ plots for $H_{0}=3$ and $|q|=0.25$.
\begin{figure}
\begin{center}
\epsfig{file=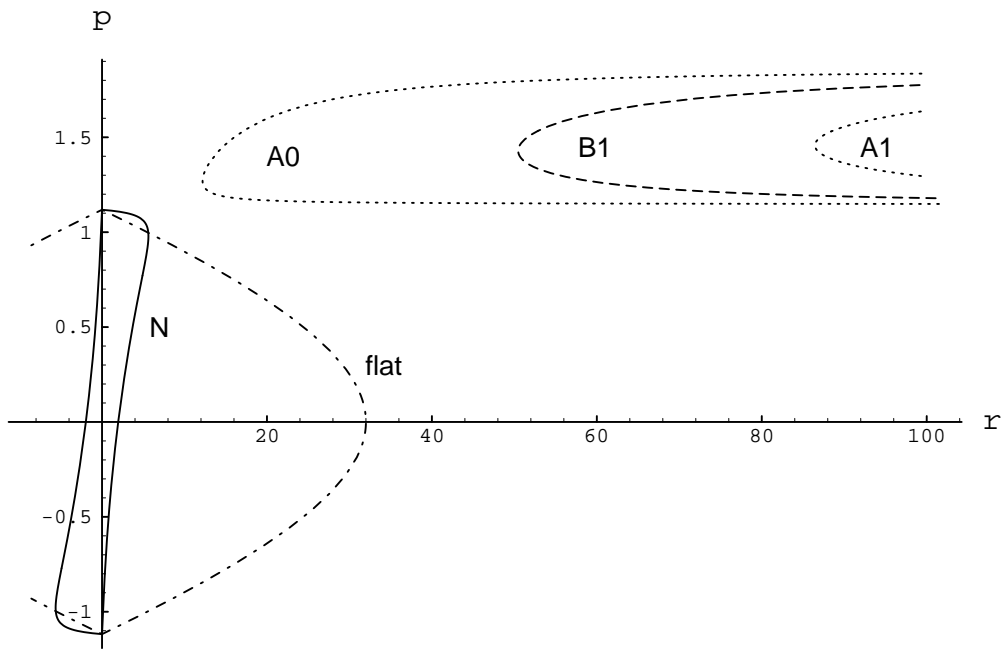,width=0.8\linewidth}
\end{center}
\caption{Phase space trajectories of the bounded and the unbounded motions 
for $H_{0}=3, m=1$ and $|q|=0.25$.}
\label{fig11}
\end{figure}
\begin{figure}
\begin{center}
\epsfig{file=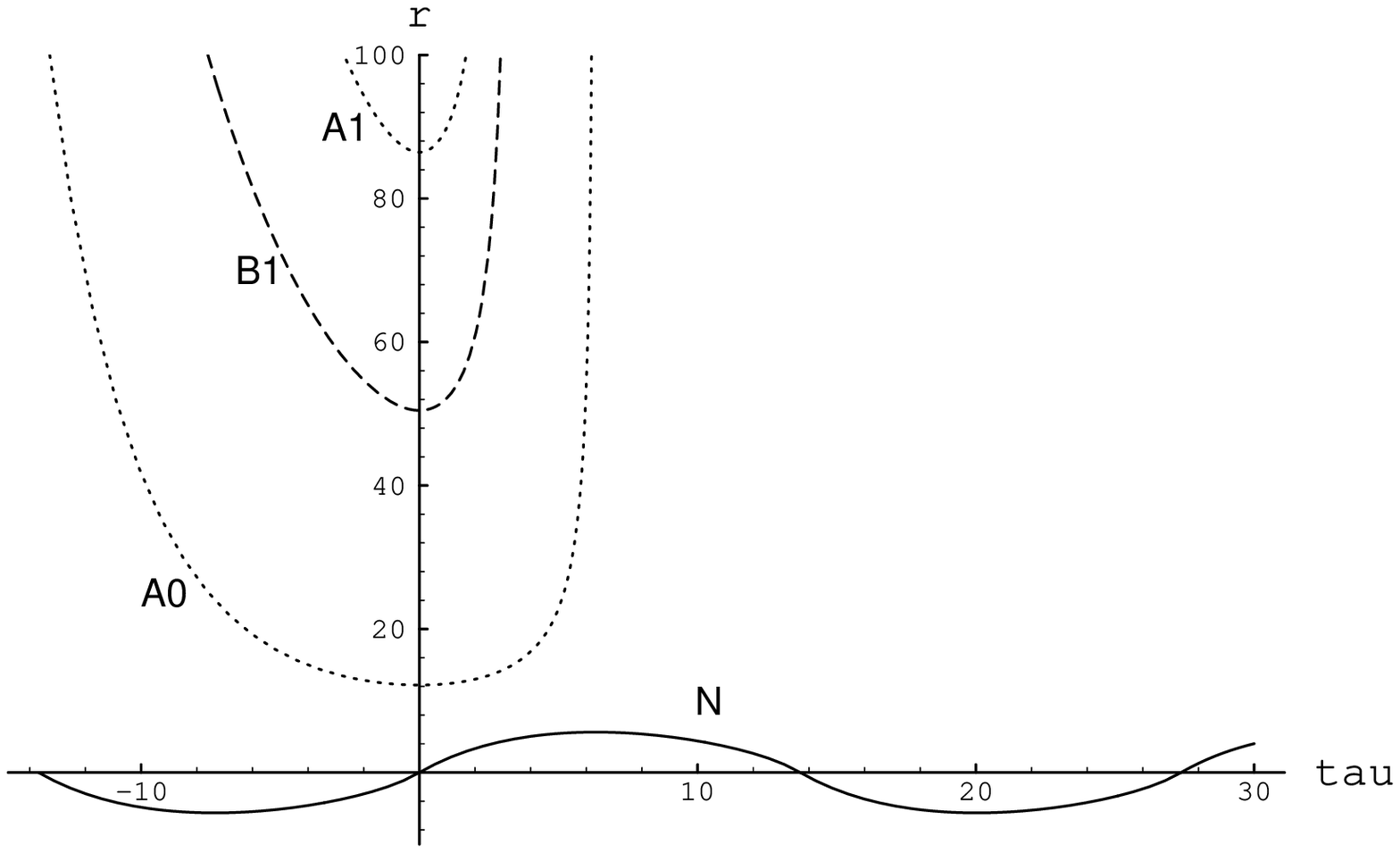,width=0.8\linewidth}
\end{center}
\caption{The $r(\tau)$ plots for the parameters $H_{0}=3, m=1$ and $|q|=0.25$.}
\label{fig12}
\end{figure}

As $|q|$ approaches $q_{c}$ the trajectories ``$N$ '' and ``$A_{0}$ '' come
close to one another, meeting at $|q|=q_{c}$, and then for $|q|$ beyond the
critical value they form new unbounded trajectories ``$N1$'' and ``$N2$'' as
shown in Fig.13. This latter case will be discussed in the next subsection.
\begin{figure}
\begin{center}
\epsfig{file=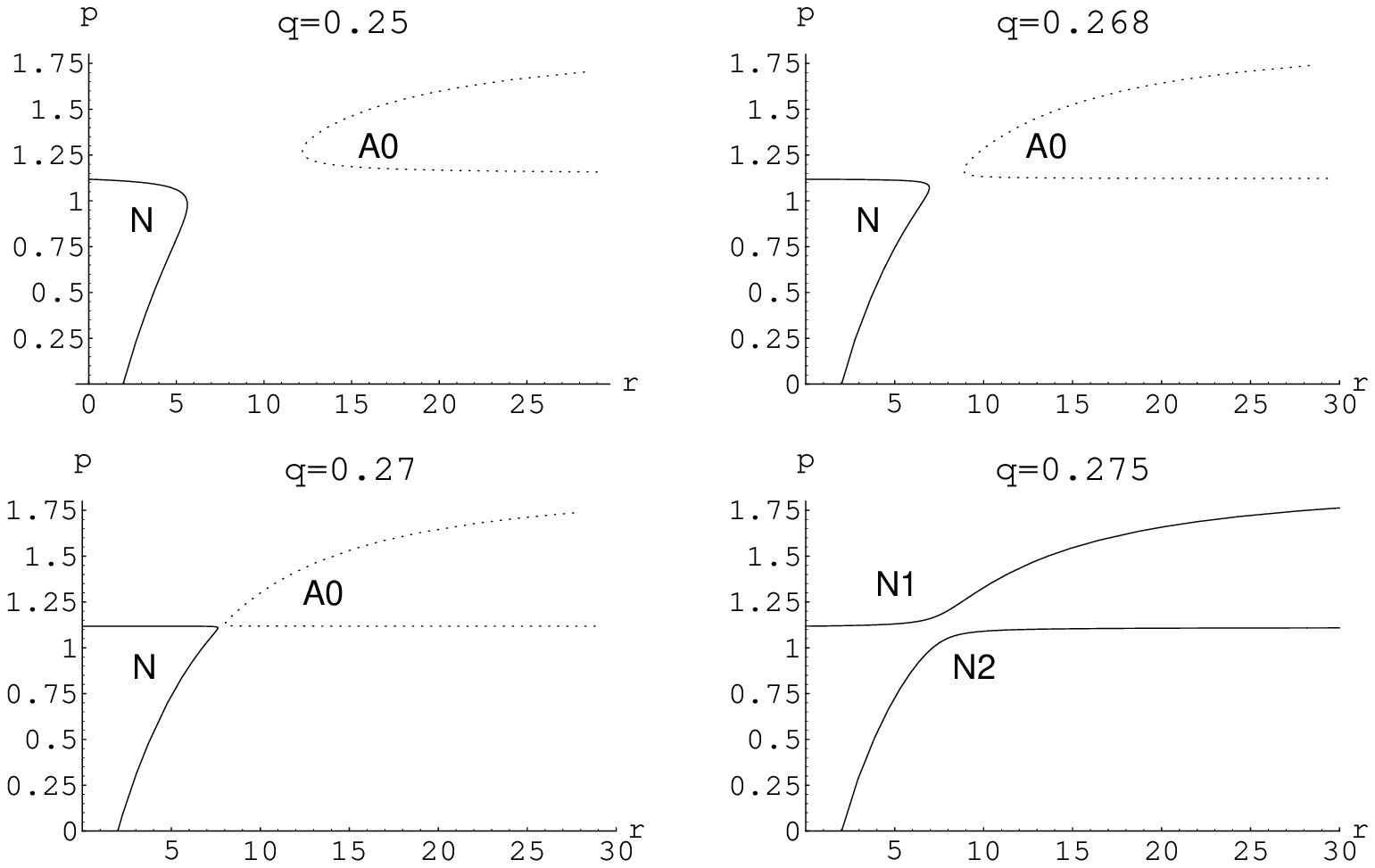,width=0.8\linewidth}
\end{center}
\caption{Transition from the bounded motion to the unbounded motion across 
$|q|=q_{c}=0.2700907567$ for $H_{0}=3$. }
\label{fig13}
\end{figure}
It should be stressed that the existence of two types of motion for fixed $H$
and $q$ is a new aspect of the relativistic gravitation theory, and has no
non-relativistic analogue.

\subsection{The repulsive case: the large $|q|$ regime ($|q|>q_{c}$)}

For $|q|$ larger than $q_{c}$ the electric repulsive force overwhelms the
attractive gravitational one and only unbounded motion is allowed. For the
case of $q_{c}<|q|<q_{d}$ the allowed value of $p$ is $-p_{0}<p<p_{1}$.
Fig.14 shows the phase space trajectories of $H_{0}=3$ and $|q|=0.3$. Here
the new unbounded trajectories $N1$ and $N2$ are realized instead of $N$ and 
$A_{0}$. The solution corresponding to the shaded area $-p_{0}<p<p_{2}$ is $%
N2$, while the solutions for $p_{2}<p<p_{1}$ are $N1$, $A_{n}$ and $%
B_{n}\;(n=1,2,\ldots )$ and they are all described by tan-type A and B. The
trajectories in $r<0$ region are obtained from those in $r>0$ by replacing
the signs of both $r$ and $p$. By comparing all these trajectories with the
analogous trajectory in flat space electrodynamics(a dot-dashed curve), we
can see how the effects of gravity deform the flat-space trajectory.
(Remember that there exist also time-reversed trajectories with $\epsilon
=-1 $.)

The phase space trajectories for the case of $|q|=2>q_{d}$ are depicted in
Fig.15. Since $p_{2}$ is smaller than $-p_{0}$ all solutions are tan-type A
and B, and a characteristic cusp appears at $r=0$ in the trajectories $N1$
and $N2$. In the figure the trajectory specified by the symbol $B01$ is a
combination of $B_{0}+B_{1}$, indicating that the solution switches between $%
B_{0}$ and $B_{1}$, namely, $B_{0}$ for $-p_{0}\leq p\leq p_{0}$ and $B_{1}$
for $p_{2}\leq p\leq -p_{0}$ and $p_{0}\leq p\leq p_{1}$. Similarly $B12$ is
composed of a combination of $B_{1}+B_{2}$.
\begin{figure}
\begin{center}
\epsfig{file=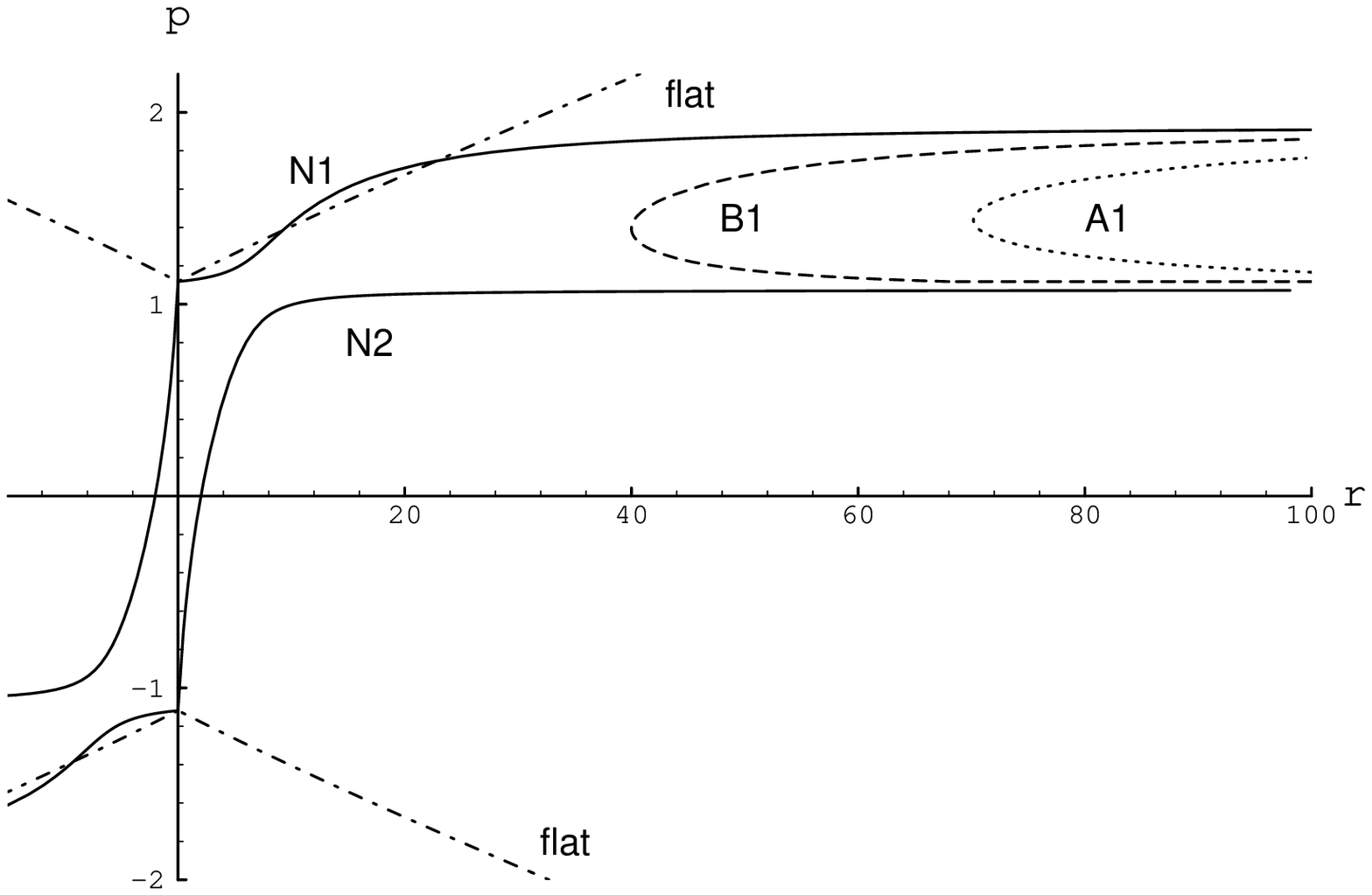,width=0.8\linewidth}
\end{center}
\caption{Phase space trajectories of the unbounded motions for $H_{0}=3$ and 
$|q|=0.3$. }
\label{fig14}
\end{figure}
\begin{figure}
\begin{center}
\epsfig{file=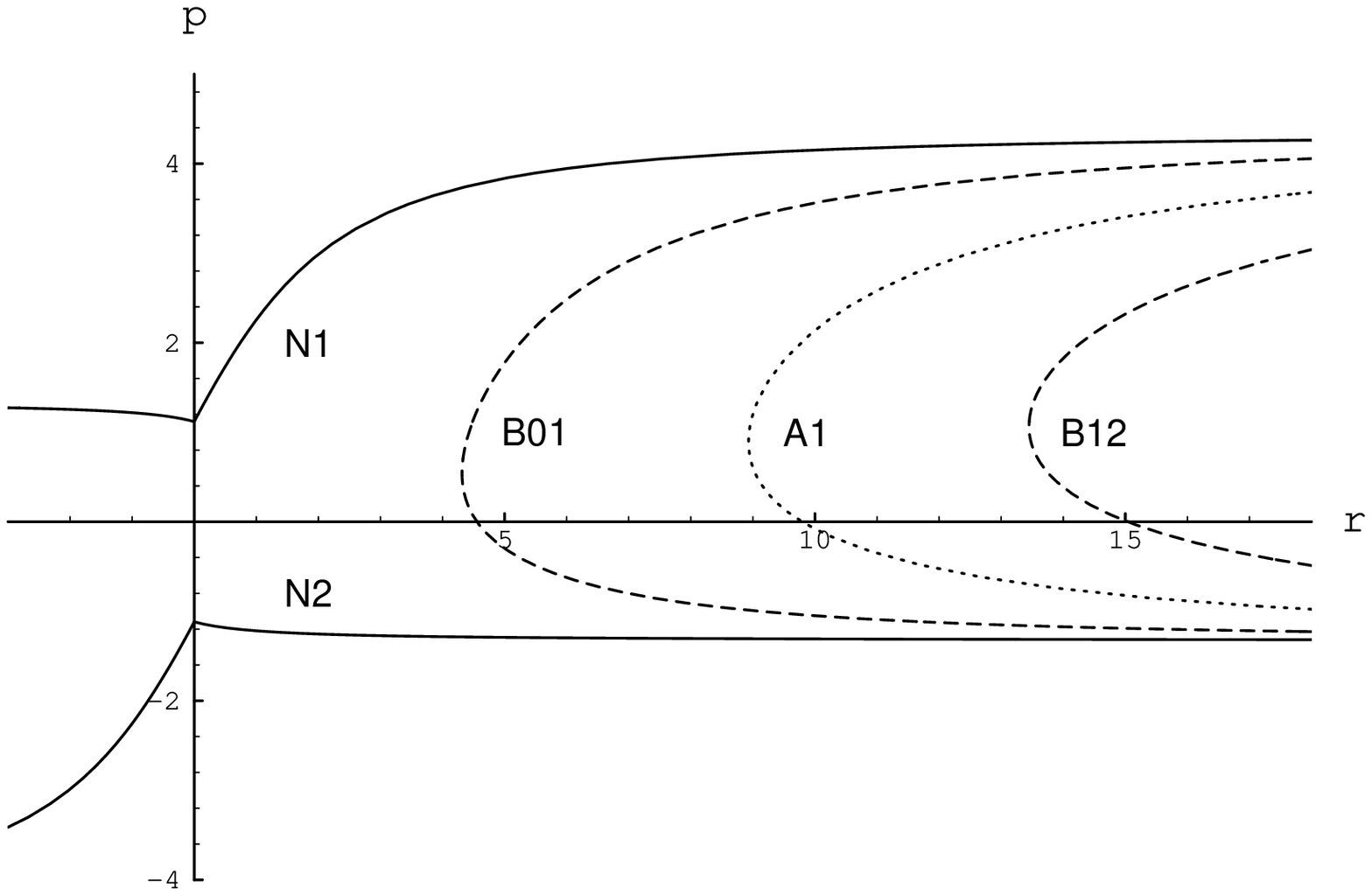,width=0.8\linewidth}
\end{center}
\caption{Phase space trajectories of the unbounded motions for $H_{0}=3$ and 
$|q|=2$.}
\label{fig15}
\end{figure}

\subsection{$H < 2m$ case for repulsive charges}

In 1+1 flat-space electrodynamics we know that for attractive charges the
total energy of the system is restricted to $H\geq 2m$, but for repulsive
charges no restriction on $H$ exists. Unbounded motion is also realized for $%
H<2m$ (the explicit solution is presented in Appendix B). It is expected
that in a general relativistic theory the restriction on $H$ is identical.
>From the $(|q|,p)$ diagram in Fig.10 we know that for $H<2m$ the shaded
area disappears and only the region of ${\cal J}^{2}<0$ remains for the
unbounded motions. In Fig.16 we show the phase space trajectories for $H=1$
and $|q|=1$. All types of unbounded motions $A_{n}$ and $B_{n}$ are
realized. The unbounded trajectories $N1$ and $N2$ that appeared in $H\geq
2m $ turn to $A_{0}$ again. As compared with the flat-space trajectory (a
dot-dashed curve) all trajectories are curved more toward the $r$-axis (due
to the additional effect of gravitational attraction) and are shifted in the
direction of the positive $p$-axis. This shift is caused by the $p$%
-dependence of the gravitational potential. As remarked previously there
exist corresponding trajectories with $\epsilon =-1$ and invariance under
time-reversal is retained.
\begin{figure}
\begin{center}
\epsfig{file=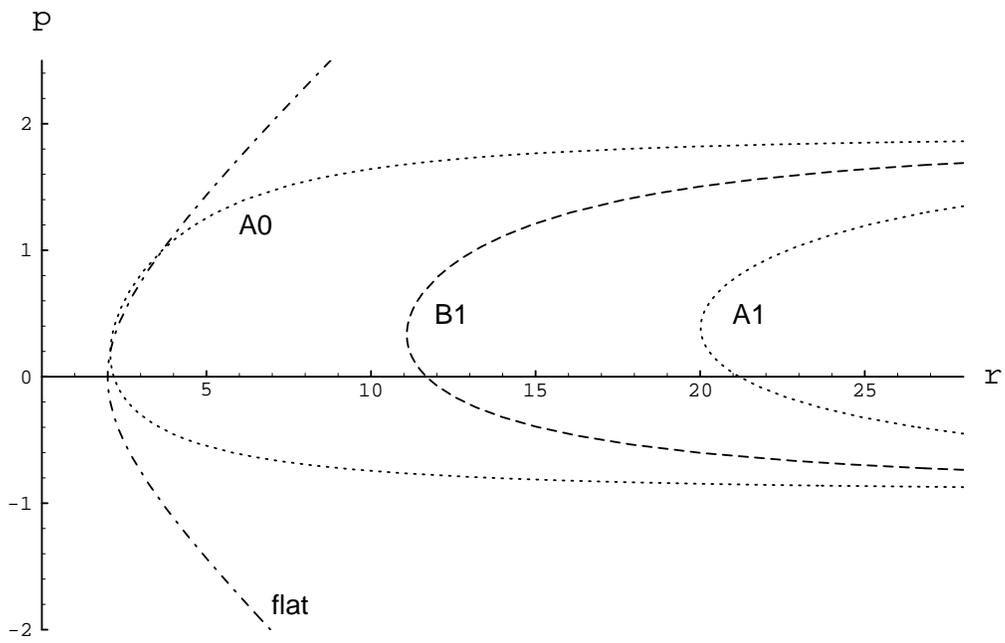,width=0.8\linewidth}
\end{center}
\caption{Phase space trajectories of the unbounded motions for $H_{0}<2m$
($m=1$, $H_{0}=1$ and $|q|=1$). }
\label{fig16}
\end{figure}

\section{ELECTRODYNAMICS WITH $\Lambda _{e}\neq 0$}

In a recent paper \cite{2bdcoslo} we presented a detailed analysis of
two-body motion with no charge, namely particle dynamics in lineal gravity.
The effects of a cosmological constant ($\Lambda _{e}=\Lambda $) may be
incorporated into a momentum-dependent potential between particles, as shown
in the two parameters expansion in terms of $\kappa $ and $\Lambda
_{e}/\kappa ^{2}$ 
\begin{eqnarray}
H &=&2\sqrt{p^{2}+m^{2}}+\frac{\kappa }{4}(\sqrt{p^{2}+m^{2}}-\epsilon 
\tilde{p})^{2}\;|r|+\frac{\kappa ^{2}}{4^{2}}(\sqrt{p^{2}+m^{2}}-\epsilon 
\tilde{p})^{3}\;r^{2}  \nonumber  \label{Happrox} \\
&&+\frac{7\kappa ^{3}}{6\times 4^{3}}(\sqrt{p^{2}+m^{2}}-\epsilon \tilde{p}%
)^{4}\;|r|^{3}-\frac{\Lambda _{e}}{2\kappa }\cdot \frac{\epsilon \tilde{p}}{%
\sqrt{p^{2}+m^{2}}}\;|r|-\frac{\Lambda _{e}}{16}\cdot \frac{\epsilon \tilde{p%
}\;m^{2}}{p^{2}+m^{2}}\;r^{2}  \nonumber \\
&&+\frac{\Lambda _{e}^{2}}{4\kappa ^{3}}\cdot \frac{\epsilon \tilde{p}}{%
(p^{2}+m^{2})^{3/2}}\;|r|+\cdot \cdot \cdot \;.
\end{eqnarray}

The exact phase space trajectories $(r(\tau ),p(\tau ))$ indicate that a
negative cosmological constant $\Lambda <0$ acts effectively as an
attractive force leading to bounded (periodic) motions, which are specified
by the tanh-type A equation (\ref{TanhA}), whereas a positive cosmological
constant acts effectively as a repulsive force. One noteworthy special
situation takes place for a particular range of negative $\Lambda $ and
small $m$: both $r(\tau )$ and the phase space trajectory have a double peak
structure. An example is shown in Figs.17 and 18 for $H_{0}=10,m=0.02$ and $%
\Lambda =-0.5$. Two particles starting at $r=0$ with initial momenta in
opposite directions depart one another, reach a maximum separation, and then
reverse direction due to the attractive force. However at some point they
reverse direction again, reaching a second maximum before returning to the
starting point. This peculiar behavior takes place due to the induced $p$%
-dependent $\Lambda $ potential combined with the gravitational attraction
and the relativistic effect.
\begin{figure}
\begin{center}
\epsfig{file=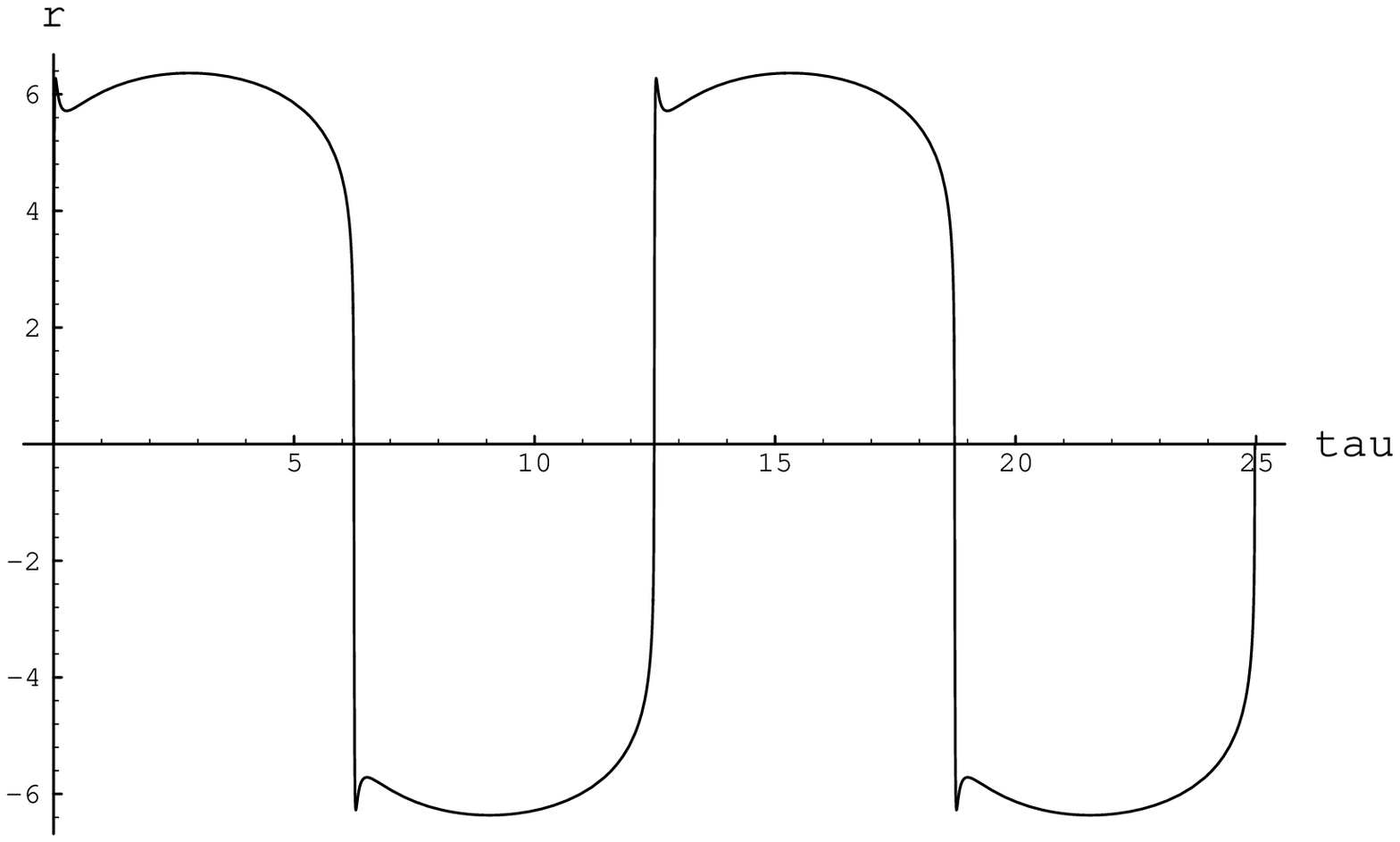,width=0.8\linewidth}
\end{center}
\caption{The $r(\tau)$ plot for $H_{0}=10, m=0.02$ and $\Lambda=-0.5$ showing a
double peak structure.}
\label{fig17}
\end{figure}
\begin{figure}
\begin{center}
\epsfig{file=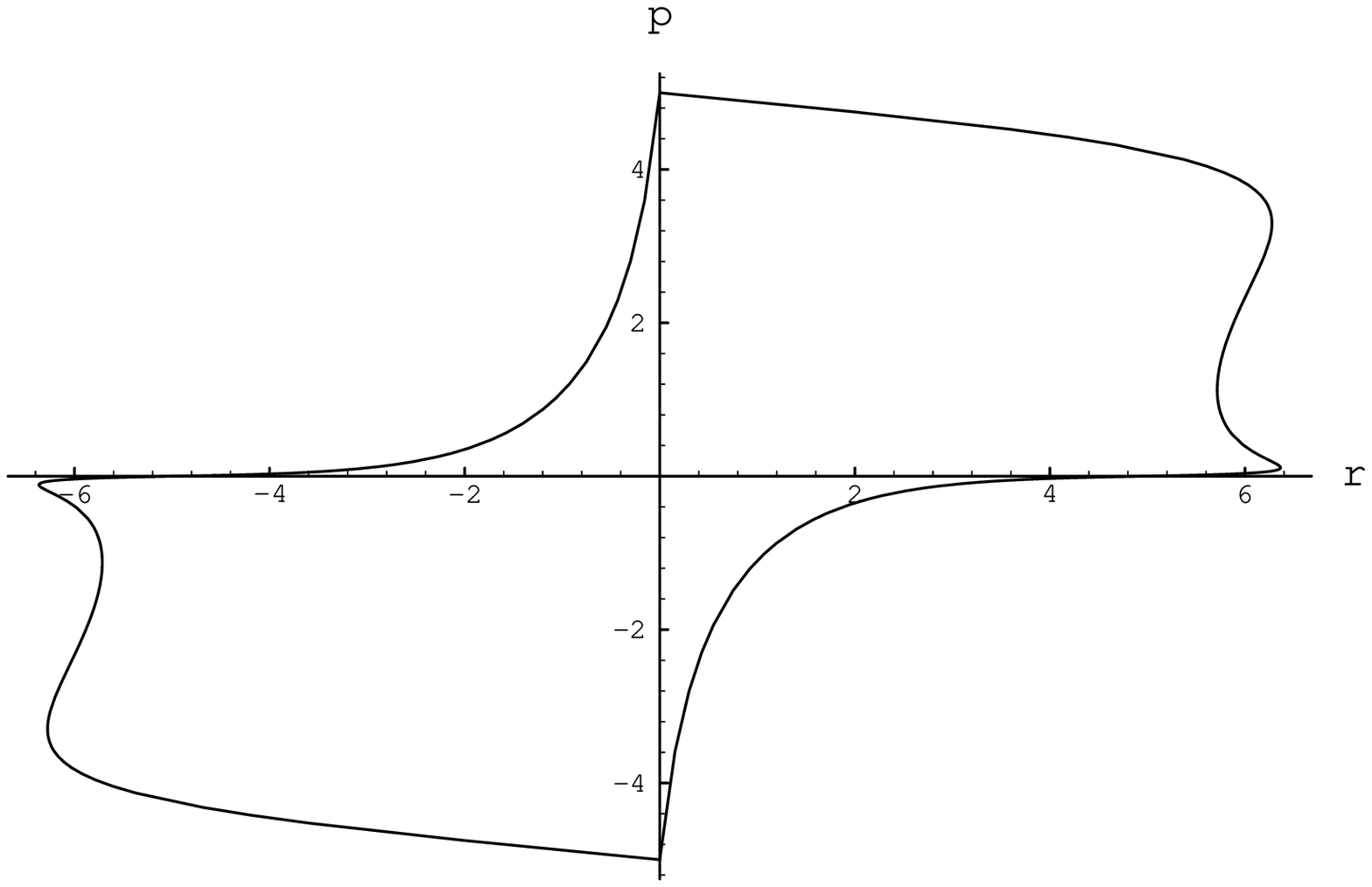,width=0.8\linewidth}
\end{center}
\caption{The phase space trajectory corresponding to the $r(\tau)$ plot in
Fig.17.}
\label{fig18}
\end{figure}
For $0<\Lambda <\Lambda _{c}=\frac{\kappa ^{2}m^{4}}{2(H^{2}-4m^{2})}$, both
bounded and unbounded motions are realized for a fixed value of $H$, as with
the motions depicted in Fig.11. For $\Lambda >\Lambda _{c}$ only unbounded
motion is realized analogous to the motions shown in Fig.14.

In the general case of electrodynamics with a cosmological constant the
dynamics of particles is governed by a combination of four factors:
gravitational attraction, the electric force between charges, the effect of
the cosmological constant and relativistic effects. The solution is
characterized by the signs of $\gamma _{m}$ and $\kappa ^{2}\left(
H_{0}-mf(\tau )+\frac{m}{f(\tau )}\right) ^{2}-8\kappa e_{1}e_{2}-8\Lambda
_{e}$. We shall investigte the motion in four combinations of the signs of $%
\Lambda _{e}$ and the charge, separately.

\subsection{$\Lambda_{e} < 0$ and attractive charges}

In this case all interactions (gravitational, cosmological and electric) are
attractive and ${\cal J}_{\Lambda }^{2}$ is positive.The motion becomes
necessarily bounded and described by the tanh-type solution. The period of
the bounded motion is 
\begin{equation}
T=\cases{ \frac{16}{\kappa m\sqrt{\gamma_{m}}}\tanh^{-1}\left(
\frac{2p_{0}\sqrt{\gamma_{m}}(1+\sqrt{\gamma_{H}})H}
{[H(1+\sqrt{\gamma_{H}})-\gamma_{e}\sqrt{p_{0}^{2}+m^{2}}]^2 -\gamma_{e}^2
p_{0}^2 -\gamma_{m} m^2}\right) &$\gamma_m>0$,\vspace{2mm}\cr \frac{32
p_{0}(1+\sqrt{\gamma_{H}})H} {\kappa
m\left\{[H(1+\sqrt{\gamma_{H}})-\gamma_{e}\sqrt{p_{0}^{2}+m^2}]^2
-\gamma_{e}^{2}p_{0}^{2}\right\}} &$\gamma_m=0$,\vspace{2mm}\cr
\frac{16}{\kappa m\sqrt{-\gamma_{m}}}\tan^{-1}\left(
\frac{2p_{0}\sqrt{-\gamma_{m}}(1+\sqrt{\gamma_{H}})H}
{[H(1+\sqrt{\gamma_{H}})-\gamma_{e}\sqrt{p_{0}^{2}+m^{2}}]^2 -\gamma_{e}^2
p_{0}^2 -\gamma_{m} m^2}\right) &$\gamma_m<0$. }
\end{equation}
For negative $\Lambda _{e}$ motion is allowed for total energy larger than $%
\sqrt{|\Lambda _{e}|/\kappa ^{2}}$. As the total energy increases both the
period and the amplitude of the motion become large and the trajectory $%
r(\tau )$ deforms just as shown in Fig.1 in $\Lambda _{e}=0$ case.

The main purpose in this sub-section is to investigate the effects of $%
\Lambda _{e}$ on the motion, especially on how the double-peak structure
appears in the system of charged particles. We find that the double peak is
caused by the interplay among the $\Lambda $ potential, gravitational
attraction and relativistic effects, and is suppressed as the attractive
force between charges becomes strong. Fig.19 shows the $r(\tau )$ plots for $%
H_{0}=10,m=0.02,|q|=0.1$ and four different values of negative $\Lambda _{e}$%
. The double peak appears for $\Lambda _{e}=-0.5$. As $\Lambda _{e}$
approaches its lower bound $-\kappa ^{2}H^{2}/8$, the form of the phase
space trajectories changes from an $S$-shaped curve to a double peaked one
and then to a diamond shape as depicted in Fig.20. In Fig.21 we trace how
the double peak in Fig.19 is affected by the value of charge $|q|$. We see
that for small values of $|q|$ the double peak structure survives, but it
disappears for large $|q|$.
\begin{figure}
\begin{center}
\epsfig{file=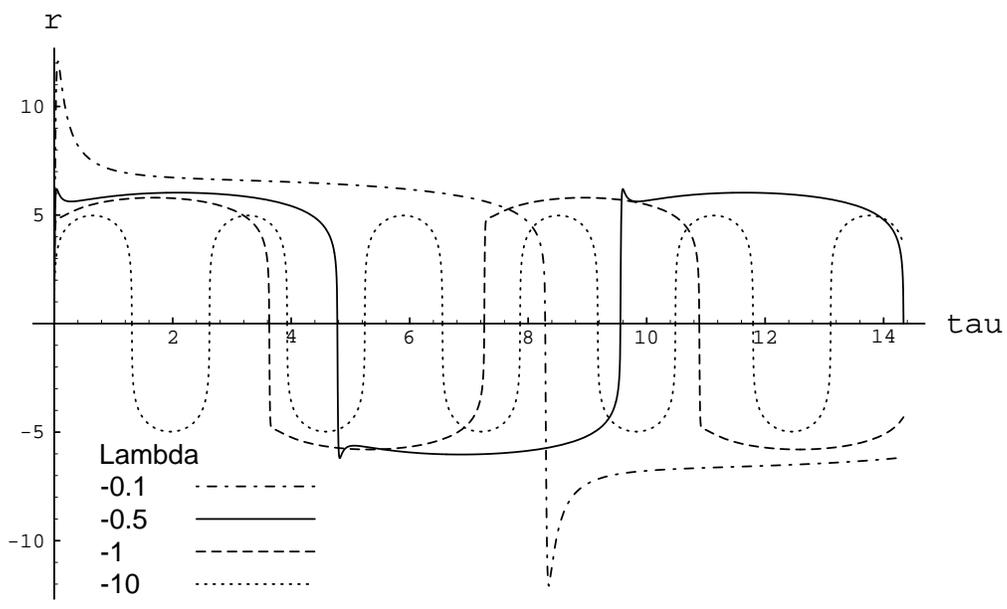,width=0.8\linewidth}
\end{center}
\caption{The $r(\tau)$ plots for $H_{0}=10, m=0.02, |q|=0.1$ and four different
values of $\Lambda_{e}$.}
\label{fig19}
\end{figure}
\begin{figure}
\begin{center}
\epsfig{file=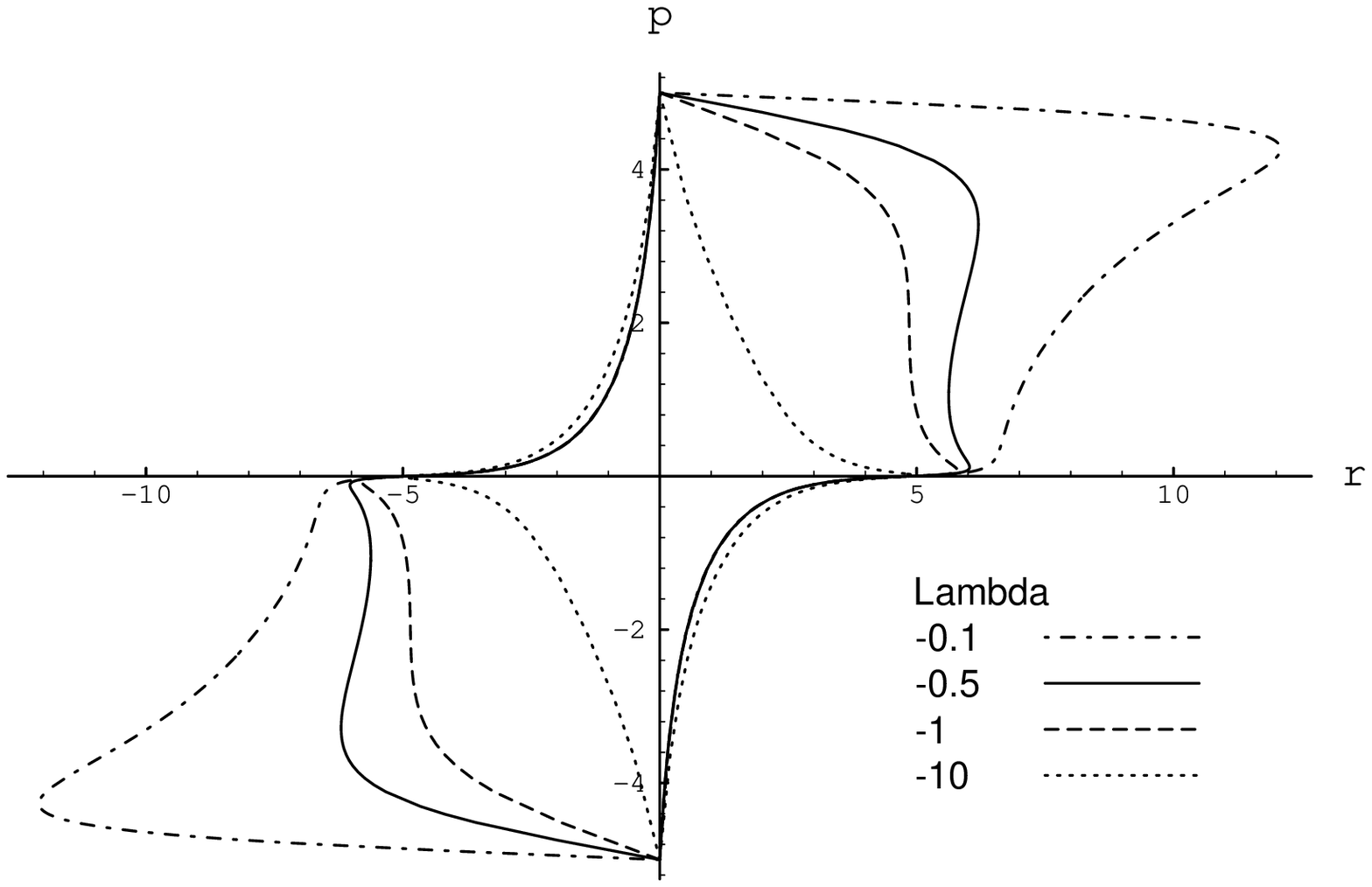,width=0.8\linewidth}
\end{center}
\caption{Phase space trajectories corresponding to the plots in Fig.19.}
\label{fig20}
\end{figure}
\begin{figure}
\begin{center}
\epsfig{file=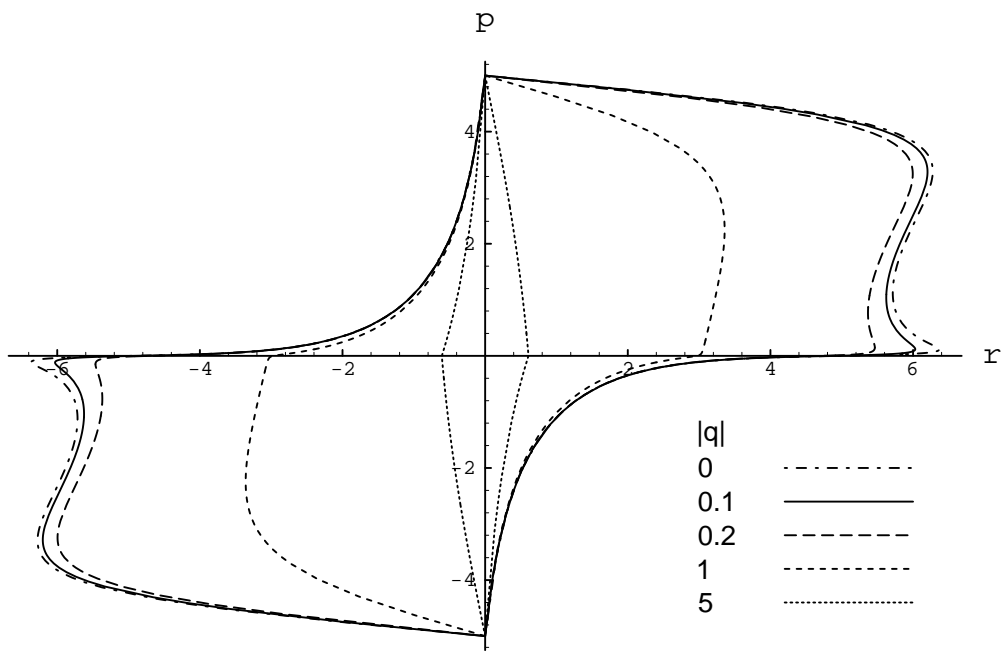,width=0.8\linewidth}
\end{center}
\caption{Phase space trajectories for $H_{0}=10, m=0.02, \Lambda_{e}=-0.5$ and
differnt values of $|q|$.}
\label{fig21}
\end{figure}

\subsection{$\Lambda_{e} < 0$ and repulsive charges}

For the case of repulsive charges ${\cal J}_{\Lambda }^{2}$ may become
negative. We can classify the solutions in terms of the $(|q|,p)$ diagram as
was used in the sub-section 5.2. The boundary of ${\cal J}_{\Lambda }^{2}=0$
is given by 
\begin{equation}
p=\sqrt{\left( \frac{H}{2}\right) ^{2}-\frac{2|\Lambda _{e}|}{\kappa ^{2}}}%
\pm \sqrt{\frac{2}{\kappa }\left( q^{2}-\frac{|\Lambda _{e}|}{\kappa }%
\right) }
\end{equation}
There are two types of $(|q|,p)$ diagram depending on whether \ the vertex $%
(q_{0}=\sqrt{\frac{|\Lambda _{e}|}{\kappa }},\sqrt{\left( \frac{H}{2}\right)
^{2}-\frac{2|\Lambda _{e}|}{\kappa ^{2}}})$ of ${\cal J}_{\Lambda }^{2}=0$
is within the region $-p_{0}\leq p\leq p_{0}:p_{0}=\sqrt{(H/2)^{2}-m^{2}}$,
or not. Fig.22 is the diagram for $m>\sqrt{2|\Lambda _{e}|}/\kappa $ where
the vertex is outside the region $-p_{0}\leq p\leq p_{0}$. The physical
region consists of the shaded area of ${\cal J}_{\Lambda }^{2}>0$ and $B>0$,
and the area of ${\cal J}_{\Lambda }^{2}<0$. The former area is the region
of tanh-type solutions and the latter is the region of tan-type solutions. A
narrow region between ${\cal J}_{\Lambda }-B=0$ (dashed line) and ${\cal J}%
_{\Lambda }=0$ in the shaded area is the region corresponding to tanh-type B
solution. The $q_{c}$ and $q_{d}$ are the values of $|q|$ at the
intersections of ${\cal J}_{\Lambda }^{2}=0$ with $p=\pm p_{0}$, namely, $%
q_{c}=\sqrt{\frac{\kappa }{2}\left\{ \sqrt{(H/2)^{2}-2|\Lambda _{e}|/\kappa
^{2}}-\sqrt{(H/2)^{2}-m^{2}}\right\} +|\Lambda _{e}|/\kappa }$ and $q_{d}=%
\sqrt{\frac{\kappa }{2}\left\{ \sqrt{(H/2)^{2}-2|\Lambda _{e}|/\kappa ^{2}}+%
\sqrt{(H/2)^{2}-m^{2}}\right\} +|\Lambda _{e}|/\kappa }$. \newline
\begin{figure}
\begin{center}
\epsfig{file=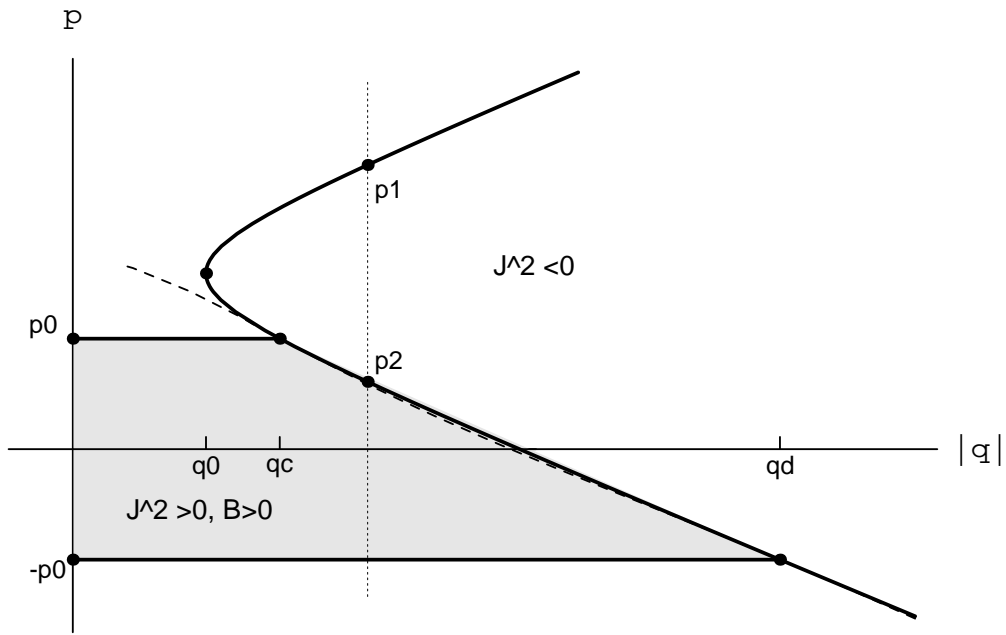,width=0.8\linewidth}
\end{center}
\caption{The $(|q|,p)$ diagram for $\Lambda _{e}<0$ and repulsive charges under
the condition $m>\sqrt{2|\Lambda _{e}|}/\kappa $. The dashed line represents 
${\cal J}_{\Lambda }-B=0$}
\label{fig22}
\end{figure}
The motions corresponding to this diagram are classified into three
categories according to the $|q|$ value: \newline
\noindent (i) $|q|\leq q_{0}$ : bounded motion, \newline
(ii) $q_{0}<|q|<q_{c}$ : both bounded and unbounded motions, \newline
(iii) $q_{c}\leq |q|$ : unbounded motion. \newline

\bigskip

\noindent For a small $|q|$ in the category (i), the attractive effect of $%
\Lambda _{e}<0$ is stronger than the repulsive effect between charges $%
(|\Lambda _{e}|>\kappa q^{2})$ and ${\cal J}_{\Lambda }^{2}$ is positive.
The phase space trajectories resemble the trajectories in Fig.4, but they
become more $S$-shaped as $|q|$ increases. For the categories (ii) and (iii)
the $(|q|,p)$ diagram is nearly the same as the diagram in Fig.10. The phase
space trajectories for (ii) are just like those in Fig.11. The trajectories
of the motion for the category (iii) are divided into two cases of $%
q_{c}\leq |q|\leq q_{d}$ and $q_{d}\leq |q|$. In the case of $q_{c}\leq
|q|\leq q_{d}$ the trajectories are analogous to those in Fig.14 in which
the $N_{2}$ trajectory is a tanh-type, while in the case of $q_{d}\leq |q|$
all trajectories are tan-type and like those in Fig.15 where $N_{1}$ and $%
N_{2}$ have cusps at $r=0$. In Fig.23 the $r(\tau )$ plot for each category
is depicted for the parameters $H_{0}=3,m=1.2$ and $\Lambda _{e}=-0.1$ with $%
q_{0}=0.316$ and $q_{c}=0.491$: $|q|=0.1,0.4$ and $0.6$ for the categories
(i), (ii) and (iii), respectively. As $|q|$ becomes large, both the period
and the amplitude of the motion increase and finally the motion becomes
unbounded, because the repulsive force between charges prevails over the
attractive forces of gravity and the cosmological constant. Fig.24 shows the
corresponding phase space trajectories.
\begin{figure}
\begin{center}
\epsfig{file=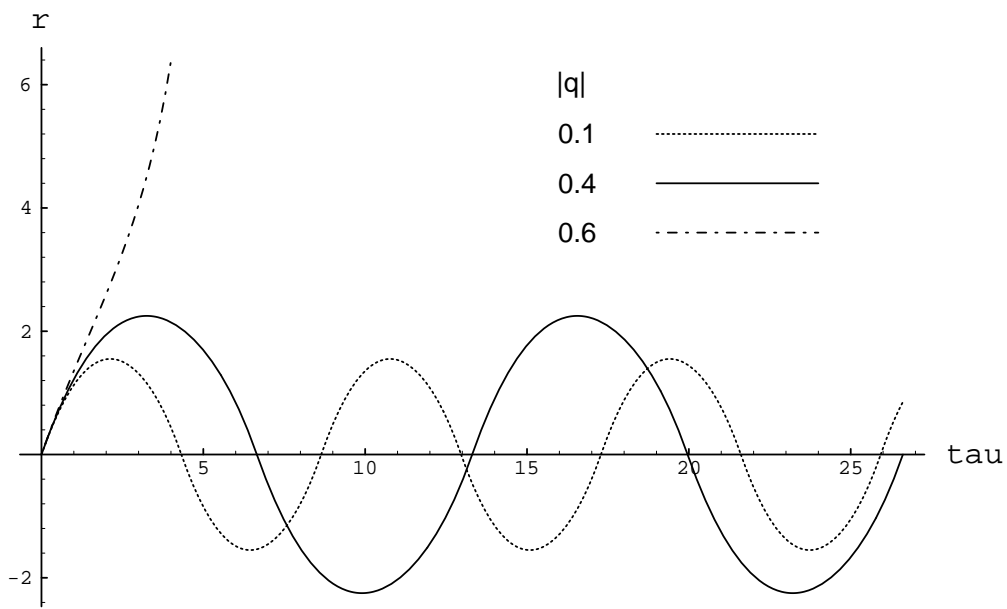,width=0.8\linewidth}
\end{center}
\caption{The $r(\tau)$ plots for the categories (i)-(iii) for the parameters 
$H_{0}=3, m=1.2$ and $\Lambda_{e}=-0.1$.}
\label{fig23}
\end{figure}
\begin{figure}
\begin{center}
\epsfig{file=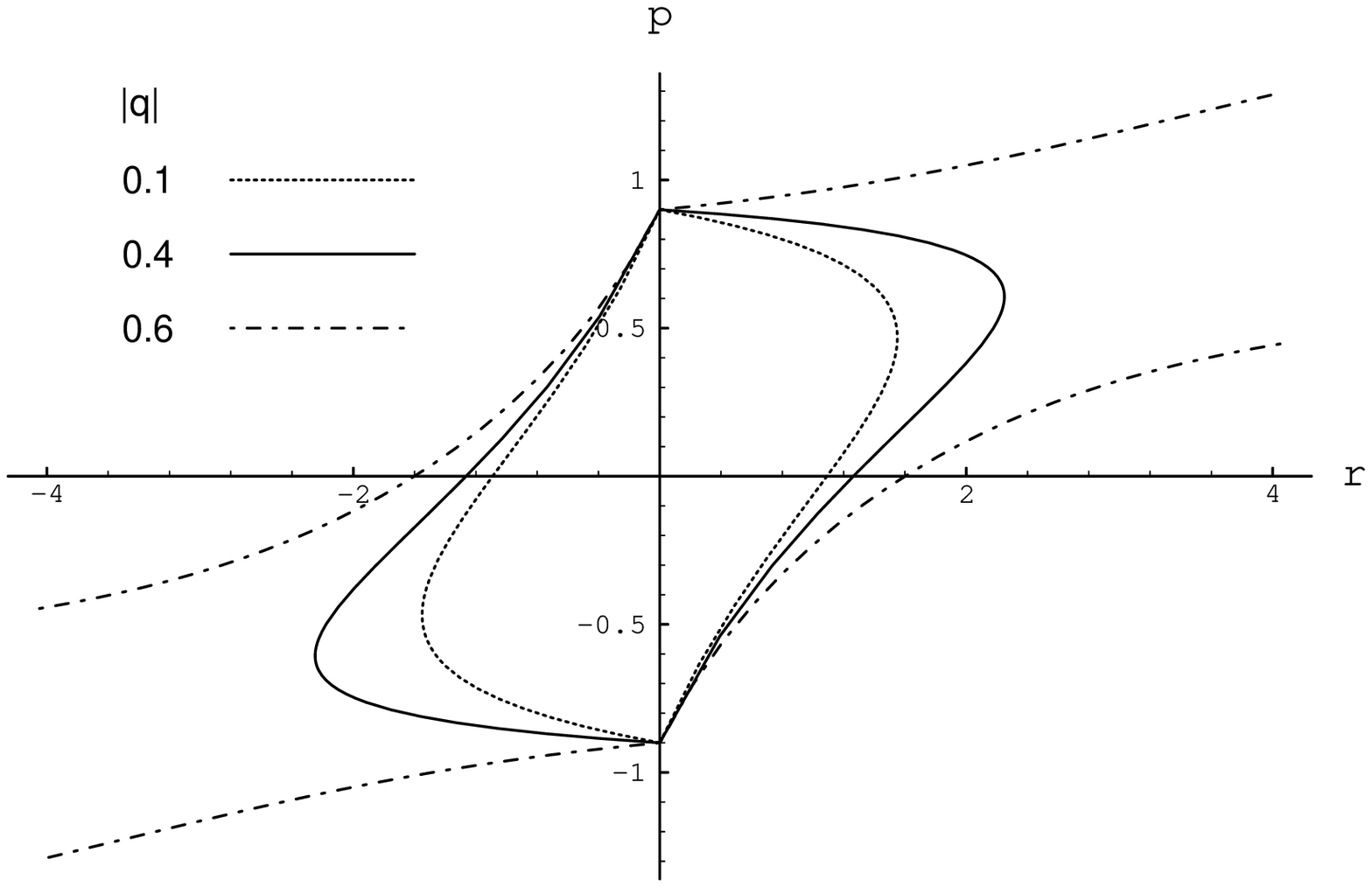,width=0.8\linewidth}
\end{center}
\caption{The phase space trajectories corresponding to the plots in Fig.23}
\label{fig24}
\end{figure}

Since the analysis in sub-section 6.1 indicates that the double peak
structure appears for negative $\Lambda _{e}$, small $m$, and small
attractive $|q|$, it can be inferred that for the repulsive charges, if $|q|$
is sufficiently small, the double peak still survives. The smallness of $m$
corresponds to the condition $m\leq \sqrt{2|\Lambda _{e}|}/\kappa $ for
which the $(|q|,p)$ diagram is given in Fig.25 and the vertex of ${\cal J}%
_{\Lambda }^{2}=0$ is within the region $-p_{0}\leq p\leq p_{0}$. The
motions are classified into two categories: \newline
\noindent (i) $|q|\leq q_{0}$ : bounded motion or unbounded motion, \newline
(ii) $q_{0}<|q|$ : unbounded motion. \newline
The physical region of the category (i) belongs to the shaded area of ${\cal %
J}_{\Lambda }^{2}>0$ and $B>0$. \newline

We can find the double peak structure for a certain range of the parameters
as shown in Fig.26, in which $H_{0}=10,m=0.02,\Lambda _{e}=-0.5$ and $%
|q|=0,\;0.1,\;0.14$.
\begin{figure}
\begin{center}
\epsfig{file=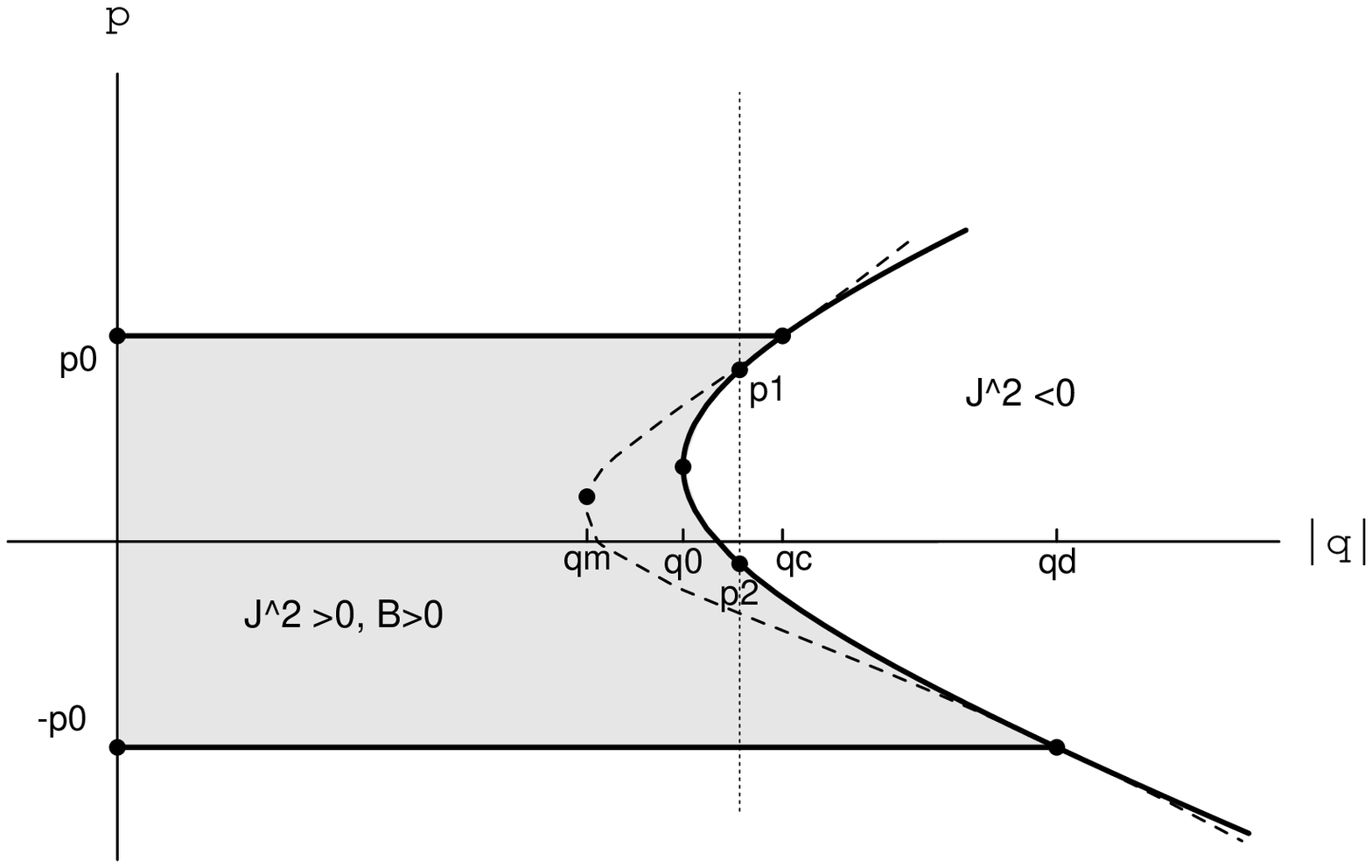,width=0.8\linewidth}
\end{center}
\caption{The $(|q|,p)$ diagram for $\Lambda _{e}<0$ and repulsive charges 
under the condition $m\leq \sqrt{2|\Lambda _{e}|}/\kappa $. The dashed line
represents ${\cal J}_{\Lambda}-B=0$}
\label{fig25}
\end{figure}
\begin{figure}
\begin{center}
\epsfig{file=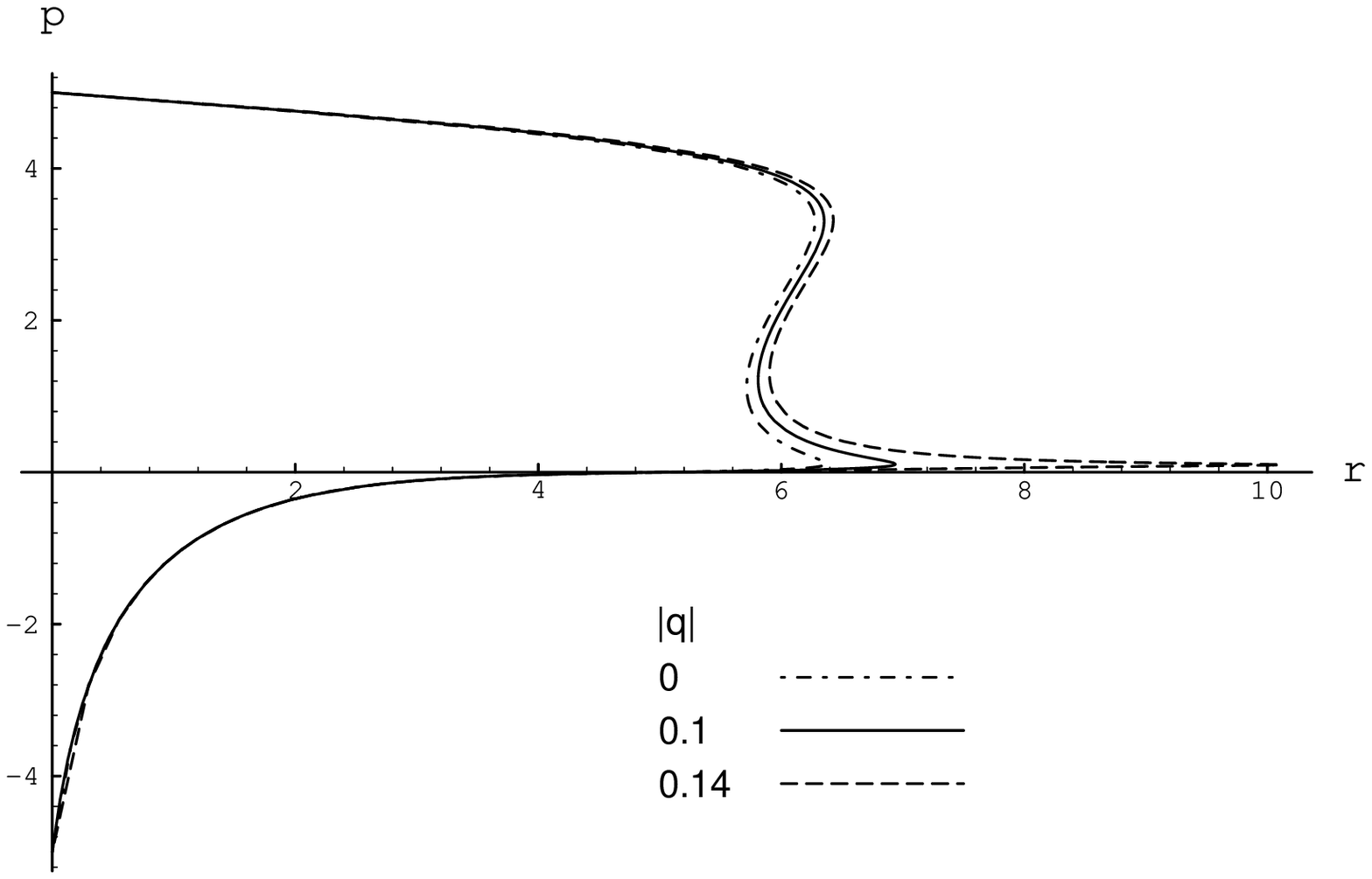,width=0.8\linewidth}
\end{center}
\caption{The $r\geq 0$ part of the phase space trajectories for $H_{0}=10,
m=0.02, \Lambda_{e}=-0.5$ and three different values of repulsive $|q|$.}
\label{fig26}
\end{figure}
The above parameters of $m,\Lambda _{e}$ and $|q|$ correpond to $\gamma
_{m}<0$. As $|q|$ exceeds $q_{m}\equiv \sqrt{m\sqrt{2|\Lambda _{e}|}-\kappa
m^{2}/2}$, the solution to $\gamma _{m}=0$, we encounter a new situation.
The maximum turning point of $r$ extends to infinity and the trajectory
splits into two nonperiodic motions as shown in Fig.27 and Fig.28.

For $|q|=q_{m}$ the asymptotic value of $p$ is $m\sqrt{(H^{2}-|\Lambda
_{e}|)/8|\Lambda _{e}|}$ and in the $(|q|,p)$ diagram of Fig.25 this
corresponds to the vertex point of a dashed-curve of ${\cal J}_{\Lambda
}-B=0 $. For $q_{m}<|q|<q_{0}$ the two asymptotic values of $p$ are the
intersections of $|q|=const$ line with ${\cal J}_{\Lambda }-B=0$ curve. The
region between these asymptotic values originally belongs to tanh-type B
solution but no trajectory exists because ${\cal J}_{\Lambda }>H-|mf-m/f|$
for the parameters in this region.
\begin{figure}
\begin{center}
\epsfig{file=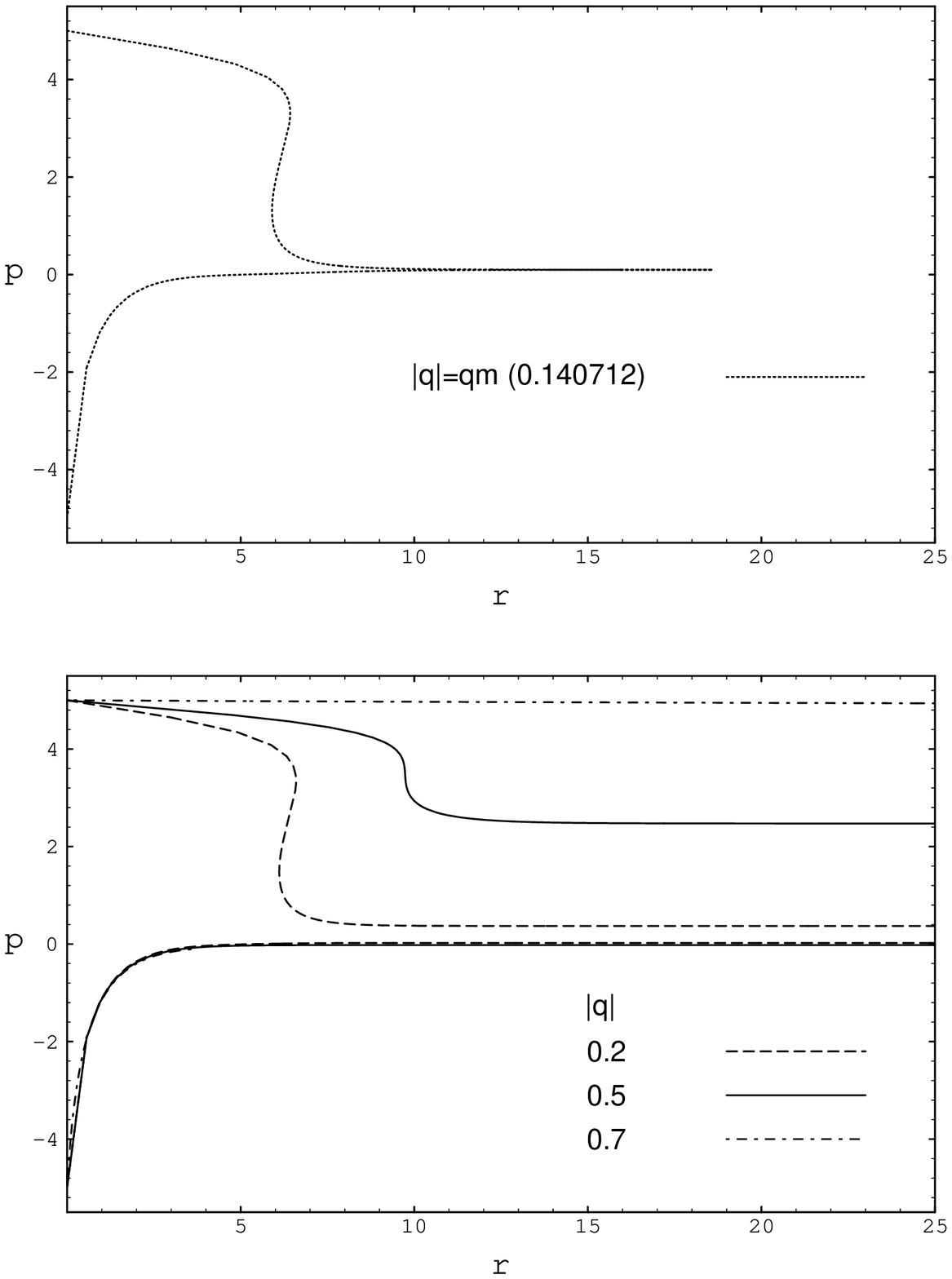,width=0.8\linewidth}
\end{center}
\caption{The $r\geq 0$ part of the phase space trajectories for $H_{0}=10,
m=0.02, \Lambda_{e}=-0.5$ and $q_{m}\leq |q|\leq q_{0}$. }
\label{fig27}
\end{figure}
\begin{figure}
\begin{center}
\epsfig{file=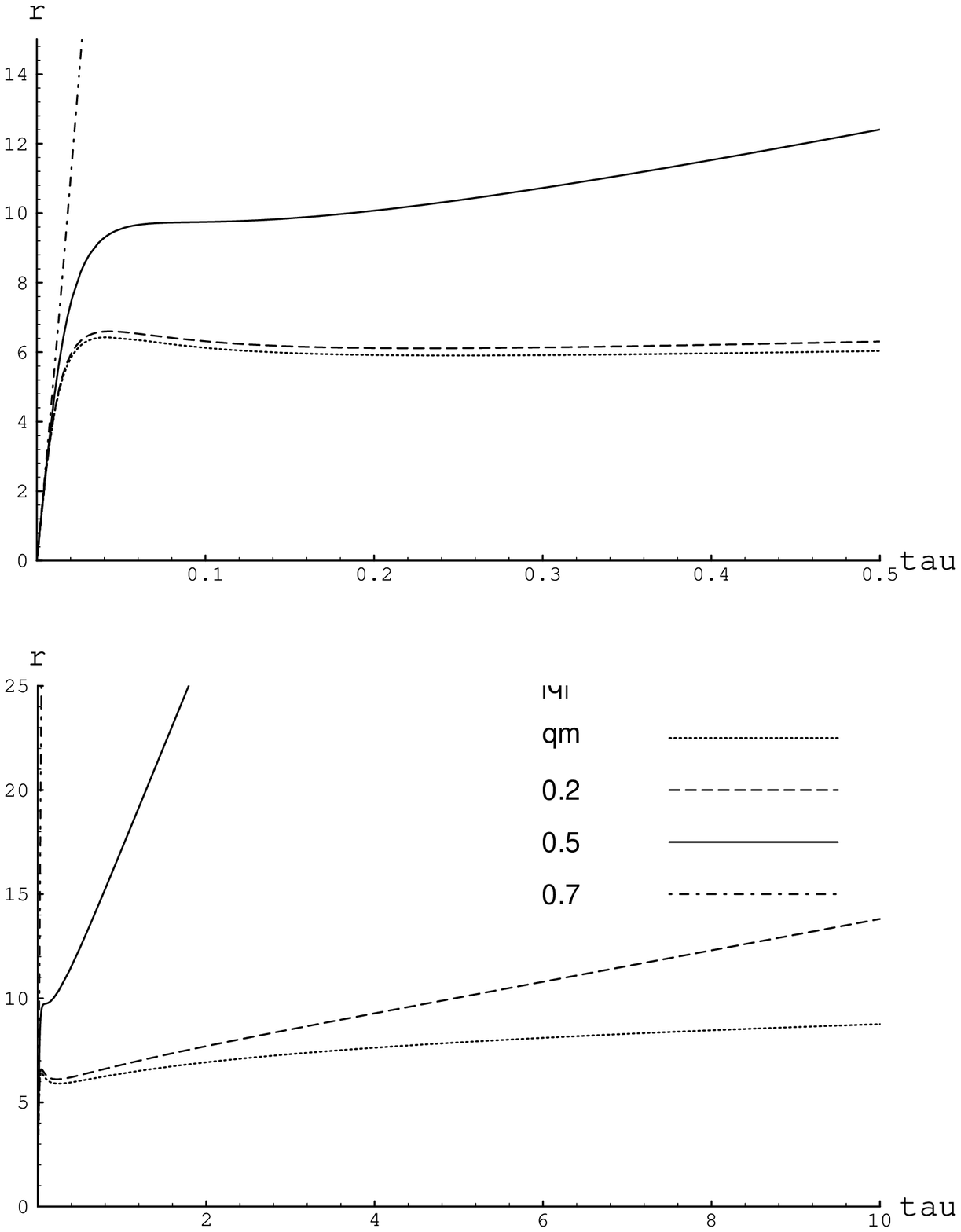,width=0.8\linewidth}
\end{center}
\caption{The $r(\tau)$ plots corresponding to the phase space trajectories in
Fig.27.}
\label{fig28}
\end{figure}

The trajectories of the motion for category (ii) are diveded into three
cases of $q_{0}<|q|<q_{c}$, $q_{c}\leq |q|\leq q_{d}$ and $q_{d}<|q|$. For $%
q_{0}<|q|<q_{c}$ under the condition $m\leq \sqrt{2|\Lambda _{e}|}/\kappa $
, a ${\cal J}_{\Lambda }^{2}<0$ region is sandwiched by the shaded areas, in
contrast to $m>\sqrt{2|\Lambda _{e}|}/\kappa $ case where the region for $%
q_{0}<|q|<q_{c}$ is separated into two parts, namely, a shaded area and a $%
{\cal J}_{\Lambda }^{2}<0$ area. A typical phase space trajectory for $%
q_{0}<|q|<q_{c}$ is shown in Fig.29 where the parameters are $%
H_{0}=3,m=0.6,\Lambda _{e}=-1$ and $|q|=1.1$. Here between two split
nonperiodic trajectories (solid curves) there appear an infinite series of
tan-type A, B solutions of the unbounded motions. For the motions of the
cases $q_{c}\leq |q|\leq q_{d}$ and $q_{d}<|q|$ the outlines of the
unbounded trajectories are easily inferred from Figs.14, 15 and 29.
\begin{figure}
\begin{center}
\epsfig{file=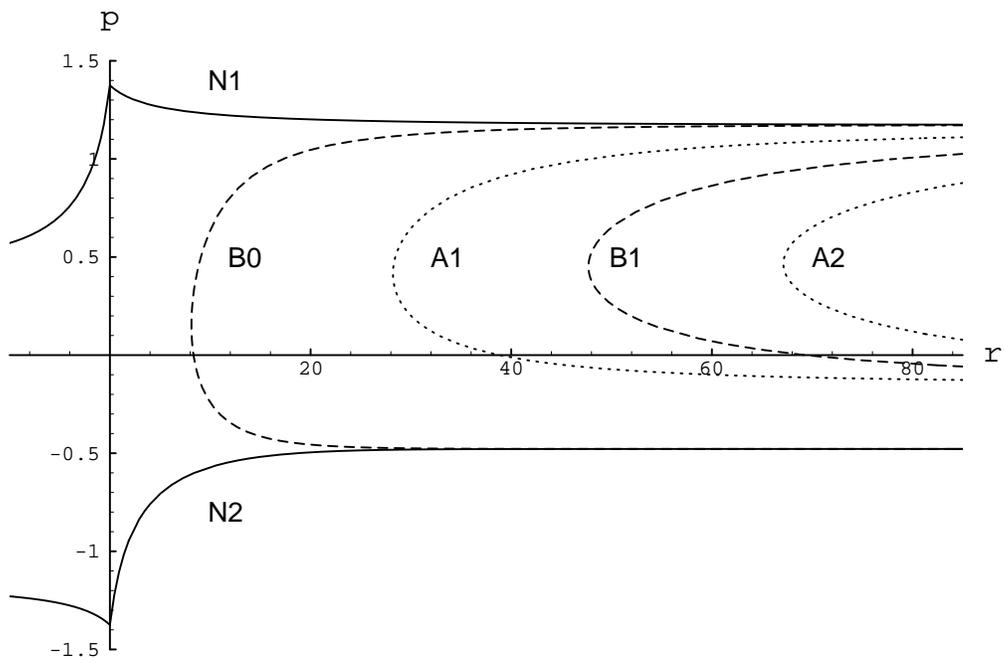,width=0.8\linewidth}
\end{center}
\caption{The phase space trajectories for $H_{0}=10, m=0.02, \Lambda_{e}=-0.5$
and $q_{0}< |q| < q_{c}$. }
\label{fig29}
\end{figure}

\subsection{$\Lambda_{e} > 0$ and attractive charges}

For the attractive charges the boundary curve of ${\cal J}_{\Lambda }^{2}=0$
is 
\begin{equation}
p=\sqrt{\left( \frac{H}{2}\right) ^{2}+\frac{2\Lambda _{e}}{\kappa ^{2}}}\pm 
\sqrt{\frac{2}{\kappa }\left( \frac{\Lambda _{e}}{\kappa }-q^{2}\right) }
\end{equation}
which opens out to the direction of $p$-axis. Fig.30 is the $(|q|,p)$
diagram for \newline
$H<\sqrt{\kappa ^{2}m^{2}/2\Lambda _{e}+4m^{2}}$. The motions for this
diagram are classified into two categories: \newline
\noindent (i) $0\leq |q|<q_{0}$ : both bounded and unbounded motions, 
\newline
(ii) $q_{0}<|q|$ : bounded motion. \newline
The phase space trajectories for category (i) are just like the trajectories
in Fig.11, while for category (ii) the motions become simply bounded as the
attractive charges overwhelm the repulsive effect of the effective
cosmological constant.
\begin{figure}
\begin{center}
\epsfig{file=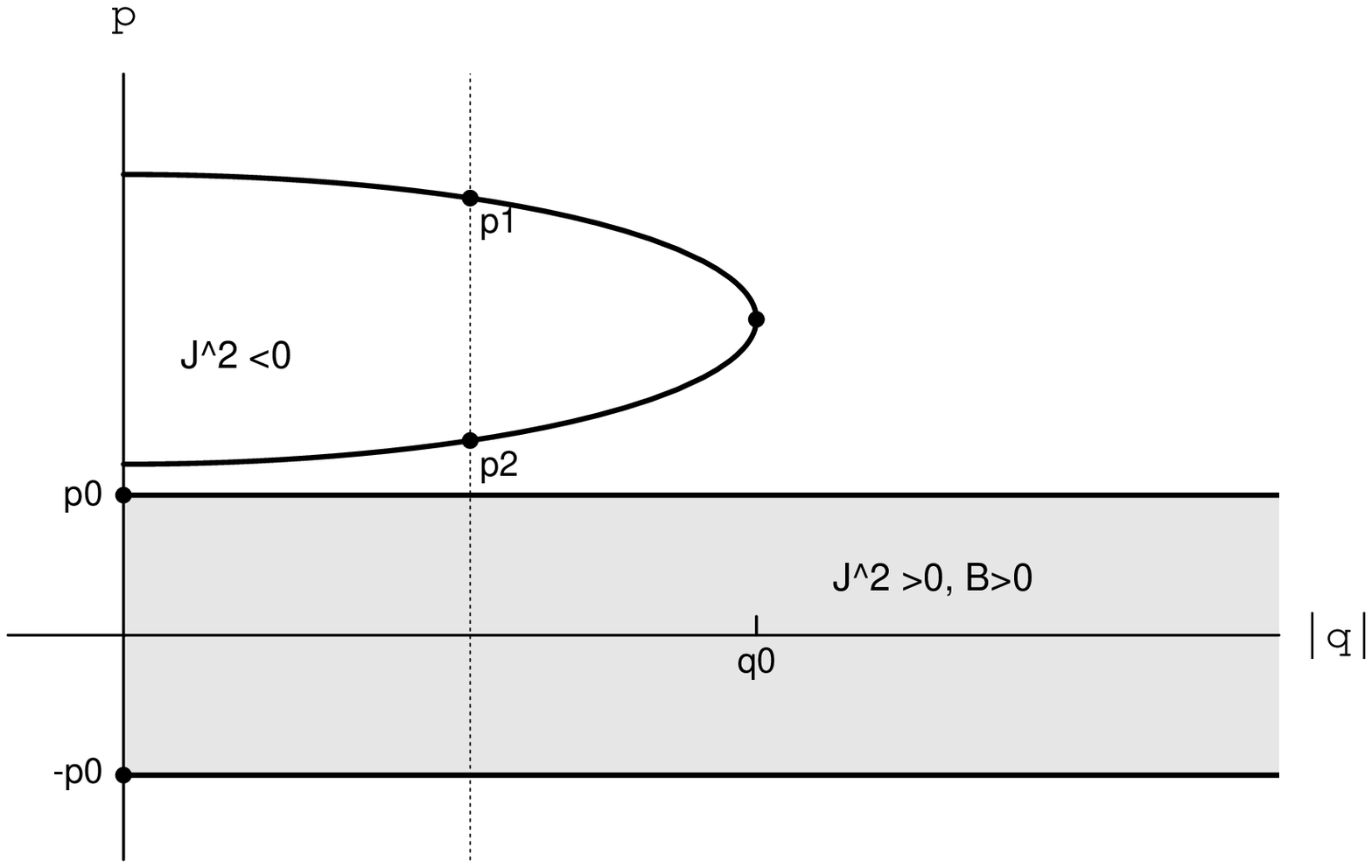,width=0.8\linewidth}
\end{center}
\caption{The $(|q|,p)$ diagram for $\Lambda _{e}>0$ and attractive charges under
the condition $H<\sqrt{\kappa ^{2}m^{2}/2\Lambda _{e}+4m^{2}}$. }
\label{fig30}
\end{figure}

Under the condition of $H\geq \sqrt{\kappa^{2}m^{2}/2\Lambda _{e}+4m^{2}}$
the situation is slightly different from the previous case, as shown in the $%
(|q|,p)$ diagram of Fig.31. In this case the motions are classified into
three categories according to the $|q|$ value: \newline
\noindent (i) $0<|q|\leq q_{c}$ : unbounded motion, \newline
(ii) $q_{c}<|q|\leq q_{0}$ : both bounded and unbounded motions, \newline
(iii) $q_{0}<|q|$ : bounded motion, \newline
where $q_{c}=\sqrt{\Lambda _{e}/\kappa -(\kappa /2)\left\{ \sqrt{%
(H/2)^{2}+2\Lambda _{e}/\kappa ^{2}}-\sqrt{(H/2)^{2}-m^{2}}\right\} ^{2}}$.
Since this $(|q|,p)$ diagram is like the inverse of{\bf \ }the diagram in
Fig.22, categories (i), (ii) and (iii) correspond to the previous categories
(iii), (ii) and (i) of $m>\sqrt{2|\Lambda _{e}|}/\kappa $ in sub-section
6.2, respectively. In Fig.32 the $r(\tau )$ plot for each category is drawn
for the parameters $H_{0}=3,m=1$ and $\Lambda_{e}=1$ with $q_{0}=1$ and $%
q_{c}=0.745$: $|q|=0.5,0.8$ and $1.5$ for the category (i), (ii) and (iii),
respectively. As $|q|$ increases, the motion changes from unbounded to
bounded and then both the period and the amplitude decrease.The phase space
trajectory for each $r(\tau )$ is easily inferred from Fig.24. \newline
\begin{figure}
\begin{center}
\epsfig{file=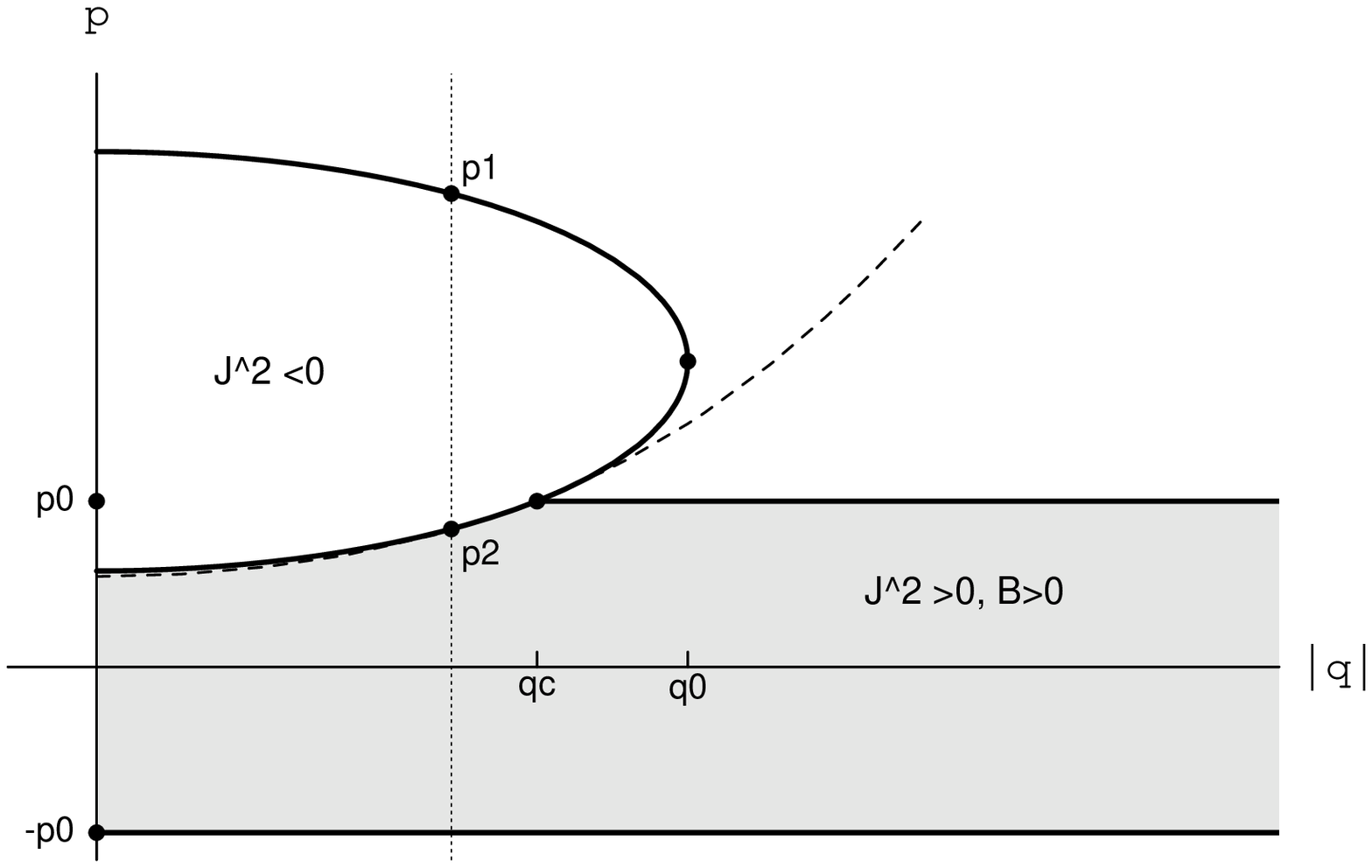,width=0.8\linewidth}
\end{center}
\caption{The $(|q|,p)$ diagram for $\Lambda _{e}>0$ and attractive charges under
the condition $H\geq \sqrt{\kappa ^{2}m^{2}/2\Lambda _{e}+4m^{2}}$. The
dashed line represents ${\cal J}_{\Lambda}-B=0$}
\label{fig31}
\end{figure}
\begin{figure}
\begin{center}
\epsfig{file=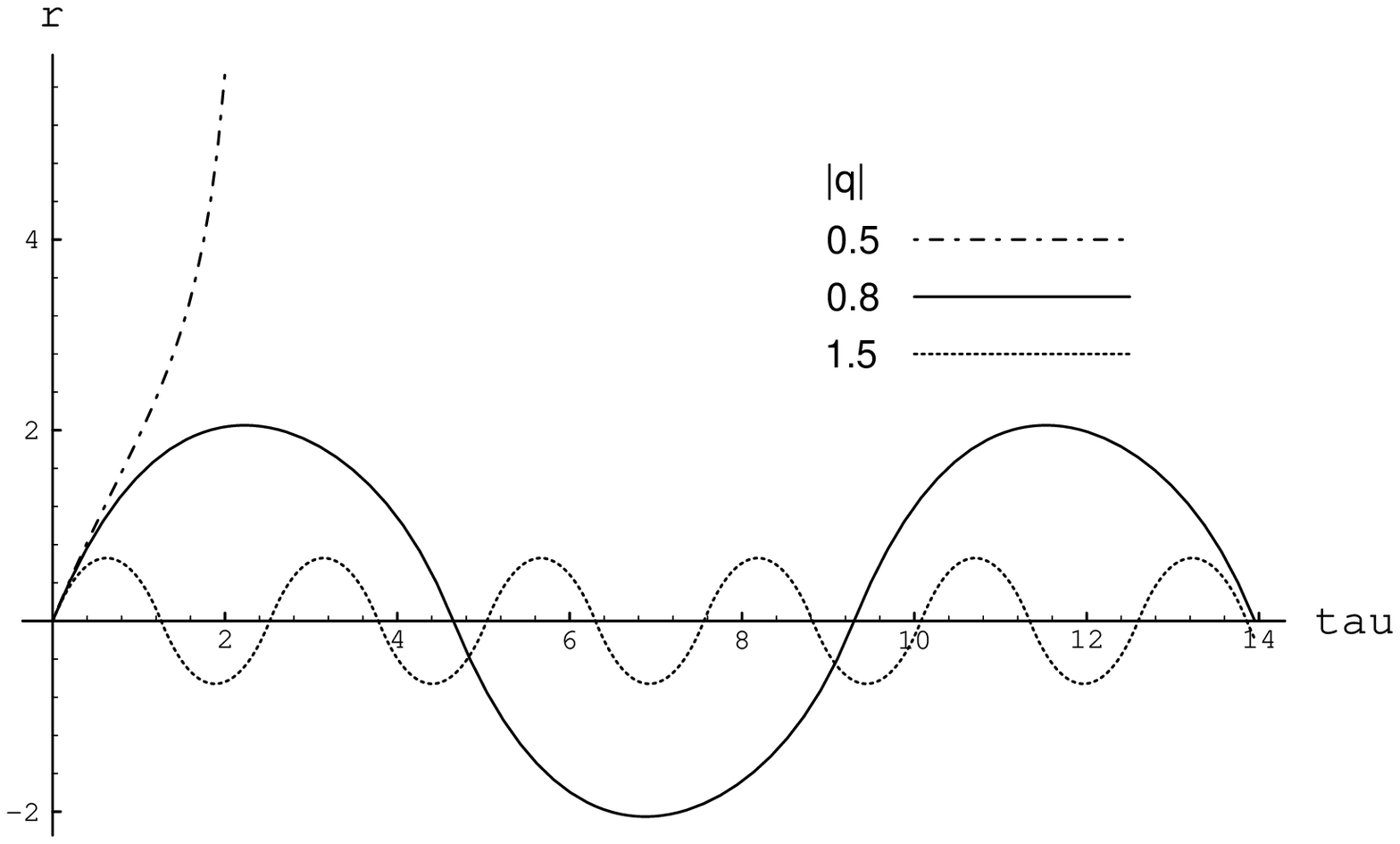,width=0.8\linewidth}
\end{center}
\caption{The $r(\tau)$ plots for the categories (i)-(iii) for the parameters 
$H_{0}=3, m=1$ and $\Lambda_{e}=1$.}
\label{fig32}
\end{figure}

\subsection{$\Lambda_{e} > 0$ and repulsive charges}

In this case, since the cosmological repulsion is commensurate with
electromagnetic repulsion, classification of the motion is analogous to that
of repulsive charges with $\Lambda _{e}=0$ in sub-section 5.2. Under the
condition $H<\sqrt{\kappa ^{2}m^{2}/2\Lambda _{e}+4m^{2}}$ the $(|q|,p)$
diagram is given by Fig.33 which has the same pattern as Fig.10. The
physical region is classified into two categories: \newline
\noindent (i) $|q|<q_{c}$ : both bounded and unbounded motions, \newline
(ii) $q_{c}\leq |q|$ : unbounded motion. \newline
The trajectories of the motion for the category (ii) are divided into two
cases of $q_{c}\leq |q|\leq q_{d}$ and $q_{d}<|q|$. The phase space
trajectories and $r(\tau )$ plots for $|q|<q_{c}$ are analogous to those in
Fig.11 and 12. For a larger $H$ of $H\geq \sqrt{\kappa ^{2}m^{2}/2\Lambda
_{e}+4m^{2}}$ the motion is always unbounded as inferred from the diagram of
Fig.34.
\begin{figure}
\begin{center}
\epsfig{file=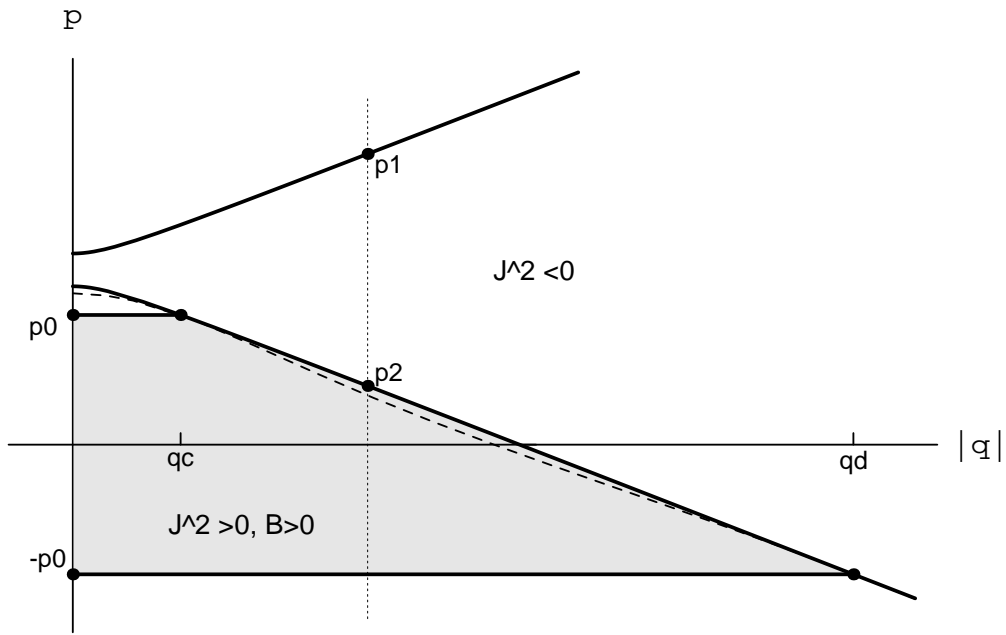,width=0.8\linewidth}
\end{center}
\caption{The $(|q|,p)$ diagram for $\Lambda _{e}>0$ and repulsive charges under
the condition $H<\sqrt{\kappa ^{2}m^{2}/2\Lambda _{e}+4m^{2}}$. The dashed
line represents ${\cal J}_{\Lambda}-B=0$}
\label{fig33}
\end{figure}
\begin{figure}
\begin{center}
\epsfig{file=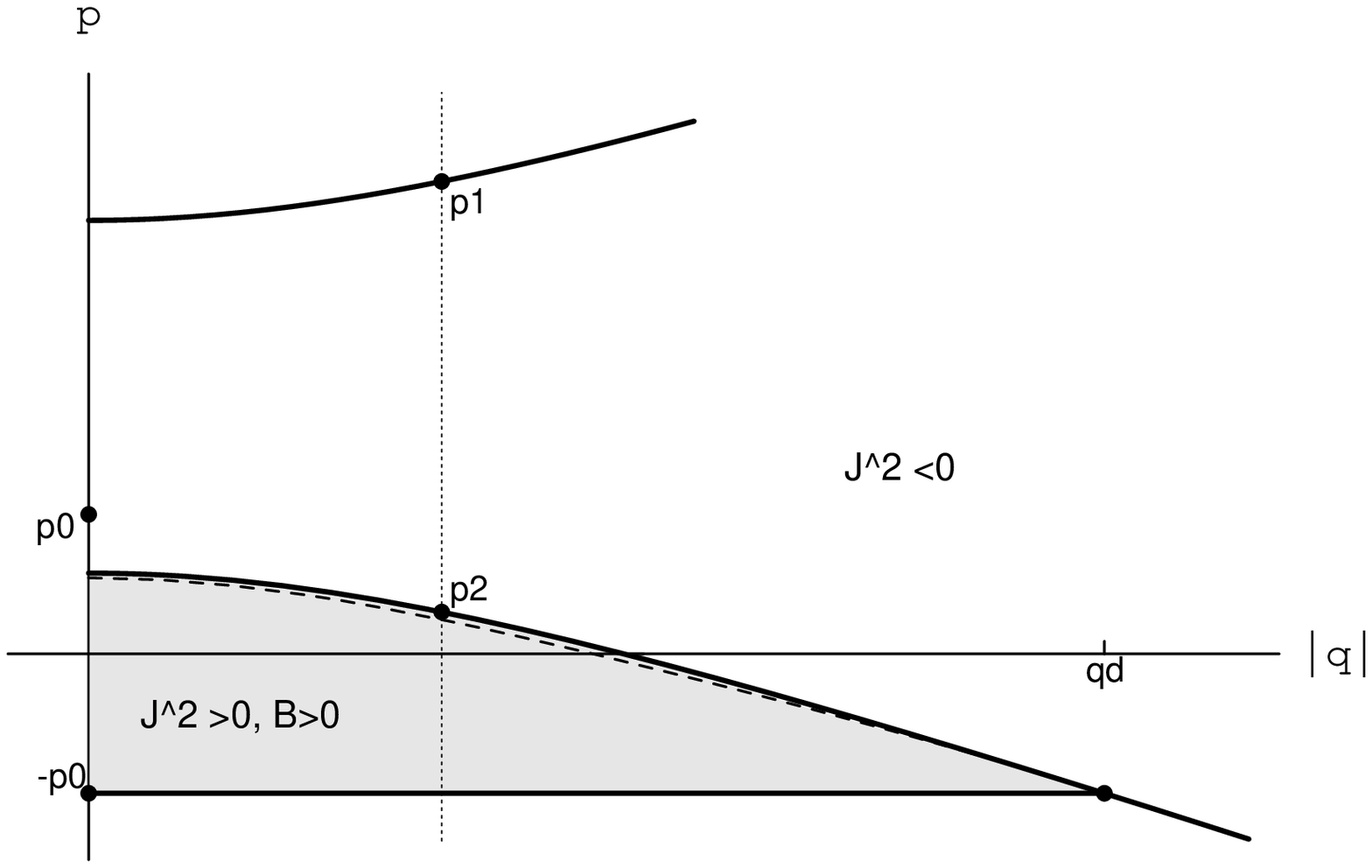,width=0.8\linewidth}
\end{center}
\caption{The $(|q|,p)$ diagram for $\Lambda_{e}>0$ and repulsive charges under
the condition $H \geq \sqrt{\kappa^2 m^2 /2\Lambda_{e} +4m^2}$. The dashed
line represents ${\cal J}_{\Lambda}-B=0$}
\label{fig34}
\end{figure}

\section{MOTION OF UNEQUAL MASSES}

What time variable is adequate for the analysis of the motion of the two
particles with unequal masses? The proper time (\ref{proper}) of each
particle is different as 
\begin{eqnarray}
d\tau _{1} &=&dt\;\frac{16YK_{0}K_{1}m_{1}}{JKM_{1}\sqrt{p^{2}+m_{1}^{2}}}\;,
\nonumber  \label{tau-2} \\
&& \\
d\tau _{2} &=&dt\;\frac{16YK_{0}K_{2}m_{2}}{JKM_{2}\sqrt{p^{2}+m_{2}^{2}}}\;,
\nonumber
\end{eqnarray}
where $K\equiv K_{+}=K_{-}$ and $Y\equiv Y_{+}=Y_{-}$. As with our analysis
of the electrically neutral case \cite{2bdcoslo}, we propose using 
\begin{equation}
d\tilde{\tau}\equiv dt\;\frac{16YK_{0}}{JK}\left( \frac{K_{1}K_{2}m_{1}m_{2}%
}{M_{1}M_{2}\sqrt{p^{2}+m_{1}^{2}}\sqrt{p^{2}+m_{1}^{2}}}\right) ^{1/2}\;,
\label{tau-3}
\end{equation}
which is symmetric with respect to $1\leftrightarrow 2$ and reduces to the
proper time (\ref{Tau}) when $m_{1}=m_{2}$.

In terms of this variable the canonical equations are expressed as 
\begin{eqnarray}
\frac{dp}{d\tilde{\tau}}&=&-\frac{1}{4\kappa}\left( \frac{%
K_{1}K_{2}M_{1}M_{2}\sqrt{p^2+m_{1}^{2}}\sqrt{p^2+m_{2}^{2}}} {m_{1}m_{2}}%
\right)^{1/2}\;,  \label{unequal-p} \\
\frac{dz_{i}}{d\tilde{\tau}}&=&(-1)^{i+1} \left( \frac{M_{1}M_{2}\sqrt{%
p^2+m_{1}^{2}}\sqrt{p^2+m_{2}^{2}}} {K_{1}K_{2}m_{1}m_{2}}\right)^{1/2}
\left\{\frac{\epsilon J}{16K_{0}} +\frac{K_{i}}{M_{i}}\left(\frac{p}{\sqrt{%
p^2+m_{i}^{2}}} -\epsilon\frac{Y}{K}\right)\right\}\;,  \nonumber
\label{unequal-z} \\
\\
\frac{dr}{d\tilde{\tau}}&=&\left( \frac{M_{1}M_{2}\sqrt{p^2+m_{1}^{2}}\sqrt{%
p^2+m_{2}^{2}}} {K_{1}K_{2}m_{1}m_{2}}\right)^{1/2}  \nonumber \\
&&\makebox[3em]{}\times\left\{\frac{\epsilon J}{8K_{0}} +\frac{K_{1}}{M_{1}}%
\left(\frac{p}{\sqrt{p^2+m_{1}^{2}}} -\epsilon\frac{Y}{K}\right) +\frac{K_{2}%
}{M_{2}}\left(\frac{p}{\sqrt{p^2+m_{2}^{2}}} -\epsilon\frac{Y}{K}%
\right)\right\}\;.  \label{unequal-r}
\end{eqnarray}
Note that $r$ still describes the proper distance between the particles at
any fixed instant.

Unlike the equal mass case, the integration $\int dp (K_{1}K_{2}M_{1}M_{2} 
\sqrt{p^2+m_{1}^{2}}\sqrt{p^2+m_{2}^{2}}\;)^{-1/2}$ can not be performed
within the framework of elementary calculus. We resort to numerical
calculation for solving above equations.

As in Sec.VI we analyze the motions by plotting $r(\tau )$ in four
combinations of the signs of $\Lambda _{e}$ and the charges. In the case of
a negative cosmological constant and attractive charges, the $r(\tau )$
plots are shown in Fig.35 for various mass ratios $m_{1}/m_{2}$ in the fixed 
$H_{0}=10,m_{2}=1,\Lambda _{e}=-1$ and $|q|=1$. When the mass ratio gets
larger than unity (the case of $m_{1}=m_{2}=1$), the gravitational
attraction is stronger and the proper distance between particles as well as
the period becomes shorter. As the mass ratio gets smaller gravity becomes
weak; however for quite small mass ratios the attractive effect due to the
cosmological constant prevails and the period changes to being short again.
The effect of the cosmological constant is seen by changing $|q|$ small but
preserving the values of other parameters. Fig.36 shows the $r(\tau )$ plots
for $|q|=0.1$. While both the proper distance and the period becomes large
as a whole, the double peak structures appear for a small mass ratio.
\begin{figure}
\begin{center}
\epsfig{file=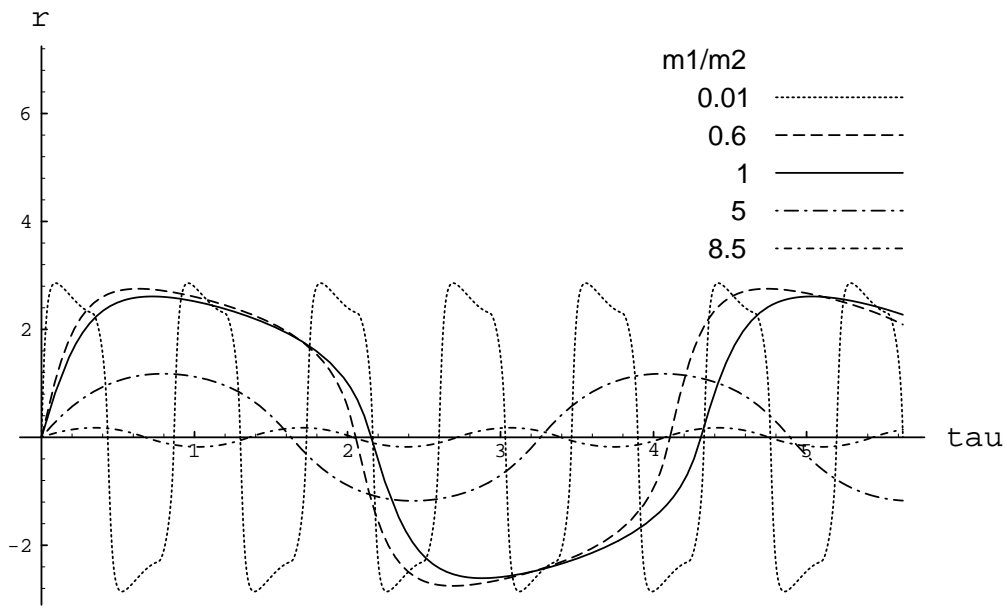,width=0.8\linewidth}
\end{center}
\caption{$r(\tau)$ plots for the different values of the mass ratio $m_{1}/m_{2}$
for $H_{0}=10, m_{2}=1, \Lambda_{e}=-1$ and $e_{1}=-e_{2}=|q|=1$.}
\label{fig35}
\end{figure}
\begin{figure}
\begin{center}
\epsfig{file=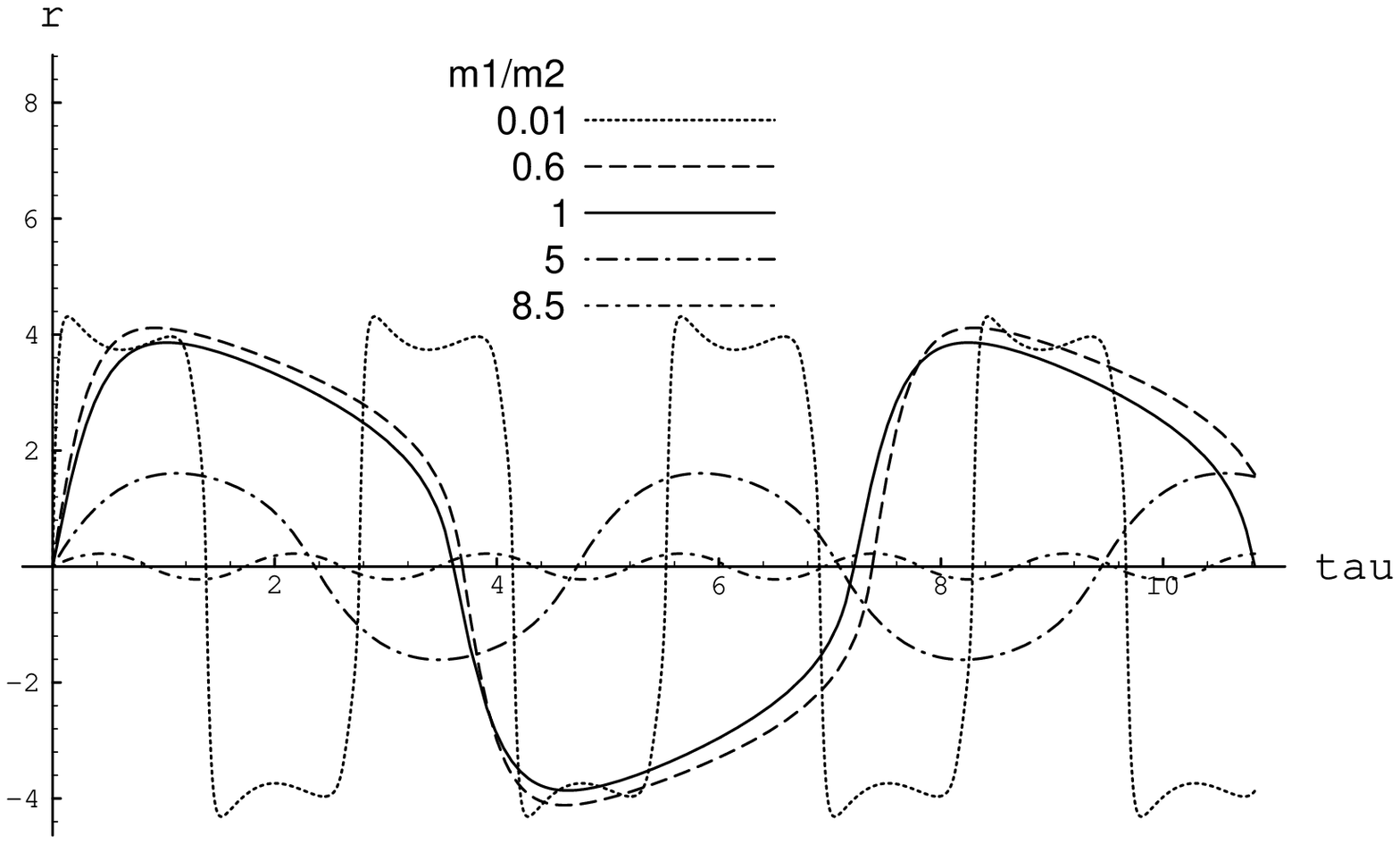,width=0.8\linewidth}
\end{center}
\caption{$r(\tau)$ plots for the same values of the parameters but $%
e_{1}=-e_{2}=|q|=0.1$.}
\label{fig36}
\end{figure}

In the case of $\Lambda _{e}<0$ and repulsive charges unbounded trajectories
appear as $|q|$ increases. An example is shown in Fig.37, in which $r(\tau )$
for a small mass ratio $m_{1}/m_{2}=0.1$ becomes unbounded because the
repulsive force between charges exceeds the weak gravity and the $\Lambda $
force.
\begin{figure}
\begin{center}
\epsfig{file=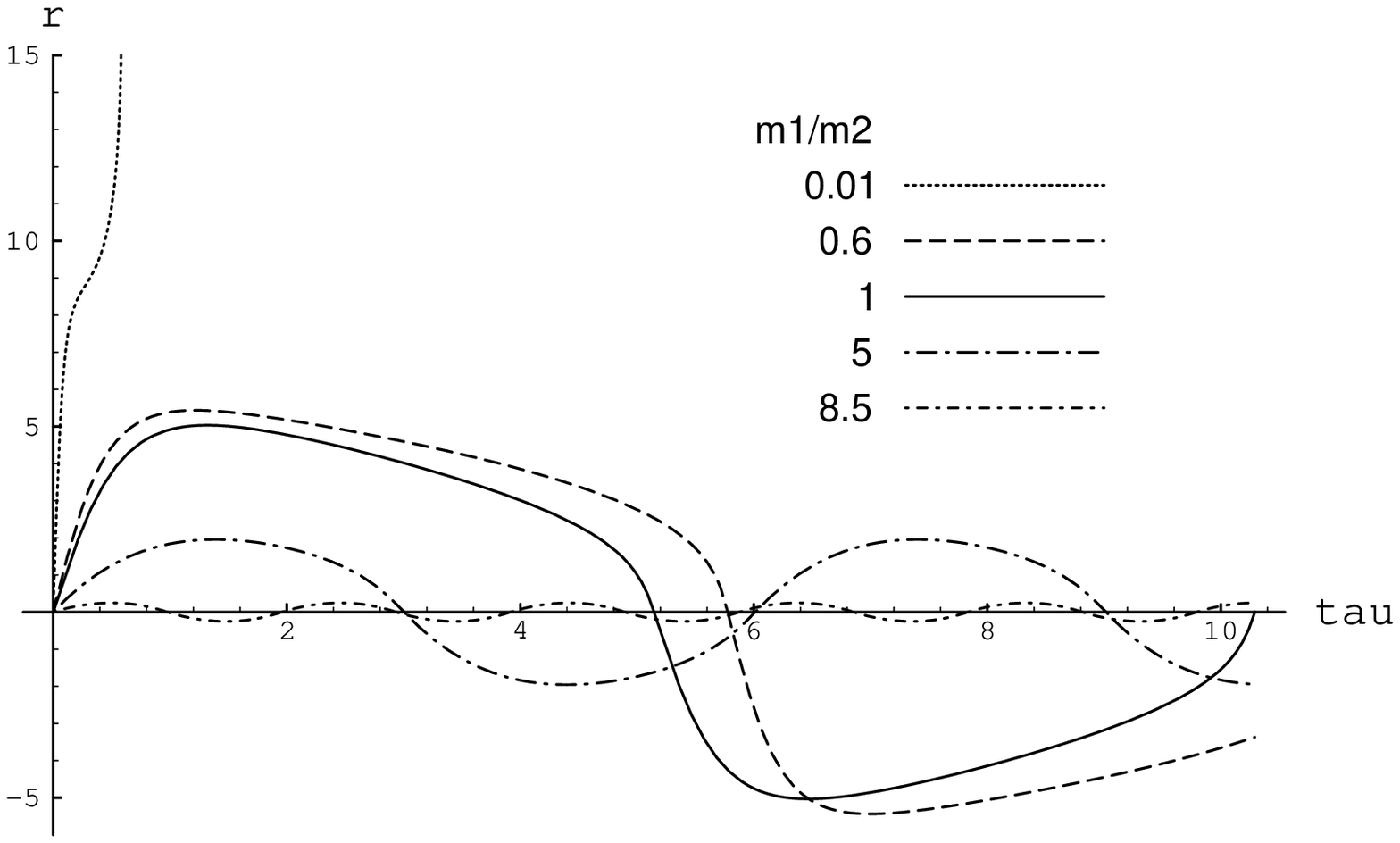,width=0.8\linewidth}
\end{center}
\caption{$r(\tau )$ plots for different values of the mass ratio $m_{1}/m_{2}$
for $H_{0}=10,m_{2}=1,\Lambda _{e}=-1$ and $e_{1}=e_{2}=|q|=0.6$. }
\label{fig37}
\end{figure}

In the case of $\Lambda _{e}>0$ and attractive charges, as we analyzed in
Sec.VI the trajectory changes from unbounded to bounded as $|q|$ increases.
In Fig.38 we plot $r(\tau )$ for various $m_{1}/m_{2}$ in the fixed $%
H_{0}=10,m_{2}=1,\Lambda _{e}=0.1$ and $|q|=0.1$, from which the strong
repulsive effect of a positive $\Lambda _{e}$ is perceived. The motions for
the case of $\Lambda _{e}>0$ and the repulsive charges are mostly unbounded
except for the case of the large masses.
\begin{figure}
\begin{center}
\epsfig{file=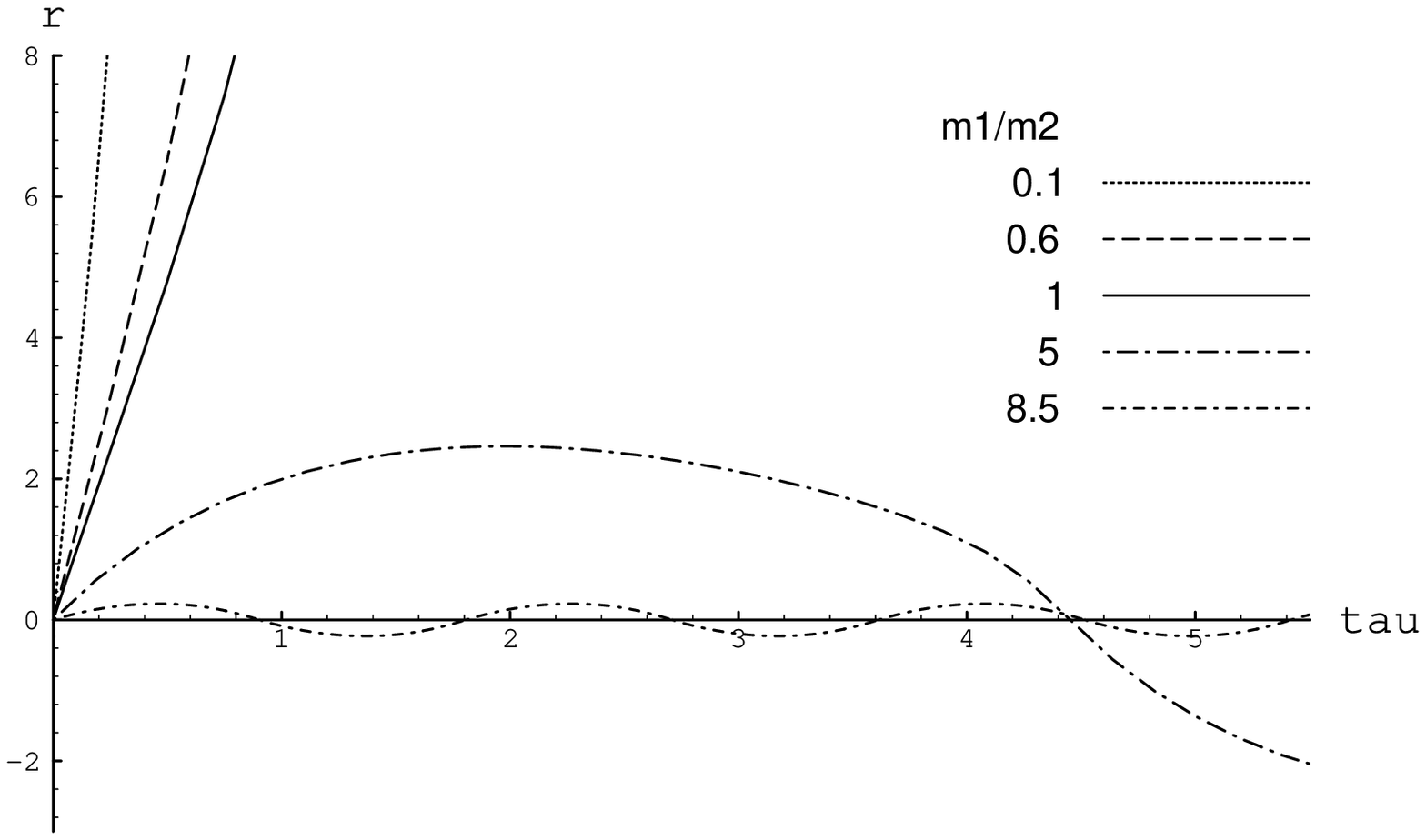,width=0.8\linewidth}
\end{center}
\caption{$r(\tau)$ plots for the different values of the mass ratio $m_{1}/m_{2}$
for $H_{0}=10, m_{2}=1, \Lambda_{e}=0.1$ and $e_{1}=-e_{2}=|q|=0.1$.}
\label{fig38}
\end{figure}

\section{STATIC BALANCE IN (1+1) DIMENSIONS}

In this section we treat the problem of static balance in (1+1) dimensions.
This problem originated in attempts to find the exact solution of the $%
N(N\geq 2)$ body system in Einstein's theory of general relativity \cite
{KSMH}. In $(3+1)$ dimensions gravitational radiation carries away energy
from a system of particles, and so exact solutions have so far been
unobtainable. One simple way to search an exact solution for $N\geq 2$ is to
balance the gravitational attraction with some extra repulsive force, a
natural candidate of which is the electric force. The first trial was
achieved by Majumdar \cite{Maj} and Papapetrou \cite{Papa} for $N=2$ and
afterwards it was generalized to $N$ bodies on a line \cite{GHA}. Their
condition for balance is 
\begin{equation}
e_{i}=\pm \sqrt{4\pi G}\;m_{i}\;,  \label{M-P}
\end{equation}
and is much more strict than the corresponding condition in Newtonian theory 
\begin{equation}
Gm_{1}m_{2}-\frac{e_{1}e_{2}}{4\pi }=0\;.  \label{Newt-b}
\end{equation}

\noindent

Since then, the question has long been raised as to why the balance
condition differs in the relativistic and non-relativistic cases. Some
people \cite{Bonn} conjectured that there should be an exact solution in
general relativity under the condition (\ref{Newt-b}). Others \cite{BO,OK}
showed in the (2nd) post-Newtonian approximation that the condition (\ref
{Newt-b}) is incompatible with the static balance condition in general
relativity. Clearly (\ref{M-P}) is a sufficient condition for static balance
but there exists no proof that (\ref{M-P}) is a necessary condition. On the
other hand a test particle analysis \cite{Bonn2} suggested that the
condition (\ref{Newt-b}) was neither necessary nor sufficient, but a
separation-dependent balance position might exist in general relativity.
Recently several numerical trials on the separation-dependence were reported 
\cite{Perry,Breton}, but so far no one has been able -in any relativistic
theory of gravity- to succeed in finding analytically an exact solution
under (\ref{Newt-b}) or another (separation-dependent) condition.

In (1+1) dimensions the absence of radiation makes it easy to fix the
balance condition in terms of the determining equation (\ref{H1}). From the
relation 
\begin{equation}
\frac{\partial H}{\partial r}=-\frac{4K_{0}K_{1}K_{2}}{\kappa J}\;,
\end{equation}
the balance condition $\partial H/\partial r=0$ leads to 
\begin{equation}
K_{1}K_{2}=(2K_{0}+2K-\kappa \sqrt{p^{2}+m_{1}^{2}})(2K_{0}+2K-\kappa \sqrt{%
p^{2}+m_{2}^{2}})=0\;,  \label{bal-1}
\end{equation}
where $K\equiv K_{\pm }=\kappa X$ and $H=\frac{4}{\kappa }K$. In deriving (%
\ref{bal-1}) we excluded the possibility that $K_{0}=0$ because it gives an
unphysical solution.

It is evident from (\ref{H1}) that the condition (\ref{bal-1}) means at the
same time $M_{1}M_{2}=0$, namely, 
\begin{equation}
(2K_{0}-2K+\kappa \sqrt{p^{2}+m_{1}^{2}})(2K_{0}-2K+\kappa \sqrt{%
p^{2}+m_{2}^{2}})=0\;,  \label{bal-2}
\end{equation}
The equations (\ref{bal-1}) and (\ref{bal-2}) lead to 
\begin{equation}
4K_{0}^{2}+(2K-\kappa \sqrt{p^{2}+m_{1}^{2}})(2K-\kappa \sqrt{p^{2}+m_{2}^{2}%
})=0  \label{bal-3}
\end{equation}
and 
\begin{equation}
K_{0}(4K-\kappa \sqrt{p^{2}+m_{1}^{2}}-\kappa \sqrt{p^{2}+m_{2}^{2}})=0\;.
\label{bal-4}
\end{equation}
The solution to (\ref{bal-4}) is 
\begin{equation}
H=\sqrt{p^{2}+m_{1}^{2}}+\sqrt{p^{2}+m_{2}^{2}}  \label{bal-H}
\end{equation}
and the insertion of this solution into (\ref{bal-3}) leads to 
\begin{equation}
\frac{\kappa }{2}(\sqrt{p^{2}+m_{1}^{2}}-\epsilon \tilde{p})(\sqrt{%
p^{2}+m_{2}^{2}}-\epsilon \tilde{p})-e_{1}e_{2}=0\;.  \label{bal-p}
\end{equation}
This condition is the force-balance condition and fixes the value of
momentum $p=p_{c}=\mbox{const}$. Under the condition (\ref{bal-p}) the two
particles move with a constant velocity. The condition (\ref{bal-p}) and the
Hamiltonian (\ref{bal-H}) indicate that the full Hamiltonian must have the
simple structure 
\begin{equation}
H=\sqrt{p^{2}+m_{1}^{2}}+\sqrt{p^{2}+m_{2}^{2}}+\left\{ \frac{\kappa }{2}(%
\sqrt{p^{2}+m_{1}^{2}}-\epsilon \tilde{p})(\sqrt{p^{2}+m_{2}^{2}}-\epsilon 
\tilde{p})-e_{1}e_{2}\right\} \;F(|r|, p)\;.  \label{H-F}
\end{equation}

Actually, in the case of $\Lambda _{e}=0$ the $\kappa $-expansion of the
Hamiltonian leads to 
\begin{eqnarray}
H &=&\sqrt{p^{2}+m_{1}^{2}}+\sqrt{p^{2}+m_{2}^{2}}-\frac{1}{2}e_{1}e_{2}\;|r|
\nonumber  \label{Hcharge} \\
&&+\frac{\kappa }{4}\left\{ \sqrt{p^{2}+m_{1}^{2}}\sqrt{p^{2}+m_{2}^{2}}%
\;|r|-\epsilon \tilde{p}(\sqrt{p^{2}+m_{1}^{2}}+\sqrt{p^{2}+m_{2}^{2}}%
\;)|r|+p^{2}|r|\right.  \nonumber \\
&&\left. \makebox[2em]{}-\frac{1}{4}e_{1}e_{2}(\sqrt{p^{2}+m_{1}^{2}}+\sqrt{%
p^{2}+m_{2}^{2}}\;)r^{2}+\frac{1}{2}\epsilon e_{1}e_{2}\tilde{p}\;r^{2}+%
\frac{1}{24}(e_{1}e_{2})^{2}|r|^{3}\right\}  \nonumber \\
&&+\frac{\kappa ^{2}}{4^{2}}\left\{ \frac{1}{2\times 4}r^{2}(\sqrt{%
p^{2}+m_{1}^{2}}+\sqrt{p^{2}+m_{2}^{2}}-2\epsilon \tilde{p}-\frac{1}{2}%
e_{1}e_{2}|r|)^{3}\right.  \nonumber \\
&&\left. +\frac{1}{3\times 4^{2}}e_{1}e_{2}\;|r|^{3}(\sqrt{p^{2}+m_{1}^{2}}+%
\sqrt{p^{2}+m_{2}^{2}}-2\epsilon \tilde{p}-\frac{1}{2}e_{1}e_{2}|r|)^{2}%
\right.  \nonumber \\
&&\left. -\frac{1}{6\times 4^{2}}r^{2}\left[ 12(\sqrt{p^{2}+m_{1}^{2}}-\sqrt{%
p^{2}+m_{2}^{2}}\;)^{2}+(e_{1}e_{2})^{2}r^{2}\right] \right.  \nonumber \\
&&\left. \makebox[5em]{}\times (\sqrt{p^{2}+m_{1}^{2}}+\sqrt{p^{2}+m_{2}^{2}}%
-2\epsilon \tilde{p}-\frac{1}{2}e_{1}e_{2}|r|)\right.  \nonumber \\
&&\left. +\frac{1}{3\times 4^{2}}e_{1}e_{2}\;|r|^{3}(\sqrt{p^{2}+m_{1}^{2}}-%
\sqrt{p^{2}+m_{2}^{2}}\;)^{2}-\frac{1}{15\times 4^{3}}%
(e_{1}e_{2})^{3}|r|^{5}\right\} +{\cal {O}}(\kappa ^{3})\;.
\end{eqnarray}
{} From (\ref{Hcharge}) we get explicitly 
\begin{equation}
F(|r|, p)=\frac{1}{2}|r|+\frac{\kappa }{2}(\sqrt{p^{2}+m_{1}^{2}}+\sqrt{%
p^{2}+m_{2}^{2}}-2\epsilon \tilde{p})\;r^{2}+{\cal {O}}(|r|^{3})\;.
\end{equation}

The solution to the condition (\ref{bal-p}) exists only for{\bf \ }$%
e_{1}e_{2}>0${\bf ; 
\begin{equation}
p_{c}=\pm \frac{|\left( \frac{\kappa }{2}\right)
^{2}m_{1}^{2}m_{2}^{2}-e_{1}^{2}e_{2}^{2}|}{\sqrt{2\kappa e_{1}e_{2}}\sqrt{(%
\frac{\kappa }{2}m_{1}^{2}+e_{1}e_{2})(\frac{\kappa }{2}m_{2}^{2}+e_{1}e_{2})%
}}\;.  \label{pc}
\end{equation}
} When the particles are initially at rest $(p_{c}=0)$, the condition (\ref
{bal-p}) becomes 
\begin{equation}
\frac{\kappa }{2}m_{1}m_{2}-e_{1}e_{2}=0\;.  \label{static}
\end{equation}
This is the condition of static balance and is identical with the condition
in Newtonian theory in (1+1) dimensions. Note that in Newtonian theory (\ref
{static}) is the force-balance condition which includes both the static case
and a uniform motion. However in the relativistic case (\ref{static})
represents only the static balance condition -- the condition of
force-balance (\ref{bal-p}) in general depends on the momentum and allows a
uniform motion in the C.I. ststem in which two particles approach or recede
with the designated constant momentum (\ref{pc}).

The above result may seem to provide suggestive evidence that in (3+1)
dimensions (\ref{Newt-b}) could be a sufficient solution. However,the post-
Newtonian Hamiltonian for the system of two charged bodies in (3+1)
dimensions is given as 
\begin{eqnarray}
H &=&m_{1}+m_{2}+\frac{\mbox{\boldmath$p$}_{1}^{2}}{2m_{1}}+\frac{%
\mbox{\boldmath$p$}_{2}^{2}}{2m_{2}}-\frac{(\mbox{\boldmath$p$}_{1}^{2})^{2}%
}{8m_{1}^{3}}-\frac{(\mbox{\boldmath$p$}_{2}^{2})^{2}}{8m_{2}^{3}}  \nonumber
\label{H-Bazan} \\
&&-\frac{Gm_{1}m_{2}}{r}\left\{ 1+\frac{3}{2}\left( \frac{\mbox{\boldmath$p$}%
_{1}^{2}}{m_{1}^{2}}+\frac{\mbox{\boldmath$p$}_{2}^{2}}{m_{2}^{2}}\right) -%
\frac{7}{2}\;\frac{(\mbox{\boldmath$p$}_{1}\cdot \mbox{\boldmath$p$}_{2})}{%
m_{1}m_{2}}-\frac{(\mbox{\boldmath$p$}_{1}\cdot \mbox{\boldmath$r$})(%
\mbox{\boldmath$p$}_{2}\cdot \mbox{\boldmath$r$})}{2m_{1}m_{2}r^{2}}\right\} 
\nonumber \\
&&+\frac{e_{1}e_{2}}{4\pi r}\left\{ 1-\frac{(\mbox{\boldmath$p$}_{1}\cdot %
\mbox{\boldmath$p$}_{2})}{2m_{1}m_{2}}-\frac{(\mbox{\boldmath$p$}_{1}\cdot %
\mbox{\boldmath$r$})(\mbox{\boldmath$p$}_{2}\cdot \mbox{\boldmath$r$})}{%
2m_{1}m_{2}r^{2}}\right\} +\frac{G^{2}m_{1}m_{2}(m_{1}+m_{2})}{2r^{2}} 
\nonumber \\
&&-\frac{G(m_{1}+m_{2})e_{1}e_{2}}{4\pi r^{2}}+\frac{%
G(m_{1}e_{2}^{2}+m_{2}e_{1}^{2})}{8\pi r^{2}}\;,
\end{eqnarray}
which is derived from Ba\.{z}a\'{n}ski Lagrangian \cite{Bazan} and also the
ADM canonical formalism. The structure of the Hamiltonian (\ref{H-Bazan}) is
rather complicated compared to (\ref{H-F}). For example in (1+1) dimensional
theory the charges appear always in a combination $e_{1}e_{2}$, while in
(3+1) dimensions there is a combination of $m_{1}e_{2}^{2}+m_{2}e_{1}^{2}$
and in higher approximations more complicated combinations of mass and
charge appear \cite{OK}. It is clear that the Hamiltonian (\ref{H-Bazan})
supports the balance condition (\ref{M-P}) for the static case, while it
does not correspond to a uniform motion. What causes this difference between
(1+1) and (3+1) dimensional theories? Is it intrinsic to the dimensionality,
or is there still a possibility for satisfying the balance condition (\ref
{Newt-b}) in (3+1) dimensions? Or is there a uniform motion in the C.I.
system in (3+1) dimensions? These are interesting open problems for further
investigation.

\section{CONCLUSIONS}

Since in (1+1) dimensions the degrees of freedom of both gravitational and
electromagnetic radiation are frozen, one expects the motion of a set of $N$
charged particles in curved-spacetime with a cosmological constant to be
described by a conservative Hamiltonian. And this is what we find to be the
case. We began by canonically reducing the charged $N$-body action to
first-order form and then for a system of two charged particles derived the
determining equation of the Hamiltonian from the matching conditions. The
canonical equations of motion given by the Hamiltonian can be solved exactly
when they are expressed in terms of the proper time, and we have given the
explicit solution. To our knowledge this is the first non-perturbative
relativistic solution of this problem.

We recapitulate the main results of this paper:

\noindent (1) In (1+1) dimensions the square of the electric field plays the
same role as the consmological constant and an overall constant part is
incorporated into the effective cosmological constant $\Lambda _{e}\equiv
\Lambda _{0}-\kappa (\sum_{a}e_{a})^{2}/4$, which induces the
momentum-dependent potential in the Hamiltonian. Effectively $\Lambda _{e}>0$
acts on the particles as a repulsive potential and $\Lambda _{e}<0$ acts as
an attractive potential. \newline
(2) The theory in the case $\Lambda _{e}=0$ is a general relativistic
electrodynamics which is an extension of flat space Newtonian
electrodynamics. For the attractive charges the motion becomes bounded
similar to $|q|=0$ case with smaller amplitude and period. For  repulsive
charges the electric force between two particles competes with the
gravitational attractive force. For $|q|<q_{c}$ and a fixed energy not only
bounded motion but also an infinite series of unbounded motions is realized.
Such multiple solutions do not exist in Newtonian theory. For $|q|>q_{c}$
all motions become unbounded. \newline
(3) The condition for static balance in  (1+1) dimensional general
relativistic electrodynamics is identical with the Newtonian balance
condition in the flat space. This simple result contrasts strikingly with
the unresolved situation in (3+1) dimensions. \newline
(4) In spacetimes with nonzero $\Lambda _{e}$ the motions are more
complicated. For a fixed total energy $H_{0}$ and $\Lambda _{e}$ the
solutions are classified in terms of $(|q|,p)$ diagram in which the region
of the parameters for the bounded and/or unbounded motions are depicted in
terms of the curves of $p=\pm p_{0},{\cal {J}}_{\Lambda }^{2}=0$ and ${\cal {%
J}}_{\Lambda }-B=0$. By drawing a $q=\mbox{const}$ line on the diagram we
can easily find what types of the motion are realized. \newline
(5) For a certain range of $\Lambda _{e}<0$ and a small $|q|$ the double
peak structure appears in $r(\tau )$ and the phase space trajectory, which
is caused by a subtle interplay amongst the momentum-dependent $\Lambda $
potential, the gravitational and the electric potentials, and relativistic
effects. \newline
(6) For unbounded motion the mutual separation of two particles diverges at
finite proper time. This is common for the cases of $\Lambda _{e}=0$ and $%
\Lambda _{e}\neq 0$. \newline
(7) In the unequal mass case the basic features of the motion are the same,
but a double peak structure appears more clearly than in the equal mass case.

A number of interesting questions arise from this work. \ First, it would be
of interest to know how the features of the motion we have found are
modified as one increases the number of bodies in the system. \ For neutral
bodies this situation is relevant in modelling stellar systems and galactic
evolution, where non-relativsitic one-dimensional self-gravitating systems
are employed as tool toward understanding such dynamics. \ For charged
bodies the relevance is most likely in terms of the physics of the early
universe, where charged black holes, strings and domain walls interact in a
highly relativistic setting. \ Second, it would be interesting to study the
circumstances under which black hole event horizons can form -- we expect
this will involve investigating the regime where eq. (\ref{cosmolimit}) is
not satisfied. A third possibility would involve generalizing our work to
other (1+1) dilatonic theories of gravity. Moreover, a full quantum
treatment of the problem would also be of considerable interest. We intend
to turn our attention to these problems in the future.

\bigskip

\section*{APPENDIX A: SOLUTION OF THE METRIC TENSOR}

Under the coordinate conditions (\ref{cc}) the field equations (\ref{e-pi}),
(\ref{e-gamma}), (\ref{e-Pi}) and (\ref{e-Psi}) become 
\begin{eqnarray}
&&\dot{\pi}+N_{0}\left\{\frac{3\kappa}{2}\pi^{2} +\frac{1}{8\kappa}%
(\Psi^{\prime})^{2} +\frac{1}{4}\left(V-\frac{\Lambda_{e}}{\kappa}\right) -%
\frac{p_{1}^{2}}{2\sqrt{p_{1}^{2}+m_{1}^{2}}}\delta(x-z_{1}(t)) -\frac{%
p_{2}^{2}}{2\sqrt{p_{2}^{2}+m_{2}^{2}}}\delta(x-z_{2}(t))\right\}  \nonumber
\\
&&+N_{1}\left\{\pi^{\prime}+p_{1}\delta(x-z_{1}(t))+p_{2}\delta(x-z_{2}(t))
\right\}+\frac{1}{2\kappa}N^{\prime}_{0}\Psi^{\prime} +N^{\prime}_{1}\pi=0
\;,  \label{eq-pi1} \\
&&\kappa \pi N_{0}+N^{\prime}_{1}=0\;,  \label{eq-N1} \\
&&\partial_{1}(\frac{1}{2}N_{0}\Psi^{\prime}+N^{\prime}_{0})=0\;,
\label{eq-N0} \\
&&\dot{\Psi}+2\kappa N_{0}\pi-N_{1}\Psi^{\prime}=0\;\;.  \label{eq-Psi1}
\end{eqnarray}
The procedure to solve these equations is just the same as that in the case
of $V=0$, which is given in the Appendix of the previous paper \cite
{2bdcoslo}.

The solutions have the same form as those of $V=0$ when they are expressed
in terms of $K_{\pm}, K_{0}, K_{1,2},{\cal M}_{1,2}, Y_{\pm}$ and $J$. For $%
r=z_{1}-z_{2}>0$ the metric tensor is determined as 
\begin{eqnarray}
N_{0(+)}(x)&=&\frac{8}{J}\left(\frac{Y_{+}}{K_{+}}+\frac{Y_{-}}{K_{-}}%
\right) \frac{K_{0}K_{1}}{{\cal M}_{1}}\;e^{K_{+}(x-z_{1})}\;,  \nonumber \\
N_{0(0)}(x)&=&\frac{1}{2K_{0}J}\left(\frac{Y_{+}}{K_{+}}+\frac{Y_{-}}{K_{-}}
\right)\left[(K_{1}{\cal M}_{1})^{\frac{1}{2}} e^{-\frac{1}{2}%
K_{0}(x-z_{1})}+(K_{2}{\cal M}_{2})^{\frac{1}{2}} e^{\frac{1}{2}%
K_{0}(x-z_{2})}\right]^{2}\;,  \nonumber \\
N_{0(-)}(x)&=&\frac{8}{J}\left(\frac{Y_{+}}{K_{+}}+\frac{Y_{-}}{K_{-}}%
\right) \frac{K_{0}K_{2}}{{\cal M}_{2}}\;e^{-K_{-}(x-z_{2})}\;,  \nonumber \\
\\
N_{1(+)}&=&\epsilon\frac{Y_{+}}{K_{+}} \left\{\frac{8}{J}\left(\frac{Y_{+}}{%
K_{+}}+\frac{Y_{-}}{K_{-}}\right) \frac{K_{0}K_{1}}{{\cal M}_{1}}%
\;e^{K_{+}(x-z_{1})}-1\right\}\;,  \nonumber \\
N_{1(0)}&=&\epsilon\left\{\frac{Y_{0}}{2JK_{0}^{2}} \left(\frac{Y_{+}}{K_{+}}%
+\frac{Y_{-}}{K_{-}}\right) \left[K_{2}M_{2}\;e^{K_{0}(x-z_{2})}-K_{1}M_{1}%
\;e^{-K_{0}(x-z_{1})} \right.\right.  \nonumber \\
&&\makebox[5em]{}\left.\left.+2K_{0}(K_{1}K_{2}M_{1}M_{2})^{1/2} \;e^{\frac{1%
}{2}K_{0}(z_{1}-z_{2})}\;x\right]+D_{0}\right\}\;,  \nonumber \\
N_{1(-)}&=&-\epsilon\frac{Y_{-}}{K_{-}} \left\{\frac{8}{J}\left(\frac{Y_{+}}{%
K_{+}}+\frac{Y_{-}}{K_{-}}\right) \frac{K_{0}K_{2}}{{\cal M}_{2}}%
\;e^{-K_{-}(x-z_{2})}-1\right\}\;,  \nonumber
\end{eqnarray}
where 
\begin{eqnarray}
D_{0}&=&-\frac{1}{2}\left(\frac{Y_{+}}{K_{+}}-\frac{Y_{-}}{K_{-}}\right) +%
\frac{1}{2J}\left(\frac{Y_{+}}{K_{+}}+\frac{Y_{-}}{K_{-}}\right) \left\{ 2%
\left[\left(\frac{Y_{0}}{K_{0}}+\frac{Y_{+}}{K_{+}}\right)K_{1} -\left(\frac{%
Y_{0}}{K_{0}}+\frac{Y_{-}}{K_{-}}\right)K_{2}\right] \right.  \nonumber \\
&&\left.-2\left[\left(\frac{Y_{0}}{K_{0}}-\frac{Y_{+}}{K_{+}}\right) \frac{1%
}{{\cal M}_{1}}-\left(\frac{Y_{0}}{K_{0}}-\frac{Y_{-}}{K_{-}}\right) \frac{1%
}{{\cal M}_{2}}\right]K_{1}K_{2} -\frac{Y_{0}}{K_{0}}K_{1}K_{2}(z_{1}+z_{2})%
\right\}\;.
\end{eqnarray}
The metric tensor for $r<0$ is simply obtained by interchanging the suffix 1
and 2. With these metric tensors and the canonical equations, the field
equations (\ref{eq-pi1}), (\ref{eq-N1}) and (\ref{eq-N0}) are proved to hold
in a whole $x$ space.

As we showed in the previous paper, to satisfy (\ref{eq-Psi1}) the dilaton
field $\Psi$ needs an extra $x$-independent function $f(t)$, which has no
effect on the dynamics of particles. After lengthy calculation Eq.(\ref
{eq-Psi1}) leads to 
\begin{eqnarray}  \label{dot-f}
\dot{f}(t)&=&-\frac{d}{dt}(K_{01}z_{1}-K_{02}z_{2}) +\frac{2}{J}\left(\frac{%
Y_{+}}{K_{+}}+\frac{Y_{-}}{K_{-}}\right)\left\{ 2K_{0}K_{1}\frac{p_{1}}{%
\sqrt{p_{1}^{2}+m_{1}^{2}}} -2K_{0}K_{2}\frac{p_{2}}{\sqrt{%
p_{2}^{2}+m_{2}^{2}}}\right.  \nonumber \\
&&\left.+\epsilon\frac{Y_{0}}{K_{0}}\left(K_{0}K_{1}+K_{0}K_{2} -\frac{%
K_{0}K_{1}K_{2}}{{\cal M}_{1}}-\frac{K_{0}K_{1}K_{2}}{{\cal M}_{2}}
\right)+4\epsilon Y_{0}K_{0}\left(\frac{K_{1}}{{\cal M}_{1}}+\frac{K_{2}}{%
{\cal M}_{2}}\right)\right\}\;\;.
\end{eqnarray}
Thus $f(t)$ is uniquely determined.

\section*{APPENDIX B: LINEAR APPROXIMATION OF THE EXACT SOLUTION IN $%
\Lambda_{e}=0$}

In this appendix we investigate how the exact solution is related to its
corresponding solution in Newtonian theory. Take the tanh-type A solution,
namely a periodic motion with $e_{1}e_{2}<0$ or small $e_{1}e_{2}>0$. For
motion starting from $r=0$ at $\tau=\tau_{0}=0$, the relationship between
the energy and the initial momentum is $H_{0}=2\sqrt{p_{0}^2+m^2}$.

In the $\kappa$ expansion, $f_{e}(\tau)$ is expressed as 
\begin{eqnarray}
f_{e}(\tau)&=&\frac{\sqrt{p_{0}^2+m^2}+\epsilon p_{0}}{m}\; e^{\frac{%
\epsilon\tau}{2m}e_{1}e_{2}}  \nonumber \\
&&-\kappa m\left[\frac{\sqrt{p_{0}^2+m^2}}{e_{1}e_{2}} \left(e^{\frac{%
\epsilon\tau}{2m}e_{1}e_{2}}-1\right) -(\sqrt{p_{0}^2+m^2}+\epsilon p_{0}) 
\frac{\epsilon\tau}{4m}\;e^{\frac{\epsilon\tau}{2m}e_{1}e_{2}}\right] + 
{\cal O}(\kappa^2)
\end{eqnarray}
and $p(\tau)$ is 
\begin{eqnarray}
p&=&\sqrt{p_{0}^2+m^2}\;\mbox{sinh}\frac{e_{1}e_{2}\tau}{2m} + p_{0}\;%
\mbox{cosh}\frac{e_{1}e_{2}\tau}{2m}  \nonumber \\
&&-\frac{\kappa m^2}{2\epsilon}\left[1+\frac{(\sqrt{p_{0}^2+m^2} -\epsilon
p_{0})^2}{m^2}\;e^{-\frac{\epsilon\tau}{m}e_{1}e_{2}}\right]  \nonumber \\
&&\makebox[2em]{}\times\left[\frac{\sqrt{p_{0}^2+m^2}}{e_{1}e_{2}} \left(e^{%
\frac{\epsilon\tau}{2m}e_{1}e_{2}}-1\right) -(\sqrt{p_{0}^2+m^2}+\epsilon
p_{0}) \frac{\epsilon\tau}{4m}\;e^{\frac{\epsilon\tau}{2m}e_{1}e_{2}}\right]
+ {\cal O}(\kappa^2)\;.
\end{eqnarray}
The trajectory $r(\tau)>0$ in the linear approximation becomes 
\begin{eqnarray}  \label{linear-r}
r(\tau)&=&- \frac{4}{e_{1}e_{2}}\left[\sqrt{p_{0}^2+m^2}\left(1-\mbox{cosh}%
\frac{e_{1}e_{2}\tau}{2m}\right)-p_{0}\;\mbox{sinh}\frac{e_{1}e_{2}\tau}{2m}%
\right]  \nonumber \\
&&-\frac{4\kappa\epsilon}{e_{1}e_{2}}\left[\sqrt{p_{0}^2+m^2}\;\mbox{sinh}%
\frac{e_{1}e_{2}\tau}{2m} + p_{0}\;\mbox{cosh}\frac{e_{1}e_{2}\tau}{2m}%
\right]  \nonumber \\
&&\makebox[5em]{}\times\left[\frac{\sqrt{p_{0}^2+m^2}(\sqrt{p_{0}^2+m^2}
-\epsilon p_{0})}{e_{1}e_{2}}\left(1-e^{-\frac{\epsilon\tau}{2m}%
e_{1}e_{2}}\right) -\frac{\epsilon m\tau}{4}\right]  \nonumber \\
&&-\frac{4\kappa}{e_{1}e_{2}}\left[\sqrt{p_{0}^2+m^2}\left(1-\mbox{cosh}%
\frac{e_{1}e_{2}\tau}{2m}\right)-p_{0}\;\mbox{sinh}\frac{e_{1}e_{2}\tau}{2m}%
\right]  \nonumber \\
&&\makebox[1em]{}\times\left\{ \frac{1}{2e_{1}e_{2}}\left[\sqrt{p_{0}^2+m^2}%
\left(1-\epsilon\;\mbox{sinh}\frac{e_{1}e_{2}\tau}{2m}\right) -\epsilon
p_{0}\;\mbox{cosh}\frac{e_{1}e_{2}\tau}{2m}\right]^2 \right.  \nonumber \\
&&\makebox[5em]{}\left.-\frac{1}{6e_{1}e_{2}}\left[\sqrt{p_{0}^2+m^2}\left(1-%
\mbox{cosh}\frac{e_{1}e_{2}\tau}{2m}\right)-p_{0}\;\mbox{sinh}\frac{%
e_{1}e_{2}\tau}{2m}\right]^2 \right\}\;\;.
\end{eqnarray}

Furthermore in the limit $\kappa \rightarrow 0$ the solution is 
\begin{eqnarray}  \label{k=0}
p&=&\sqrt{p_{0}^2+m^2}\;\mbox{sinh}\frac{e_{1}e_{2}\tau}{2m} + p_{0}\;%
\mbox{cosh}\frac{e_{1}e_{2}\tau}{2m}\;,  \nonumber \\
\\
r&=&- \frac{4}{e_{1}e_{2}}\left[\sqrt{p_{0}^2+m^2}\left(1-\mbox{cosh} \frac{%
e_{1}e_{2}\tau}{2m}\right)-p_{0}\;\mbox{sinh}\frac{e_{1}e_{2}\tau}{2m}\right]%
\;\;.  \nonumber
\end{eqnarray}
We can transform it into an expression in terms of the original time
coordinate $t$, by using the relation 
\begin{equation}
t=\frac{2\sqrt{p_{0}^2+m^2}}{e_{1}e_{2}}\;\mbox{sinh}\frac{e_{1}e_{2} \tau}{%
2m} +\frac{2p_{0}}{e_{1}e_{2}}\;\left(\mbox{cosh}\frac{e_{1}e_{2} \tau}{2m}%
-1\right)
\end{equation}
which is obtained by integrating (\ref{Tau}). Then 
\begin{eqnarray}
p(t)&=&p_{0}+\frac{e_{1}e_{2}}{2}t\;, \\
r(t)&=&\frac{4}{e_{1}e_{2}}\left\{ \sqrt{(p_{0}+\frac{e_{1}e_{2}}{2}t)^2+m^2}%
-\sqrt{p_{0}^2+m^2}\right\}\;\;.
\end{eqnarray}
This solution is identical with that derived from the Hamiltonian for
flat-space electrodynamics in (1+1) dimensions: $H=2\sqrt{p^2+m^2}%
-e_{1}e_{2}|r|/2$. The same solution can be applied to the system of
arbitrary charges as far as $H_{0} \geq 2m$.

On the other hand, for the repulsive charges $(e_{1}e_{2}>0)$ there exists
the solution for $H_{0} < 2m$, which denotes an unbounded motion and is
given by 
\begin{eqnarray}
p(t)&=&\frac{e_{1}e_{2}}{2}t\;, \\
r(t)&=&\frac{4}{e_{1}e_{2}}\;\left\{\sqrt{\frac{(e_{1}e_{2})^2}{4}t^2 + m^2}
- \frac{H_{0}}{2}\right\}\;.
\end{eqnarray}
This solution is also derived from the tan-type A solution in the limit $%
\kappa\rightarrow 0$.

Another limit of the solution (\ref{linear-r}) is to take $e_{a}\rightarrow
0 $, which leads to 
\begin{equation}
r(\tau)=\frac{2p_{0}}{m}\;\tau - \kappa \frac{\sqrt{p_{0}^2+m^2}}{4}\;\tau^2
\;\;.
\end{equation}
In the non-relativistic approximation (keeping the terms to the lowest order
of $p_{0}/m$) we get the solution for the motion in Newtonian gravity in
(1+1) dimensions: 
\begin{equation}
r(t)=\frac{2p_{0}}{m}\;t- \frac{\kappa m}{4}\;t^2 \;\;.
\end{equation}

\section*{APPENDIX C: CAUSAL RELATIONSHIPS BETWEEN PARTICLES}

One of the striking features we found in the analysis of the two body motion
is that in the repulsive trajectories $N_{1},N_{2},A_{n}$ and $B_{n}$ the
particles reach the asymptotic regime ($r\rightarrow \pm \infty ,p=%
\mbox{finite}$) at some finite proper time $\tau _{\infty }$. For example
for $N_{1}$ trajectory it is 
\begin{equation}
\tau _{\infty }=\frac{4}{\kappa m\sqrt{\gamma _{m}}}\mbox{log}\left( \frac{%
H(1+\sqrt{\gamma _{m}})-\left\{ p_{+}+\sqrt{p_{+}^{2}+m^{2}}\right\} (\sqrt{%
\gamma _{m}}+\gamma _{e})}{\left[ H(1+\sqrt{\gamma _{m}})+\left\{ p_{+}+%
\sqrt{p_{+}^{2}+m^{2}}\right\} (\sqrt{\gamma _{m}}-\gamma _{e})\right] \eta }%
\right)
\end{equation}
with $p_{+}=\frac{1}{2\kappa }\left\{ \sqrt{\kappa ^{2}H^{2}+8\Lambda _{e}}+%
\sqrt{8\Lambda _{e}+8e_{1}e_{2}}\right\} $.

We can understand this feature in a simple flat-space model. Consider the
2-velocity 
\begin{equation}
u^{\mu }=(f(\sigma \tau ),\sqrt{f^{2}(\sigma \tau )-1})\quad \mbox{with}%
\quad d\tau ^{2}=dt^{2}-dx^{2}\;,
\end{equation}
where $f(\sigma \tau )$ is some function. This 2-velocity means 
\begin{equation}
\frac{dx}{dt}=\frac{\sqrt{f^{2}-1}}{f}\;,
\end{equation}
and leads to the 2-acceleration 
\begin{equation}
a^{\mu }=\frac{du^{\mu }}{d\tau }=f^{\prime }\sigma (1,\frac{f}{\sqrt{%
f^{2}(\sigma \tau )-1}})\;,
\end{equation}
where $f^{\prime }=df(\tau )/d\tau $. There exist functions $f$ such that $%
f\rightarrow \infty $ as $\tau \rightarrow \tau _{\infty}$, $\tau _{\infty }$
being finite; the particle thus becomes lightlike in a finite amount of
proper time, but an infinite amount of coordinate time $t=\int_{\tau_{i}}^{%
\tau}d\tau f(\sigma \tau )$. An example would be $f=\sec (\sigma \tau )$.
The acceleration is not constant, but increases as a function of proper
time, diverging at $\tau =\tau _{\infty }\equiv \frac{\pi }{2\sigma }$. This
is the situation we encounter for the unbounded motions of two charged
particles (or of two neutral particles with sufficiently large positive
cosmological constant).

To see the causal relationships between particles and the space-time
structure we try to pursue the path of light signal emitted from particle 2
in the metric given in Appendix A. The path $x(t)$ of the light is governed
by $d\tau =0$, which reads 
\begin{equation}
\left( \frac{dx}{dt}\right) ^{2}+2N_{1}\frac{dx}{dt}-(N_{0}^{2}-N_{1}^{2})=0%
\;.
\end{equation}
The equation of the light signal emitted inward (directed to particle 1) is 
\begin{equation}
\frac{dx}{dt}%
=N_{0}(x(t),z_{1}(t),z_{2}(t),p(t))-N_{1}(x(t),z_{1}(t),z_{2}(t),p(t))\;,
\label{eq1light}
\end{equation}
and the light emitted outward is described by 
\begin{equation}
\frac{dx}{dt}%
=-N_{0}(x(t),z_{1}(t),z_{2}(t),p(t))-N_{1}(x(t),z_{1}(t),z_{2}(t),p(t))\;.
\label{eq2light}
\end{equation}

Let's take first a typical bounded motion in the case of $\Lambda _{e}=0$
and the attractive charges for the parameters of $H_{0}=3,m=1$ and $q=1$. In
Fig.39 the trajectories of light signals emitted from particle 2 at various
times $T$ are plotted. The two particles are always in causal contact,
because the inward light signal from particle 2 reaches particle 1. We see
one striking feature for the case of $\Lambda _{e}=0$, namely that the
outward light pulse has a constant velocity $c$, since the equations (\ref
{eq1light}) and (\ref{eq2light}) become $dx/dt=\pm 1$, where the plus and
the minus signs correspond to the light emitted outward from the particle 1
and 2, respectively.
\begin{figure}
\begin{center}
\epsfig{file=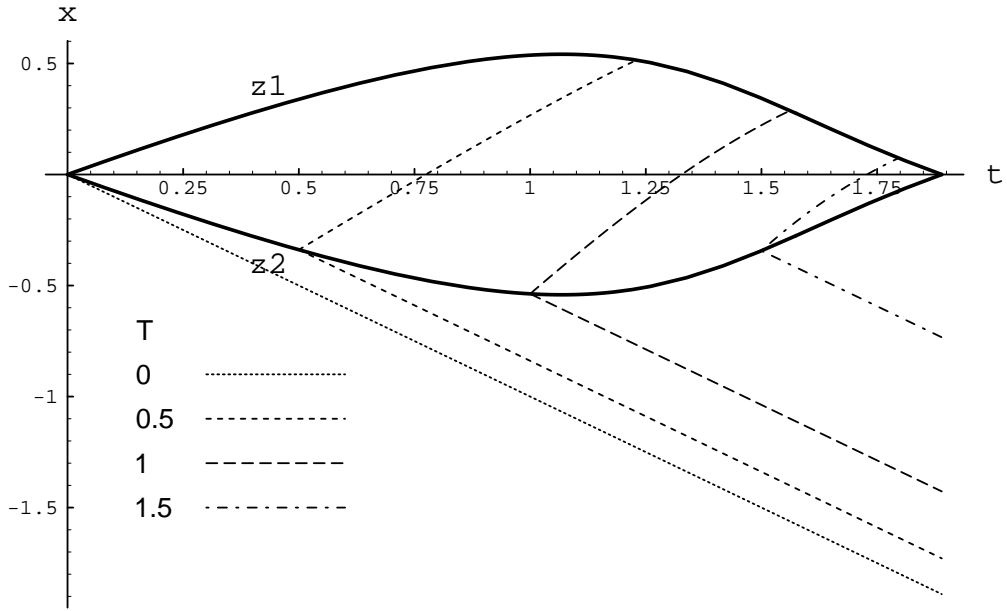,width=0.8\linewidth}
\end{center}
\caption{The trajectories of light signals emitted from the particle 2 for the
case of $\Lambda_{e}=0$.}
\label{fig39}
\end{figure}

Next we look into the effect of the cosmological constant on the path of the
light signal. Fig.40 shows the trajecories of light for the case of a
negative cosmological constant $\Lambda _{e}=-1$ and the same values of the
parameters $H_{0},m$ and $q$. The outward light behaves as if it is
subjected to a repulsive force, in significant contrast to the particles'
trajectories on which the cosmological constant acts effectively as an
attractive force as we analyzed in Sec.VI. On the other hand, while the
positive cosmological constant causes a repulsive force between particles,
the outward light behaves as if it undergoes an attractive force and
approaches a kind of horizon given by the line $%
N_{0}(x,z_{1}(t),z_{2}(t),p(t))\pm N_{1}(x,z_{1}(t),z_{2}(t),p(t))=0$ which
is shown in Fig.41 as a narrow solid line for the case of $\Lambda _{e}=0.5$.
\begin{figure}
\begin{center}
\epsfig{file=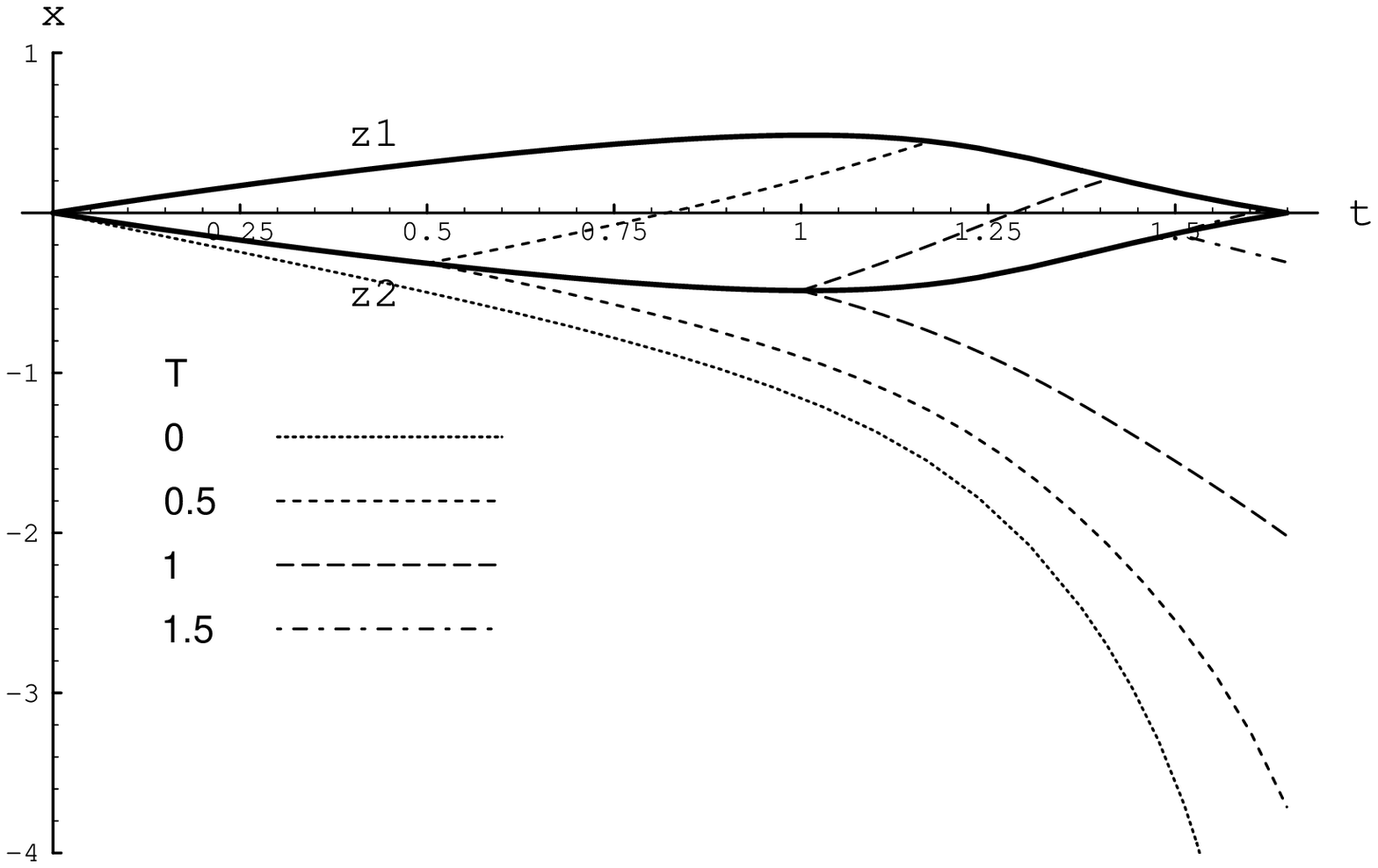,width=0.8\linewidth}
\end{center}
\caption{The trajectories of light signals emitted from the particle 2 for the
case of $\Lambda_{e}=-1$.}
\label{fig40}
\end{figure}
\begin{figure}
\begin{center}
\epsfig{file=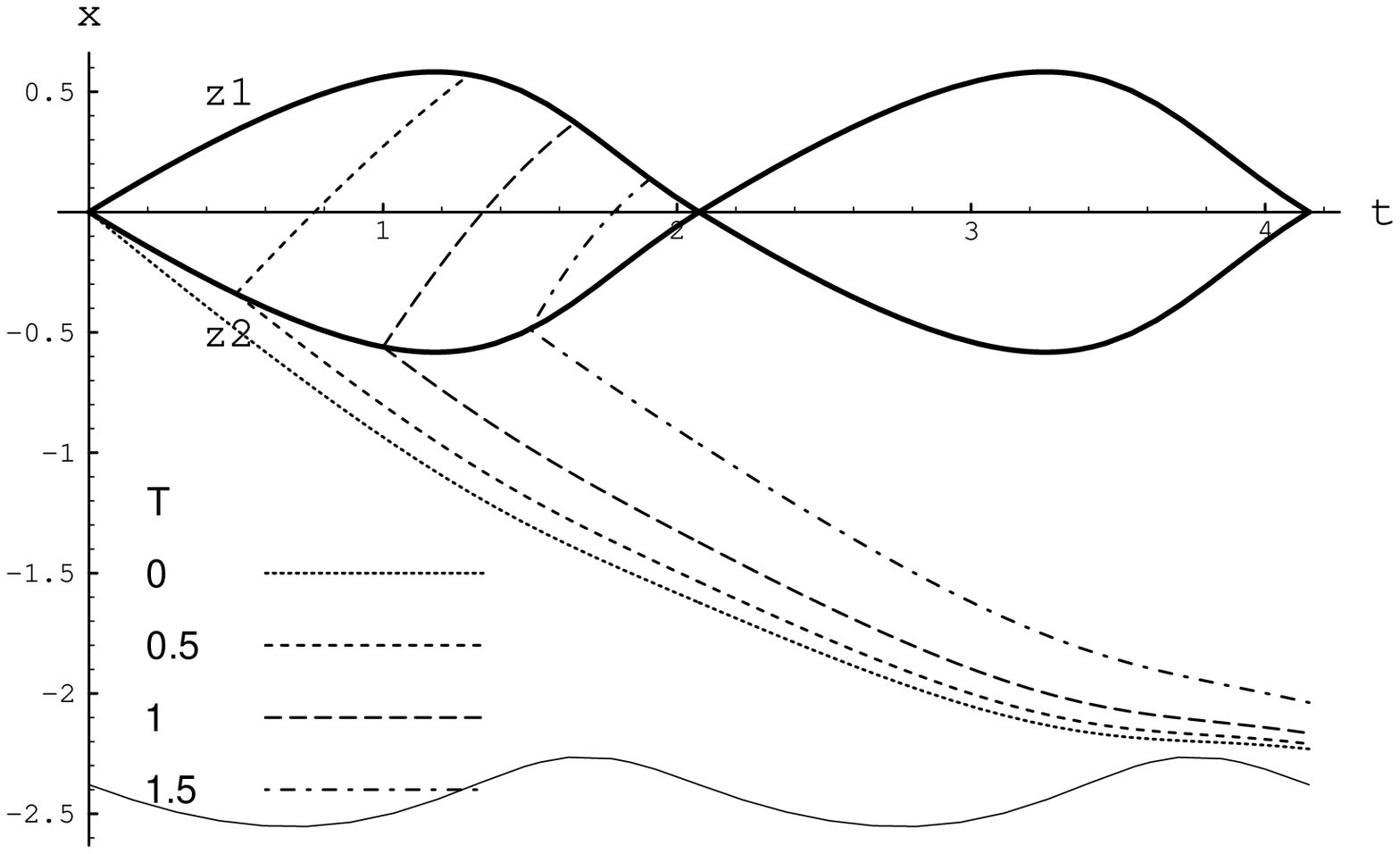,width=0.8\linewidth}
\end{center}
\caption{The trajectories of light signals emitted from the particle 2 for the
case of $\Lambda_{e}=0.5$. The narrow solid line denotes $N_{0}(x, z_{1}(t),
z_{2}(t), p(t)) + N_{1}(x, z_{1}(t),z_{2}(t), p(t))=0$.}
\label{fig41}
\end{figure}

When the cosmological constant exceeds a critical value, the particles'
motion becomes unbounded and the light signals emitted from the particle 2
exhibit new characteristics as shown in Fig.42 for the case of $\Lambda
_{e}=2.5$. For small $T(T<0.9)$, the particles are in causal contact (the
inward dotted curve), but for $T\approx 0.9$ the signal just barely catches
up with particle 1, which is almost in light-like motion (the inward dashed
curve). For $T=2$ the inward world line $x(t)$ is parallel to $z_{1}(t)$ at
large $t$ and in the outward direction it goes nearly on the same trajectory
with the particle 2. For large $T$ ($T>2$) the particles are out of causal
contact with each other : a light ray sent from particle 2 toward particle 1
receives a strong repulsive effect and ultimately reverses direction,
following behind particle 2.
\begin{figure}
\begin{center}
\epsfig{file=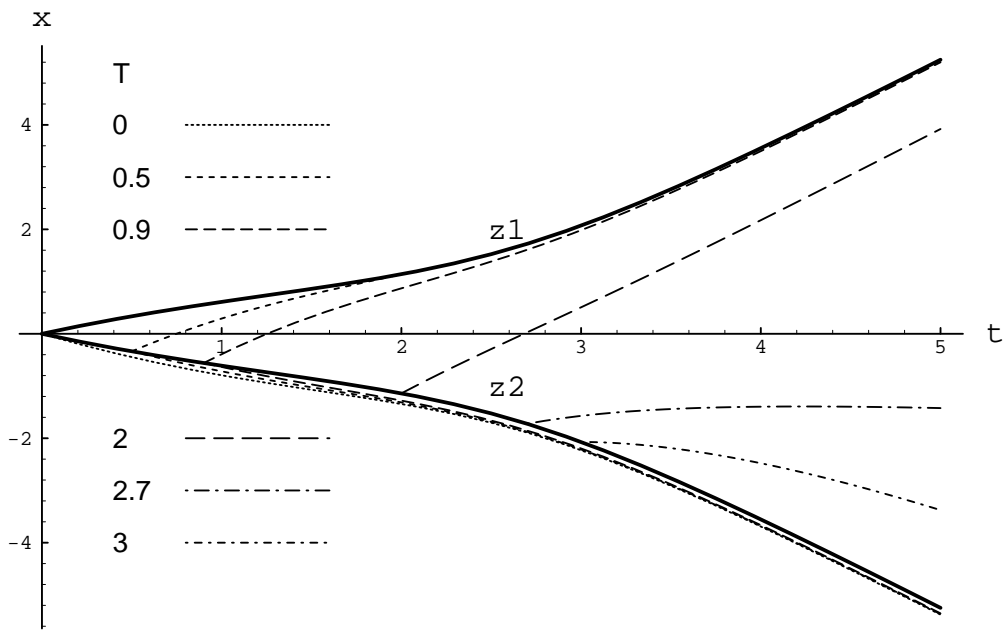,width=0.8\linewidth}
\end{center}
\caption{The trajectories of light signals emitted from the particle 2 for the
case of $\Lambda_{e}=2.5$.}
\label{fig42}
\end{figure}

{\bf \bigskip Acknowledgements}

This work was supported in part by the Natural Sciences and Engineering\
Research Council of Canada.

\end{document}